\documentclass[a4paper,onecolumn,11pt,unpublished]{quantumarticle}
\pdfoutput=1
\usepackage[utf8]{inputenc}
\usepackage[english]{babel}
\usepackage[T1]{fontenc}
\usepackage{amsmath}

\usepackage{tikz}
\usepackage{lipsum}

\usepackage[numbers]{natbib}
\usepackage{amssymb}
\usepackage{amsthm}
\usepackage{graphicx}
\usepackage{hyperref}
\usepackage{braket}
\usepackage{color}
\usepackage{bm}
\usepackage{float}
\usepackage{qcircuit}
\usepackage{subcaption}
\usepackage{longtable}
\usepackage[toc]{multitoc}
\makeatletter
\newcount\col@number
\makeatother

\newtheorem{theorem}{Theorem}
\newtheorem{lemma}{Lemma}

\theoremstyle{definition}
\newtheorem{problem}{Problem}
\newtheorem{assumption}{Assumption}
\newtheorem{remark}{Remark}

\newcommand{\Ex}{\mathbb{E}}

\newcommand{\D}{\mathcal{D}}
\newcommand{\F}{\mathcal{F}}
\newcommand{\E}{\mathcal{E}}

\newcommand{\R}{\mathbb{R}}

\newcommand{\Z}{\mathcal{Z}}
\newcommand{\cj}{^\dagger}
\newcommand{\RE}{\text{Re}}
\newcommand{\IM}{\text{Im}}

\newcommand{\diag}{\text{diag}}
\newcommand{\dist}{\text{dist}}
\newcommand{\pmin}{\bar{p}_{\text{min}}}
\newcommand{\ptail}{\bar{p}_{\text{tail}}}

\begin{document}

\title{Quantum Spectral Estimation with Multiple Initial States: Improved Error and Cost Scaling}

\author{Lingwei Zhang}
\email{zlwum@umich.edu}
\author{Zhiyan Ding}
\email{zyding@umich.edu}
\affiliation{Department of Mathematics, University of Michigan, Ann Arbor, MI 48109, USA}

\maketitle

\begin{abstract}
We study quantum spectrum and eigenstate property estimation using multiple initial states and generalized Hadamard tests. Building on the QFAMES (quantum filtering and analysis of multiplicities in eigenvalue spectra) framework, we introduce a new method that samples evolution times independently for each initial-state pair. Under the stated assumptions, when statistical error determines the sampling cost, the total Hamiltonian simulation cost required for fixed-accuracy eigenvalue and observable estimation is independent of the number of initial states, up to logarithmic factors. This improves scalability when many initial states are needed to explore the target eigenspaces. The method also reduces the cost of multiplicity recovery and retains Heisenberg-limited scaling for eigenvalue estimation. Numerical experiments support the predicted statistical error scalings.
\end{abstract}

\renewcommand{\contentsname}{Outline}
\setcounter{tocdepth}{2}
\begingroup
\small
\tableofcontents
\endgroup
\clearpage


\section{Introduction}

Given a Hamiltonian $H$, characterizing its spectrum and the physical properties of its eigenstates constitutes a fundamental task in quantum physics and quantum computation. Spectral information has a wide range of applications. For example, the density of states determines the partition function and hence equilibrium thermodynamic quantities \cite{SummerChiaracaneMitchisonGoold2024DOS,SchrodiSilviTschirsichFazioMontangero2017DOS}, while eigenvalue multiplicities, and in particular ground-state degeneracies, provide important signatures of quantum phases and topological order \cite{Wen2013TopologicalOrder,Wen2017TopologicalPhases}. In quantum chemistry, excited-state energies and observables are needed to predict optical spectra, reaction processes, and energy- and charge-transfer phenomena \cite{SantagatiEtAl2018Eigenstates}. Moreover, expectation values and matrix elements of observables in individual many-body eigenstates form the basis of the eigenstate thermalization hypothesis and the study of quantum chaos \cite{DAlessioKafriPolkovnikovRigol2016ETH}. Two fundamental spectral characterization problems are of particular interest:

\begin{itemize}
\item \emph{Density-of-states (DOS) estimation}: estimating the spectral density of $H$, including its eigenvalues and their multiplicities.
\item \emph{Eigenstate characterization}: estimating observables associated with an eigenstate or, more generally, with a degenerate eigenspace.
\end{itemize}

Although these two problems are fundamental in quantum physics, extracting the relevant information from a quantum system is challenging because the dimension of the Hilbert space grows exponentially with the system size. For example, DOS estimation has been shown to be $\#\mathrm{BQP}$-complete—the counting analogue of QMA—in the worst case~\cite{BrownFlammiaSchuch2011DOS}. Thus, efficient algorithms can generally be expected only under additional assumptions on the physically relevant Hamiltonians and the available initial states.

Recently, a quantum algorithm called quantum filtering and analysis of multiplicities in eigenvalue spectra (QFAMES)~\cite{QFAMES} was proposed to address these two questions. Similar to previous quantum phase and eigenvalue estimation algorithms~\cite{QMEGS,DingLin2023,Ding2023simultaneous,ni2023lowdepth_2,ni2023lowdepth}, QFAMES can efficiently estimate the dominant eigenvalues associated with the initial states, namely, eigenvalues whose corresponding eigenspaces have significant overlap with the initial states. More importantly, QFAMES can also provably and efficiently estimate the multiplicities of these dominant eigenvalues, as well as expectation values of observables restricted to the corresponding degenerate eigenspaces. These additional capabilities substantially broaden its scope of applicability compared with previous eigenvalue estimation algorithms.

Let $H=\sum_{m=1}^M \lambda_m |E_m\rangle \langle E_m|$ be the spectral decomposition of $H$, and let $\{\ket{\phi_l}\}_{l\in[L]}$ and $\{\ket{\psi_r}\}_{r\in[R]}$ be two sets of initial states. For each eigenstate $\ket{E_m}$, define its aggregate overlap amplitude with the initial states by
\begin{equation}
p_m
:=
\left(\sum_{l=1}^L \left|\langle \phi_l|E_m\rangle\right|^2\right)^{1/2}
\left(\sum_{r=1}^R \left|\langle \psi_r|E_m\rangle\right|^2\right)^{1/2}.
\end{equation}
We denote by $\mathcal{D}$ the set of dominant indices, characterized by the dominance condition $ p_{\min}:=\min_{m\in\mathcal{D}}p_m
\gtrsim
p_{\mathrm{tail}}:=\sum_{m\notin\mathcal{D}}p_m; $ see Section~\ref{sec:problem_setup} for a precise formulation. Under this condition, QFAMES can efficiently perform the following three tasks:
\begin{enumerate}
\item Estimate the dominant eigenvalues $\{\lambda_m\}_{m\in\mathcal{D}}$;

\item Estimate the multiplicity of each dominant eigenvalue, including the case of near-degenerate eigenvalues satisfying $\lambda_m\approx\lambda_{m'}$ for distinct $m,m'\in\mathcal{D}$;

\item Estimate expectation values of observables restricted to the corresponding degenerate or near-degenerate eigenspaces.
\end{enumerate}
Although QFAMES achieves favorable complexity in terms of the total Hamiltonian simulation time, its dependence on the numbers of initial states, $L$ and $R$, remains unsatisfactory. In particular, for the first task of estimating the dominant eigenvalues, the complexity upper bound established in~\cite{QFAMES} contains a factor of $LR$ as the numbers of initial states increase. This dependence can become a significant bottleneck, since using a large collection of initial states is often necessary to ensure sufficient overlap with, and hence adequate exploration of, the target eigenspaces. Indeed, as demonstrated in~\cite[Proposition~C1]{QFAMES}, sufficiently many initial states may be required to guarantee that the relevant eigenspaces are well represented. This inspires us to ask the following question, which is the focus of this work:

\emph{How can we design a more efficient algorithm with a significantly improved scaling in the numbers of initial states? }

In this work, we answer this question by introducing a new data-generation strategy that substantially reduces the cost of QFAMES and improves its error scaling with respect to the numbers of initial states, $L$ and $R$. We sample evolution times independently for each initial-state pair, whereas the original QFAMES algorithm shares the same times across all pairs. The theoretical results established in this work can be summarized as follows:

\begin{itemize}

\item For observable estimation with $L=R$, the total Hamiltonian simulation cost bound for achieving $\epsilon$-accuracy is independent of $L$ and $R$, up to logarithmic factors, in the idealized parameter regime given by Theorem~\ref{thm:observable_informal}.

More specifically, with the other algorithmic parameters fixed and $L=R$, the statistical term in our observable error bound scales as $1/\sqrt{LR}$, up to logarithmic factors, whereas the corresponding shared-sampling bound does not decrease with $L$ and $R$.

\item For the eigenvalue estimation task, the cost of achieving $\epsilon$-accuracy can also be made independent of $L$ and $R$ in practically relevant scenarios.

\item For multiplicity estimation, our total Hamiltonian simulation cost bound depends linearly on $\max(L,R)$, up to logarithmic factors, in the parameter regime of Theorem~\ref{thm:multiplicity_informal}, compared with the $LR$ dependence of the original QFAMES bound.

\end{itemize}

In the parameter regimes specified in Section~\ref{sec:main_results}, increasing the numbers of initial states need not increase the total Hamiltonian simulation cost of fixed-accuracy eigenvalue and observable estimation. Independent sampling reduces the statistical error as $L$ and $R$ grow, allowing the number of time samples per initial-state pair to decrease and offset the cost of measuring more pairs. Moreover, a larger collection of initial states provides richer information about the target degenerate eigenspaces, which can further improve multiplicity and observable estimation. Our numerical experiments support the predicted statistical error trends and illustrate how independent time sampling can improve the practical performance of QFAMES.

The remainder of the paper is organized as follows. In Section~\ref{sec:problem_setup}, we formally introduce the problem setup and the underlying assumptions. In Section~\ref{sec:preliminary}, we review the QFAMES framework, introduce the necessary notation, and state the assumptions used in our analysis. In Section~\ref{sec:main_results}, we present our main theoretical results, including the improved error scaling with respect to the numbers of initial states. An informal analysis is provided in Section~\ref{sec:intuitive_analysis} to give intuition for the main results. In Section~\ref{sec:numerical_results}, we provide numerical experiments that support the predicted statistical error trends. Finally, in Section~\ref{sec:discussion}, we conclude the paper with a discussion of potential future directions.


\subsection{Related work}

Quantum phase estimation is a fundamental problem in quantum computation and a central tool for estimating Hamiltonian eigenvalues. Foundational QPE algorithms~\cite{Kitaev1995QPE,DobsicekEtAl2007QPE} achieve Heisenberg-limited scaling: for a fixed success probability and a fixed nonzero overlap with the target eigenstate, the total Hamiltonian evolution time scales as $\widetilde{\mathcal{O}}(\epsilon^{-1})$ for energy precision $\epsilon$, which is optimal in its dependence on precision up to logarithmic factors~\cite{Giovannetti2011}. Recent algorithms build on the single-ancilla Hadamard-test framework by combining carefully chosen evolution times with classical signal processing to reduce circuit requirements while retaining this scaling. Quantum complex exponential least squares (QCELS)~\cite{DingLin2023} and robust phase estimation~\cite{ni2023lowdepth} achieve Heisenberg-limited total simulation cost, with a reduced maximum evolution time when the initial state has sufficiently large overlap with the target eigenstate.

For simultaneous estimation of several eigenvalues, MM-QCELS fits a sum of complex exponentials to the measured signal~\cite{Ding2023simultaneous}, while adaptive robust multiple-phase estimation combines signal processing with successive amplification of the evolution time~\cite{ni2023lowdepth_2}. Quantum multiple eigenvalue Gaussian filtered search (QMEGS)~\cite{QMEGS} instead locates peaks in a Gaussian-filtered signal; it achieves Heisenberg-limited scaling without a spectral gap assumption and permits further depth reductions under suitable gap and overlap conditions. A related filtering approach was previously proposed in~\cite{zeng2022universalquantumalgorithmiccooling}, where Fourier representations of cooling functions, including Gaussian filters, are combined with real-time evolution and classical postprocessing to estimate eigenenergies and eigenstate observables. For eigenenergy estimation, these methods typically use the scalar correlation $\langle\psi|e^{-iHt}|\psi\rangle$, whose spectral weights aggregate the overlaps within each eigenspace and therefore do not, in general, determine its dimension. QFAMES~\cite{QFAMES} extends Gaussian filtering to matrices of transition amplitudes between two families of initial states, enabling multiplicity and eigenspace-observable estimation under appropriate overlap and separation assumptions. The present work develops this framework by improving its dependence on the sizes of these families.

An important class of quantum eigenvalue estimation methods is based on subspace diagonalization. Examples include quantum filter diagonalization~\cite{ParrishMcMahon2019QFD}, multireference quantum Krylov methods~\cite{StairHuangEvangelista2020Krylov}, and the nonorthogonal quantum eigensolver~\cite{BaekEtAl2023NOQE}. These methods construct a subspace from trial states or their time-evolved counterparts, measure Hamiltonian and overlap matrix elements on a quantum computer, and solve a generalized eigenvalue problem classically to approximate ground- and excited-state energies. Nearly linearly dependent basis states can make the overlap matrix ill-conditioned and amplify measurement errors; truncation-based regularization and its error guarantees have been studied in~\cite{EpperlyLinNakatsukasa2022QSD,Kirby2024KrylovErrors}.

Another closely related quantum eigenvalue estimation approach is observable dynamic mode decomposition (ODMD), introduced by~\cite{ShenEtAl2025ODMD}. ODMD arranges measurements of time-evolution overlaps into two time-shifted Hankel matrices, fits a linear evolution map by least squares, and extracts energies from its eigenphases. Singular-value truncation regularizes the fit, providing a practical means of suppressing noise. Its interpretation as an approximation of the Koopman operator on a space of observables connects quantum spectral estimation with classical dynamical systems and subspace methods. Shen et al. subsequently proposed multiobservable dynamic mode decomposition (MODMD)~\cite{ShenEtAl2026MODMD}, which uses classical shadows to estimate many observable-dependent time signals and enlarge the accessible signal subspace. This extension improves access to excited-state energies and establishes spectral approximation guarantees. MODMD illustrates how richer measurement data can improve spectral reconstruction with modest additional sampling cost under its shadow-measurement assumptions. Our work studies a complementary resource question: how the choice of evolution times for each pair of initial states, together with the number of initial states, affects the statistical error and Hamiltonian simulation cost of recovering eigenvalues, multiplicities, and observables within degenerate eigenspaces.

\section{Problem Setup}\label{sec:problem_setup}

In this section, we formally introduce the setup for the DODS estimation problem and the observable measurement problem in the degenerate eigenspace. Let $H$ be a Hamiltonian with the spectral decomposition
\begin{equation}\label{eq:H_decomp}
    H = \sum_{m=1}^M \lambda_m |E_m\rangle \langle E_m|.
\end{equation}
Assume black-box access to controlled time evolution $e^{-iHt}$ for any $t \in \R$, as well as two families of (controlled) state-preparation unitaries $\{U_l\}_{l \in [L]}$ and $\{V_r\}_{r \in [R]}$ satisfying $U_l |0\rangle=\ket{\phi_l}$ and $V_r |0 \rangle = \ket{\psi_r}$. For the set of initial states $\{\ket{\phi_l}\}_{l \in [L]}$ and $\{\ket{\psi_r}\}_{r \in [R]}$, we define overlap matrices $\Phi \in \mathbb{C}^{L \times M}$ and $\Psi \in \mathbb{C}^{R \times M}$ element-wise as
\begin{equation}\label{eq:defn_Phi_Psi}
\begin{aligned}
    &\Phi_{l,m} := \langle \phi_l | E_m \rangle \quad \forall l \in [L], m \in [M], \\
    &\Psi_{r,m} := \langle \psi_r | E_m \rangle \quad \forall r \in [R], m \in [M].
\end{aligned}
\end{equation}
For each eigenstate $|E_m\rangle$, its overlap amplitude with the initial states is characterized by
\begin{equation}\label{eq:defn_p_m}
    p_m := \|\Phi_{:,m}\|_2\|\Psi_{:,m}\|_2.
\end{equation}
The multiset of \textit{dominant eigenvalues}, indexed by $\D$, consists of those eigenstates that satisfy the \textit{dominance condition},
\begin{equation}\label{eq:defn_dom_cond}
    \left( p_{\text{min}}:= \min_{m \in \D} p_m \right) \gtrsim \left( p_{\text{tail}} := \sum_{m \notin \D} p_m \right).
\end{equation}
We group the dominant eigenvalues into $I$ well-separated clusters, with eigenstates in cluster $i$ indexed by $\D_i$. Its spectral interval is $\mathcal{I}_i := [\min_{m \in \mathcal{D}_i} \lambda_m, \, \max_{m \in \mathcal{D}_i} \lambda_m]$, and $\lambda_i^*$ denotes its midpoint. We require each cluster to contain all eigenstates whose eigenvalues lie in its spectral interval, i.e., $\mathcal D_i=\{m\in[M]:\lambda_m\in\mathcal I_i\}$. Accordingly, $\D = \bigcup_{i \in [I]} \D_i$ denotes the index set of all dominant eigenstates, and $\D^c$ denotes its complement. The cluster sizes satisfy $|\D|=\sum_{i \in [I]} |\D_i|$. For an exactly degenerate cluster, $\lambda_i^*$ is its eigenvalue and $|\D_i|$ is its multiplicity.

\begin{problem}[Density of Dominant Eigenstates (DODS) Estimation]\label{prob:dods}
With high probability, for each $i \in [I]$, estimate the dominant cluster center $\lambda_i^*$ within $\epsilon$-accuracy and exactly determine its cluster size $|\D_i|$.
\end{problem}

Let $O$ be an observable. For a target dominant eigenvalue cluster $\mathcal{D}_i$ centered at $\lambda_i^*$ with cluster size $|\D_i|$, we define the projected observable matrix restricted to this (nearly) degenerate subspace as:
\begin{equation}\label{eq:defn_proj_observable}
    O_{\D_i} := \big( \langle E_k | O | E_{k'} \rangle \big)_{k,k' \in \mathcal{D}_i}.
\end{equation}

\begin{problem}[Degenerate Eigenstate Property Estimation]
\label{prob:observable}
With high probability, for each $\D_i$, estimate the eigenvalues $\lambda_k^O$ of $O_{\D_i}$ within $\epsilon_O$-accuracy.
\end{problem}

To successfully resolve these two problems, our algorithms rely on a few structural assumptions regarding the physical system and the initial states, which are formally defined in Section~\ref{subsec:assumptions}. Intuitively, we require the following:
\begin{itemize}
    \item \textbf{Cluster Separation (Assumption~\ref{assum:cluster_sep}):} We assume well-separated target dominant eigenvalue clusters to control cross-cluster interference.
    \item \textbf{Dominant Signal (Assumption~\ref{assum:dom_threshold}):} We assume a sufficiently strong dominant signal relative to the background tail residuals.
    \item \textbf{Uniform Overlap (Assumption~\ref{assum:uniform_overlap}):} The set of initial states should relatively uniformly explore the target eigensubspace, ensuring that no specific eigen-direction is hidden or missed.
\end{itemize}

\begin{remark}
    In practice, satisfying the uniform overlap condition and ensuring that the initial states fully span the degenerate eigensubspace often necessitates a large number of initial states (i.e., large $L$ and $R$). This practical requirement is the primary motivation for our work: we must rigorously analyze how the estimation error scales with respect to the number of initial states, which we will address in the subsequent sections.
\end{remark}


\section{Preliminaries and Notation}\label{sec:preliminary}

\subsection{QFAMES Framework}\label{subsec:QFAMES_framework}

In this section, we briefly review the QFAMES algorithms for Problems~\ref{prob:dods} and~\ref{prob:observable}. For a detailed step-by-step implementation, we refer the reader to Appendix~\ref{app:review}.

\paragraph{Step 1: Data generation.} To extract spectral information, we first construct a filtered data matrix $G(\theta)$ that approximates the ideal noiseless signal $\F(\theta)$, defined as:
\begin{equation}\label{eq:defn_F}
    \F(\theta) := \Phi \cdot \diag \left( \left\{ \exp\left(-(\theta - \lambda_m)^2 T^2\right) \right\}_{m \in [M]} \right) \cdot \Psi\cj.
\end{equation}
Here, $T$ is the filter time scale, which is comparable with the maximal evolution time used in the Hamiltonian simulations. Specifically, for each pair $(l,r)\in [L]\times [R]$, we approximate
\begin{equation}\label{eqn:defn_G}
\begin{aligned}
\mathcal{F}_{l,r}(\theta)&=\bra{\phi_l} e^{-(\theta-H)^2 T^2} \ket{\psi_r} \approx \frac{1}{N}\sum^N_{n=1} \bra{\phi_l} \exp(i\left(\theta-H\right)t_{l,r,n}) \ket{\psi_r}\\
&\approx G_{l,r}(\theta):=\frac{1}{N}\sum^N_{n=1} Z_{l,r}(t_{l,r,n})\exp(i\theta t_{l,r,n}),
\end{aligned}
\end{equation}
where $N$ is the number of time samples per initial-state pair, $\{t_{l,r,n}\}_{n=1}^N$ are the random Hamiltonian evolution times sampled from a proper distribution and the random variable $Z_{l,r}(t_{l,r,n})$ is generated by generalized Hadamard tests satisfying $\Ex[Z_{l,r}(t_{l,r,n})\mid t_{l,r,n}] = \bra{\phi_l} e^{-iHt_{l,r,n}} \ket{\psi_r}$. In~\cite{QFAMES}, the Hamiltonian evolution times $\{t_n\}_{n=1}^N$ are universally shared across all initial-state pairs $(l,r)$, i.e., $t_{l,r,n} = t_n$ for all $l \in [L]$ and $r \in [R]$, where $\{t_n\}_{n=1}^N$ are independently sampled from the truncated Gaussian distribution $a_T^{\mathrm{trunc}}$ defined in Eq.~\eqref{eq:truncated_gaussian_density}.

\paragraph{Step 2: Eigenvalue estimation.} From Eq.~\eqref{eq:defn_F}, it is evident that the dominant eigenvalues $\{\lambda_i^*\}_{i \in [I]}$ correspond to the approximate local maxima of $\|\F(\theta)\|_F$, and thus the peaks of $\|G(\theta)\|_F$ are empirical estimators for these dominant eigenvalues.

We locate the dominant eigenvalues by iteratively searching for the maximum peaks of $\|G(\theta)\|_F$ and blocking their surrounding neighborhoods to avoid redundant discoveries. This search and block method was first introduced in~\cite{QMEGS} and later extended in~\cite{QFAMES} to the matrix setting. This provides a set of candidate peaks $\{\theta^*_i\}_{i=1}^{\tilde{I}}$ corresponding to the set of dominant eigenvalues $\{\lambda_i^*\}_{i \in [I]}$.

\paragraph{Step 3: Multiplicity estimation.} For each identified peak $\theta_i^*$, $G(\theta_i^*)\approx \mathcal{F}(\lambda_i^*)$ is expected to suppress contributions from other eigencomponents and isolate, across the initial states, the components associated with the dominant cluster $\mathcal{D}_i$. The effective rank of $G(\theta_i^*)$ then provides an empirical estimate of the dimension of the dominant eigenspace captured by the initial states, and thus serves as an approximation to the multiplicity $|\mathcal{D}_i|$.

To determine the multiplicity $|\mathcal{D}_i|$, we perform the singular value decomposition
\begin{equation}\label{eq:defn_G_SVD}
    G(\theta_i^*)=U_i \Sigma_i V_i\cj
\end{equation}
and define
\begin{equation}
    m_i=\# \{j \mid \left(\Sigma_i\right)_{j,j} > \tau\}
\end{equation}
for some given threshold $\tau$, which gives the estimated multiplicity associated with $\theta_i^*$. If $m_i=0$, then $\theta_i^*$ corresponds to a spurious dominant eigenvalue and should be discarded.

\paragraph{Step 4: Observable estimation.} To extract the expectation values of a physical observable $O$, we construct an observable-weighted data matrix $G^O(\theta)$. Similar to $G(\theta)$, $G^O(\theta)$ approximates the ideal observable signal $\F^O(\theta)$, defined as:
\begin{equation}\label{eq:defn_F_O}
    \F^O(\theta) := \Phi \cdot \diag\big(\{e^{-(\theta-\lambda_k)^2 T^2 / 2}\}_{k \in [M]}\big) \cdot O_{[M]} \cdot \diag\big(\{e^{-(\theta-\lambda_{k'})^2 T^2 / 2}\}_{k' \in [M]}\big) \cdot \Psi\cj,
\end{equation}
Similar to $\mathcal{F}_{l,r}(\theta)$, each entry of $\F^O(\theta)$ can be estimated using generalized Hadamard tests with the observable $O$ inserted in the middle of the circuit. Thus, $G^O(\theta^*_i)$, as an approximate estimator for $\F^O(\lambda_i^*)$, is expected to isolate the observable information associated with the dominant cluster $\mathcal{D}_i$. The eigenvalues of the projected observable $O_{\D_i}$ can then be estimated by solving the generalized eigenvalue problem between a truncated version of $G^O(\theta^*_i)$ and $G(\theta^*_i)$.


Recall that we have obtained the SVD $G(\theta_i^*)=U_i \Sigma_i V_i\cj$ from Eq.~\eqref{eq:defn_G_SVD} and can thus restrict the matrices $G(\theta_i^*)$ and $G^O(\theta_i^*)$ to the large singular value subspace with cutoff threshold $\tau$ to get rid of the spurious singular directions. This yields the truncated matrices:
\begin{equation}\label{eq:defn_G_tilde}
    \tilde{G}(\theta_i^*) = (U_i')^\dagger \cdot G(\theta_i^*) \cdot V_i'=\Sigma_i',
\end{equation}
\begin{equation}\label{eq:defn_G_O_tilde}
    \tilde{G}^O(\theta_i^*) = (U_i')^\dagger \cdot G^O(\theta_i^*) \cdot V_i',
\end{equation}
where we denote $U_i':=(U_i)_{:,[m_i]}$, $V_i':=(V_i)_{:,[m_i]}$ and $\Sigma_i':=(\Sigma_i)_{[m_i],[m_i]}$ for simplicity. The rest is to solve the generalized eigenvalue problem:
\begin{equation}\label{eq:gen_eigenvalue_empirical}
    \tilde{G}^O(\theta^*_i) w = \lambda^O \tilde{G}(\theta^*_i) w.
\end{equation}

As established in~\cite[Section V]{QFAMES}, if we ignore noise, tail contributions, and contributions from other clusters, when restricted to the dominant cluster $\mathcal{D}_i$, the overlap matrices can be factorized using the empirical singular vectors and invertible matrices $P_U$ and $P_V$, such that $\Phi_{:,\mathcal{D}_i} = U_i' P_U$ and $\Psi_{:,\mathcal{D}_i} = V_i' P_V$. Consequently, the projected matrices take the approximate forms $\tilde{G}^O(\theta_i^*) \approx P_U D O_{\mathcal{D}_i} D P_V^\dagger$ and $\tilde{G}(\theta_i^*) \approx P_U D^2 P_V^\dagger$, where $D = \diag(\{e^{-(\theta_i^* - \lambda_k)^2 T^2 / 2}\}_{k \in \mathcal{D}_i})$. Solving the generalized eigenvalue problem in Eq.~\eqref{eq:gen_eigenvalue_empirical} effectively cancels out the invertible transformations $P_U D$ and $D P_V^\dagger$, ensuring that the generalized eigenvalues $\lambda^O$ recover the spectrum of the projected observable $O_{\D_i}$.

\subsection{Data Decomposition and Signal Model}
Recall the filtered data matrices $G(\theta)$ and $G^O(\theta)$, along with the noiseless matrices $\F(\theta)$ and $\F^O(\theta)$, as introduced in Section~\ref{subsec:QFAMES_framework}. We define the error matrices as
\begin{equation}\label{eq:defn_E}
    \E(\theta) := G(\theta) - \F(\theta),\quad \E^O(\theta) = G^O(\theta) - \F^O(\theta)\,.
\end{equation}
The evolution times are sampled from the truncated Gaussian distribution in Eq.~\eqref{eq:truncated_gaussian_density} to bound the circuit depth. Thus, we define the truncated signals as
\begin{equation}\label{eq:defn_F_trunc}
    \F^{\mathrm{trunc}}(\theta):=\Ex[G(\theta)],\quad \F^{O,\mathrm{trunc}}(\theta):=\Ex[G^O(\theta)].
\end{equation}
The error matrix decomposes into a zero-mean statistical error and a deterministic truncation bias:
\begin{equation}\label{eq:defn_E_stat_bias}
    \E(\theta)=\underbrace{G(\theta)-\F^{\mathrm{trunc}}(\theta)}_{\E^{\mathrm{stat}}(\theta)}+\underbrace{\F^{\mathrm{trunc}}(\theta)-\F(\theta)}_{\E^{\mathrm{bias}}(\theta)}.
\end{equation}
We define $\E^{O,\mathrm{stat}}$ and $\E^{O,\mathrm{bias}}$ analogously, so $\E^O=\E^{O,\mathrm{stat}}+\E^{O,\mathrm{bias}}$. By definition, $\Ex[\E^{\mathrm{stat}}(\theta)]=\Ex[\E^{O,\mathrm{stat}}(\theta)]=0$. Note that under the \textit{Shared Time Sampling} scheme, the statistical errors in different matrix entries need not be independent because they share the sampled times $\{t_n\}_{n=1}^N$.

In particular, for the truncation bias, when $\sigma \ge 1$, Lemma~\ref{lemma:truncation_bias} gives, uniformly in $\theta$,
\begin{equation}\label{eq:truncation_bias_bound}
\begin{aligned}
    \|\E^{\mathrm{bias}}(\theta)\|_F&\le\mathcal{O}\left(\sqrt{LR}e^{-\sigma^2/4}\right),\\
    \left\|\frac{d}{d\theta}\E^{\mathrm{bias}}(\theta)\right\|_F&\le\mathcal{O}\left(T\sqrt{LR}e^{-\sigma^2/4}\right).
\end{aligned}
\end{equation}
The same bounds hold for $\E^{O,\mathrm{bias}}$ with an additional factor $\|O\|_2$.

To further analyze the localized behavior of the filter around a specific eigenvalue cluster, we partition $\F(\theta)$ into three distinct components:
\begin{equation}\label{eq:defn_FABC}
\F(\theta) = A(\theta) + B(\theta) + C(\theta).
\end{equation}
Let $\pi_\theta$ denote the index of the dominant cluster that is closest to the current filter parameter $\theta$. The components are rigorously defined as follows:
\begin{itemize}
    \item \textbf{Target Cluster Signal ($A(\theta)$):} The localized signal contribution arising from the dominant eigenvalues within the target cluster $\mathcal{D}_{\pi_\theta}$:
    \begin{equation}\label{eq:defn_A}
        A(\theta) = \Phi_{:,\mathcal{D}_{\pi_\theta}} \cdot \diag\left(\left\{\exp\left(-(\theta - \lambda_m)^2T^2\right)\right\}_{m \in \mathcal{D}_{\pi_\theta}}\right) \cdot \left(\Psi_{:,\mathcal{D}_{\pi_\theta}}\right)\cj
    \end{equation}

    \item \textbf{Tail Residuals ($B(\theta)$):} The background contribution from the non-dominant eigenvalues outside all major clusters ($m \in \mathcal{D}^c$):
    \begin{equation}\label{eq:defn_B}
        B(\theta) = \Phi_{:,\mathcal{D}^c} \cdot \diag\left(\left\{\exp\left(-(\theta - \lambda_m)^2T^2\right)\right\}_{m \in \mathcal{D}^c}\right) \cdot \left(\Psi_{:,\mathcal{D}^c}\right)\cj
    \end{equation}

    \item \textbf{Cross-Cluster Interference ($C(\theta)$):} The spectral cross-talk or interference caused by all other dominant clusters $\mathcal{D}_i$ where $i \neq \pi_\theta$:
    \begin{equation}\label{eq:defn_C}
        C(\theta) = \sum_{i \neq \pi_\theta} \Phi_{:,\mathcal{D}_i} \cdot \diag\left(\left\{\exp\left(-(\theta - \lambda_m)^2T^2\right)\right\}_{m \in \mathcal{D}_i}\right) \cdot \left(\Psi_{:,\mathcal{D}_i}\right)\cj
    \end{equation}
\end{itemize}
In the observable estimation problem, we decompose $\F^O(\theta)$ in the same way:
\begin{equation}\label{eq:defn_FABC_O}
    \F^O(\theta)=A^O(\theta) + B^O(\theta) + C^O(\theta).
\end{equation}
Here, $A^O$ contains terms with both indices in the target cluster; $B^O$ contains all terms with at least one tail index; $C^O$ contains the remaining dominant terms, including terms coupling different clusters.

When $\theta$ is sufficiently close to a dominant eigenvalue, both $B(\theta)$ and $C(\theta)$ are expected to be small and can therefore be treated as perturbations of the target cluster matrix $A(\theta)$. Together with the statistical errors and truncation biases in Eq.~\eqref{eq:defn_E_stat_bias}, these terms constitute the perturbations in the data matrices $G(\theta)$ and $G^O(\theta)$.

\subsection{Underlying Assumptions}\label{subsec:assumptions}
Our theoretical guarantees use three structural conditions on the physical system and algorithmic setup.

First, we assume that the target dominant eigenvalue clusters are well separated from one another. Let
\begin{equation}\label{eq:defn_delta}
    \delta := \max_i\max_{m,n\in\mathcal D_i}|\lambda_m-\lambda_n|
\end{equation}
denote the maximum spectral width among all individual dominant clusters, and let
\begin{equation}\label{eq:defn_Delta}
    \Delta := \min_{i\ne j}\dist(\mathcal I_i,\mathcal I_j)
\end{equation}
denote the minimum spectral distance between any two distinct dominant clusters.

\begin{assumption}[Cluster Separation]\label{assum:cluster_sep}
The minimum spectral distance should be significantly larger than the maximum cluster width. Specifically,
\begin{equation}
    \Delta \gg \delta.
\end{equation}
\end{assumption}
The quantitative separation required of $\Delta$ and $\delta$ is specified by the parameter conditions in Eqs.~\eqref{eq:thm_eigenvalue_cond1} and~\eqref{eq:thm_eigenvalue_cond2}.

Second, we provide a quantitative characterization of the dominance condition in Eq.~\eqref{eq:defn_dom_cond}. Recall the overlap matrices $\Phi$ and $\Psi$ from Eq.~\eqref{eq:defn_Phi_Psi} and $p_m$ for each eigenstate $|E_m\rangle$ defined in Eq.~\eqref{eq:defn_p_m}. Because $p_m$ scales with the numbers of initial states $L$ and $R$, we normalize it to define the \textit{Normalized Overlap Amplitude}, allowing us to decouple the numbers of initial states from the physical overlap properties:
\begin{equation}\label{eq:defn_p_m_avg}
    \bar{p}_m := \frac{p_m}{\sqrt{LR}} = \frac{\|\Phi_{:,m}\|_2 \|\Psi_{:,m}\|_2}{\sqrt{LR}}.
\end{equation}
This quantity represents the average overlap amplitude of the eigenstate $|E_m\rangle$ with respect to the initial states $\{\ket{\phi_l}\}_{l \in [L]}$ and $\{\ket{\psi_r}\}_{r \in [R]}$. Analogously, based on this normalized metric, we define the \textit{Minimum Dominant Overlap} $\pmin$ and the \textit{Tail Overlap} $\ptail$ as follows:
\begin{equation}\label{eq:defn_p_min_tail}
    \pmin := \min_{m \in \mathcal{D}} \bar{p}_m, \quad \ptail := \sum_{m \in \mathcal{D}^c} \bar{p}_m.
\end{equation}
Given the normalization of the initial states, it naturally follows that $\pmin \in (0, 1]$ and $\ptail \in [0, 1)$. The dominance condition can then be reformulated in terms of these normalized quantities as $\pmin \gtrsim \ptail$ in the following assumption.


\begin{assumption}[Dominant Overlap Threshold]\label{assum:dom_threshold}
The minimum dominant overlap $\pmin$ and the tail overlap $\ptail$ satisfy the rigorous domination constraint:
\begin{equation}
    \pmin \ge C_p \ptail,
\end{equation}
for some sufficiently large constant $C_p > 1$.
\end{assumption}

Finally, our analysis assumes sufficiently uniform coverage of each dominant eigenspace by the initial states. We quantify this through the following conditioning bounds on the overlap matrices $\Phi_{:,\D_i}$ and $\Psi_{:,\D_i}$ for each dominant cluster $\D_i$.

Following~\cite[Eq.~(C1)]{QFAMES}, for $X\in\mathbb{C}^{n\times d}$ with $n\ge d$, let $s_1(X)\ge\cdots\ge s_d(X)\ge0$ denote its singular values and define
\begin{equation}\label{eq:defn_s_avg}
    s_{\min}(X):=s_d(X),
    \qquad
    s_{\mathrm{avg}}(X)
    :=\left(\frac1d\sum_{k=1}^d s_k(X)^2\right)^{1/2}
    =\frac{\|X\|_F}{\sqrt d}.
\end{equation}

\begin{assumption}[Uniform Overlap Condition]
\label{assum:uniform_overlap}
For every dominant cluster $\D_i$, assume that $|\D_i|\le\min\{L,R\}$ and that $\Phi_{:,\D_i}$ and $\Psi_{:,\D_i}$ have full column rank. There exists a constant $\chi>0$ such that, for every $i\in[I]$,
\begin{equation}
    s_{\min}(\Phi_{:,\D_i})
    >\frac{s_{\mathrm{avg}}(\Phi_{:,\D_i})}{1+\chi},
    \qquad
    s_{\min}(\Psi_{:,\D_i})
    >\frac{s_{\mathrm{avg}}(\Psi_{:,\D_i})}{1+\chi}.
\end{equation}
\end{assumption}
Assumption~\ref{assum:uniform_overlap} provides a quantitative condition for stable recovery in our analysis. Failure of its conditioning bound alone does not imply impossibility of recovery. Rank-deficient overlap matrices can obstruct recovery of the full target eigenspace; see~\cite[Proposition~C1]{QFAMES}.


\section{Main Results}\label{sec:main_results}

In our paper, we consider two distinct time-sampling strategies to generate $G$:
\begin{itemize}
    \item \textbf{Shared Time Sampling:} The Hamiltonian evolution times $\{t_n\}_{n=1}^N$ are universally shared across all initial-state pairs $(l,r)$, i.e., $t_{l,r,n} = t_n$ for all $l \in [L]$ and $r \in [R]$. This is the time-sampling scheme used in the original QFAMES algorithm~\cite{QFAMES}.
    \item \textbf{Independent Time Sampling:} The Hamiltonian evolution times $\{t_{l,r,n}\}_{n=1}^N$ are generated stochastically and independently for each initial-state pair $(l,r)$.
\end{itemize}

In this section, we present the core theoretical results of our work. Specifically, we provide a rigorous error analysis for the existing algorithms addressing the two problems: Problem~\ref{prob:dods} (Theorems~\ref{thm:eigenvalue_error} and~\ref{thm:multiplicity}) and Problem~\ref{prob:observable} (Theorem~\ref{thm:observable}).

We prove that, under the new \textit{Independent Time Sampling} scheme, the estimation error scales more favorably with respect to the number of initial states $L$ and $R$, compared to the original \textit{Shared Time Sampling} scheme. For fixed-accuracy cluster-center estimation in Problem~\ref{prob:dods} and observable estimation in Problem~\ref{prob:observable}, the total Hamiltonian simulation cost bounds are independent of $L$ and $R$, up to logarithmic factors, in the parameter regimes specified below. Completing Problem~\ref{prob:dods} also requires cluster-size recovery, whose cost bound retains a dependence on $\max(L,R)$. Here, the cost is measured by the total Hamiltonian simulation time, defined as
\begin{equation}\label{eq:defn_T_total}
T_{\rm total}=2\sum_{l=1}^L\sum_{r=1}^R\sum_{n=1}^N |t_{l,r,n}|.
\end{equation}

Here, the constant $2$ is associated with the separate
real-part and imaginary-part Hadamard tests. For observable estimation, the total cost additionally includes the absolute evolution times of both segments in each observable sample; see Eq.~\eqref{eq:numerics_observable_cost}. This changes only a constant factor when the spectral and observable estimators use the same sample count and comparable time scales.

Below, we provide the informal versions of our main theorems, highlighting the scaling with respect to $T_{\rm total}$. The rigorous theoretical guarantees, including the exact choice of algorithmic parameters and strict probability bounds, are deferred to Appendix~\ref{app:thms}. The numerical experiments in Section~\ref{sec:numerical_results} support the predicted statistical error trends and illustrate the mechanisms behind the improved bounds. Throughout this section and the corresponding rigorous proofs, we assume that the three structural conditions (Assumptions~\ref{assum:cluster_sep},~\ref{assum:dom_threshold}, and~\ref{assum:uniform_overlap}) introduced in Section~\ref{subsec:assumptions} hold. For the informal resource scalings below, we additionally assume $\chi=\mathcal O(1)$ and require the chosen algorithmic parameters to satisfy the conditions specified in Sections~\ref{subsec:formal_eigenvalue}--\ref{subsec:formal_observable} and the corresponding formal theorems.

\subsection{Scaling of Eigenvalue Error}

We first characterize the eigenvalue estimation error under the two time-sampling architectures.

\begin{theorem}[Eigenvalue Error Scaling, Informal]
\label{thm:eigenvalue_error_informal}
Assume $0<\ptail<1$. Given $L$ and $R$, to achieve an $\epsilon$-accuracy in estimating the dominant cluster centers with high probability, the maximal evolution time scales as $T_{\max} = \widetilde{\mathcal{O}}\left(\frac{\ptail}{\pmin \epsilon}\right)$, and the total simulation time $T_{\rm total}$ required scales as follows:
\begin{enumerate}
    \item \textbf{Independent Time Sampling:}
    \begin{equation}
        T_{\rm total}=\widetilde{\mathcal{O}}\left(\max\left\{\frac{1}{\ptail\pmin\epsilon},\frac{\sqrt{LR}}{\pmin^2\epsilon},\frac{LR \ptail}{\pmin\epsilon}\right\}\right)
    \end{equation}
    In particular, when $\sqrt{LR}$ is smaller than the ratio between $\pmin$ and $\ptail$, i.e., $\sqrt{LR} \leq \pmin \ptail^{-1}$, the total simulation time achieves the favorable scaling $T_{\rm total}=\widetilde{\mathcal{O}}\left(1/(\ptail\pmin\epsilon)\right)$ independent of $L$ and $R$.
    \item \textbf{Shared Time Sampling:}
    \begin{equation}
        T_{\rm total} = \widetilde{\mathcal{O}}\left(
        \frac{LR}{\ptail\pmin \epsilon}
    \right).
    \end{equation}
\end{enumerate}
\end{theorem}

For fixed $\delta>0$, the theorem's conditions limit how small $\epsilon$ can be, affecting only very fine precision for sufficiently narrow clusters; the stated scalings still hold within the admissible range.

The rigorous formulation of this result, including exact parametric dependencies, is detailed in Section~\ref{subsec:formal_eigenvalue} and Theorem~\ref{thm:eigenvalue_error}.

\begin{remark}
There are three comments regarding Theorem~\ref{thm:eigenvalue_error_informal}:
\begin{itemize}
  \item In the case of the \textit{Independent Time Sampling} scheme, when the number of initial states is not too large, namely, $\sqrt{LR}\leq \pmin\ptail^{-1}$, the total simulation time achieves the favorable scaling $T_{\rm total}=\widetilde{\mathcal O}\left(\frac{1}{\ptail\pmin\epsilon}\right)$, which is independent of $L$ and $R$. This regime is often the practically relevant one, since the tail eigenvectors typically have only very small overlap with the initial states.

On the other hand, for sufficiently large $L$ and $R$ within the theorem's parameter regime, the stated cost bound is dominated by the term proportional to $LR$, although its dependence on $\ptail$ and $\pmin$ remains more favorable than that of the \textit{Shared Time Sampling} bound. This transition reflects our sufficient parameter choices: the tail term in the error bound is controlled by choosing $T_{\rm max}=\widetilde{\mathcal O}(\ptail/(\pmin\epsilon))$, while increasing $L$ and $R$ initially allows fewer samples per pair. Once the sample count reaches the imposed minimum $N=2$, this compensation stops, and the cost for this choice of evolution time grows proportionally to $LR$, up to logarithmic factors. This is a property of the stated sampling prescription, rather than a general complexity lower bound.

  \item According to Assumption~\ref{assum:dom_threshold}, we have $\ptail/\pmin\lesssim1$, so it suffices to take $T_{\max}=\widetilde{\mathcal O}(1/\epsilon)$. Furthermore, when $\ptail/\pmin\rightarrow0$, the prefactor of $1/\epsilon$ in $T_{\max}$ can be much smaller than 1, up to logarithmic factors, which corresponds to the highly desirable short-depth regime observed in recent literature~\cite{QMEGS,QFAMES,DingLin2023,Ding2023simultaneous,ni2023lowdepth_2,ni2023lowdepth}.

  Moreover, although we choose $T_{\max}=\widetilde{\mathcal O}(\ptail/(\pmin\epsilon))$ in Theorem~\ref{thm:eigenvalue_error_informal}, which leads to a $T_{\rm total}$ dependence inversely proportional to $\ptail$, the formal result in Theorem~\ref{thm:eigenvalue_error} provides greater flexibility in balancing $T_{\max}$ and $T_{\rm total}$. In particular, under the independent scheme, when the numbers of initial states are not too large, namely, $\sqrt{LR}\lesssim c^{-1}$, one may choose $T_{\max}=\widetilde{\mathcal O}(c\epsilon^{-1})$ and hence $T_{\rm total}=\widetilde{\mathcal O}\left((c\pmin^2)^{-1}\epsilon^{-1}\right)$, for any constant $c>\ptail/\pmin$, provided that the remaining conditions in Section~\ref{subsec:formal_eigenvalue} and Theorem~\ref{thm:eigenvalue_error} hold. This yields a flexible trade-off between the maximum evolution time and the total simulation time, which can be particularly useful in practical settings where the maximum allowable evolution time is constrained by hardware limitations.

  \item We note that the result for the \textit{Shared Time Sampling} scheme has the same scaling as~\cite[Theorem 2]{QFAMES}: the apparent change from $\mathcal{O}(L^2R^2)$ to $\mathcal{O}(LR)$ comes from our normalization of $\pmin$ and $\ptail$, which are not normalized in that reference. The shared-sampling cost bound retains an $LR$ factor; this upper bound does not exclude improved scaling for particular problem structures. In Section~\ref{subsec:intuitive_eigenvalue}, we provide an intuitive explanation for this dependence, which is further supported by numerical experiments in Section~\ref{subsec:numerics_eigenvalue}.

\end{itemize}
\end{remark}

\subsection{Resolving Exact Multiplicities}

Based on Ref.~\cite[Algorithm 1]{QFAMES}, recovering the exact multiplicity fundamentally relies on distinguishing the singular values of the dominant target signal from those generated by irrelevant signals and background noise.

Under the \textit{Shared Time Sampling} scheme, correlated time-sampling fluctuations can grow with the same $\sqrt{LR}$ factor as the target signal. Increasing $L$ and $R$ therefore need not improve the signal-to-noise ratio. In contrast, under the \textit{Independent Time Sampling} scheme, the statistical perturbations are independent across initial-state pairs, allowing their spectral norm to grow more slowly than the target signal. This leads to an improved multiplicity recovery guarantee.

\begin{theorem}[Multiplicity Recovery, Informal]
\label{thm:multiplicity_informal}
Assume that the normalized singular-value gap $\Delta_{\mathrm{phys}}$ defined in Eq.~\eqref{eq:defn_physical_gap} satisfies $\Delta_{\mathrm{phys}}=\Theta(\pmin)$. Given $L$ and $R$ satisfying $L,R\lesssim\pmin^{-2}$, under the \textit{Independent Time Sampling} scheme, to perfectly recover the exact multiplicities for all valid dominant clusters and mathematically reject spurious noise-induced clusters with high probability, the total simulation time required scales as:
\begin{equation}
    T_{\rm total} = \widetilde{\mathcal{O}}\left(\frac{\max(L,R)}{\pmin^2 \Delta}\right).
\end{equation}
In contrast, under the \textit{Shared Time Sampling} scheme, the total simulation time required scales as $T_{\rm total} = \widetilde{\mathcal{O}}\left(\frac{LR}{\pmin^2 \Delta}\right)$.
\end{theorem}

We provide the complete thresholding conditions in Eqs.~\eqref{eq:thm_multiplicity_tau} and~\eqref{eq:thm_multiplicity_tau_interval}, the formal recovery guarantee in Theorem~\ref{thm:multiplicity}, and its proof in Appendix~\ref{app:proofs_multiplicity}. The shared-sampling scaling follows from the perturbation analysis in Section~\ref{subsec:intuitive_multiplicity} and retains the $LR$ dependence of the original QFAMES bound~\cite[Theorem 2]{QFAMES}. The numerical results in Section~\ref{subsec:numerics_multiplicity} illustrate the different statistical contamination produced by the two sampling schemes.

\subsection{Scaling of Observable Error}

Finally, we analyze the complexity of estimating physical observables restricted to the degenerate subspace under the \textit{Independent Time Sampling} scheme. For simplicity, the following informal theorem presents the bound in the most common setting $L=R$; the full result is given in Theorem~\ref{thm:observable}.

\begin{theorem}[Observable Error Scaling, Informal]
\label{thm:observable_informal}
Assume exact degeneracy ($\delta=0$), a positive inter-cluster gap ($\Delta>0$), $\pmin=\Omega(1)$ and no initial-state support on tail eigenvalues ($\Phi_{:,\mathcal D^c}=\Psi_{:,\mathcal D^c}=0$), with the remaining technical conditions of the idealized setting as in Section~\ref{subsec:formal_observable}, including Eq.~\eqref{eq:thm_observable_cond}.

Given $L=R$ and $0<\epsilon_O\lesssim\|O\|_2/L$, under the \textit{Independent Time Sampling} scheme, to achieve an $\epsilon_O$-accuracy for estimating the eigenvalues of the projected observable $O_{\D_i}$ with high probability, the total simulation time required scales as:
\begin{equation}
    T_{\rm total} = \widetilde{\mathcal{O}}\left(\frac{\|O\|_2^2}{\epsilon_O^2 \Delta}\right).
\end{equation}
In contrast, under the \textit{Shared Time Sampling} scheme, the total simulation time required scales as $T_{\rm total}=\widetilde{\mathcal O}\left(
    \frac{LR\|O\|_2^2}{\epsilon_O^2\Delta}
\right)$.
\end{theorem}

A fully formalized version of Theorem~\ref{thm:observable_informal} is established in Theorem~\ref{thm:observable}.
\begin{remark} There are two comments regarding Theorem~\ref{thm:observable_informal}:
\begin{itemize}
    \item Similar to the eigenvalue estimation case in Theorem~\ref{thm:eigenvalue_error_informal}, the total simulation time $T_{\rm total}$ is independent of the numbers of initial states $L$ and $R$ under the \textit{Independent Time Sampling} scheme, provided that $L=R$ and $\epsilon_O$ is sufficiently small relative to $\|O\|_2/\sqrt{LR}$. Thus, for a fixed estimation accuracy, this independence holds within the corresponding range of $L$ and $R$. For the \textit{Shared Time Sampling} scheme, the perturbation-based cost bound retains a factor of $LR$. Section~\ref{subsec:intuitive_observable} explains this dependence, and Section~\ref{subsec:numerics_observable} illustrates a correlated-state configuration in which the corresponding shared-time fluctuations remain visible.

    \item Different from the eigenvalue estimation problem, standard sampling-based observable estimation does not achieve the Heisenberg-limited scaling of $\mathcal{O}(1/T_{\rm total})$ with respect to the total simulation time $T_{\rm total}$. In the usual setting where the observable expectation is estimated from independent measurement outcomes, the statistical error is limited by the standard quantum limit~\cite{Giovannetti2011}, scaling as $\mathcal{O}(1/\sqrt{T_{\rm total}})$ when the cost per sample is constant. In contrast, eigenvalue information can be coherently accumulated as a phase during Hamiltonian evolution, allowing an error scaling of $\mathcal{O}(1/T_{\rm total})$.

\end{itemize}
\end{remark}


\section{Intuitive Analysis}\label{sec:intuitive_analysis}

In this section, we provide an intuitive analysis of the theoretical bounds established in Theorem~\ref{thm:eigenvalue_error_informal}, Theorem~\ref{thm:multiplicity_informal}, and Theorem~\ref{thm:observable_informal}. Note that the total Hamiltonian simulation time satisfies
\begin{equation}
    T_{\rm total}=\Theta(LRNT)
\end{equation}
with high probability, where $T$ is the parameter used in the data generation procedure Eq.~\eqref{eq:defn_F}. The truncation also gives $T_{\max}\le \sigma T$. Therefore, to characterize the scaling of $T_{\rm total}$, it suffices to analyze how the estimation error depends on the parameters $T$, $L$, $R$, and $N$.

Throughout this section, we assume $\chi=\mathcal O(1)$ and choose $\sigma$ at the logarithmic scale required by Eq.~\eqref{eq:truncation_precision}. All algorithmic parameters satisfy the conditions specified in Sections~\ref{subsec:formal_eigenvalue}--\ref{subsec:formal_observable} and the corresponding formal theorems. We explicitly identify the additional parameter regimes in which their error bounds give the resource scalings stated in Section~\ref{sec:main_results}.

\subsection{Analysis of Theorem~\ref{thm:eigenvalue_error_informal}}\label{subsec:intuitive_eigenvalue}

Recall the ideal signal $\F(\theta)$ and the decomposition $\E=\E^{\mathrm{stat}}+\E^{\mathrm{bias}}$ in Eq.~\eqref{eq:defn_E_stat_bias}. Then
\begin{equation}\label{eq:intuitive_eigenvalue_G}
    G(\theta)
    =\F^{\mathrm{trunc}}(\theta)+\E^{\mathrm{stat}}(\theta)
    =A(\theta)+B(\theta)+C(\theta)
    +\E^{\mathrm{bias}}(\theta)+\E^{\mathrm{stat}}(\theta),
\end{equation}
where $A(\theta)$, $B(\theta)$, and $C(\theta)$ are defined in Eqs.~\eqref{eq:defn_A},~\eqref{eq:defn_B}, and~\eqref{eq:defn_C}, respectively. The truncation bias and its derivative are exponentially small by Eqs.~\eqref{eq:lemma_truncation_bias} and~\eqref{eq:lemma_truncation_bias_derivative}. Under Eq.~\eqref{eq:truncation_precision}, these contributions are already included in the total-error estimates of Lemmas~\ref{lemma:error_term} and~\ref{lemma:cross_terms}.

Suppose $\theta_i^*$ is a selected peak and $\lambda_{\pi_i}^*$ is the center of its corresponding dominant cluster $\mathcal D_{\pi_i}$. Let $\tilde\theta_i^*$ be a nearest grid point to $\lambda_{\pi_i}^*$, so that $T|\tilde\theta_i^*-\lambda_{\pi_i}^*|\le q/2$. The localization and blocking argument in Appendix~\ref{app:proofs_eigenvalue} ensures that this grid point remains available when $\theta_i^*$ is selected. By the optimality of $\theta_i^*$,
\begin{equation}\label{eq:intuitive_eigenvalue_GFE}
\begin{aligned}
    0
    &\le
    \|G(\theta_i^*)\|_F^2
    -\|G(\tilde\theta_i^*)\|_F^2
    =X+Y+Z,\\
    X&:=
    \|\F(\theta_i^*)\|_F^2
    -\|\F(\tilde\theta_i^*)\|_F^2,\\
    Y&:=
    2\RE\langle\F(\theta_i^*),\E(\theta_i^*)\rangle
    -2\RE\langle\F(\tilde\theta_i^*),
                  \E(\tilde\theta_i^*)\rangle,\\
    Z&:=
    \|\E(\theta_i^*)\|_F^2
    -\|\E(\tilde\theta_i^*)\|_F^2.
\end{aligned}
\end{equation}

Let $\delta_\theta:=T|\theta_i^*-\lambda_{\pi_i}^*|$ and $U:=\|\Phi_{:,\mathcal D_{\pi_i}}(\Psi_{:,\mathcal D_{\pi_i}})\cj\|_F$. Lemma~\ref{lemma:bound_U} gives $\frac{\sqrt{LR}\pmin}{1+\chi}\le U\le\sqrt{LR}$. By Eqs.~\eqref{eq:A_decomp},~\eqref{eq:A1_bound}, and~\eqref{eq:A2_bound}, the target signal satisfies
\begin{equation}\label{eq:intuitive_eigenvalue_A}
    \|A(\theta_i^*)\|_F
    =
    e^{-\delta_\theta^2}U
    +\mathcal O(\sqrt{LR}\ptail^2).
\end{equation}
Thus, within the localized neighborhood, the decrease in the squared target signal is of order $U^2\delta_\theta^2$, up to the grid and cluster-width errors.

According to Eq.~\eqref{eq:B_bound}, the tail residue term satisfies
\begin{equation}\label{eq:intuitive_eigenvalue_B}
    \|B(\theta)\|_F\le\sqrt{LR}\ptail.
\end{equation}
Because of the filtering effect, $C(\theta)$ can be exponentially suppressed by $\Delta T$, where $\Delta$ is defined in Eq.~\eqref{eq:defn_Delta}. Specifically, Eq.~\eqref{eq:C_bound} gives
\begin{equation}
    \|C(\theta)\|_F
    \le\sqrt{LR}e^{-\Delta^2T^2/4}
    \le\frac14\sqrt{LR}\ptail
\end{equation}
under Eq.~\eqref{eq:thm_eigenvalue_cond1}.

Combining these terms and their local variations, Eq.~\eqref{eq:X_bound} yields
\begin{equation}\label{eq:intuitive_eigenvalue_X}
    X
    \le
    -\left(e^{-q^2/2}-e^{-2\delta_\theta^2}\right)U^2+
    \mathcal O(\sqrt{LR}\ptail U)
    T|\theta_i^*-\tilde\theta_i^*|
    +\mathcal O(LR\ptail^2).
\end{equation}
The individual contributions are bounded in Eqs.~\eqref{eq:X1_bound}--\eqref{eq:X3_bound}. In particular, the interaction between the target signal and the tail contributes a term linear in the distance between the two evaluation points.

Next, we study the terms involving $\E(\theta)$. Under the \textit{Independent Time Sampling} scheme, the statistical part of the cross term $Y$ aggregates independent samples across initial-state pairs, while under the \textit{Shared Time Sampling} scheme, these samples are correlated through the shared times $\{t_n\}$. Applying Lemma~\ref{lemma:cross_terms} and Eqs.~\eqref{eq:lipschitz_A}--\eqref{eq:lipschitz_C}, as in Eq.~\eqref{eq:Y_bound}, gives
\begin{equation}\label{eq:intuitive_eigenvalue_Y}
\begin{aligned}
    Y
    &=
    2\RE\langle
        \F(\theta_i^*)-\F(\tilde\theta_i^*),
        \E(\theta_i^*)
    \rangle+
    2\RE\langle
        \F(\tilde\theta_i^*),
        \E(\theta_i^*)-\E(\tilde\theta_i^*)
    \rangle\\
    &\le
    \begin{cases}
        \displaystyle
        \widetilde{\mathcal O}\left(\frac{U}{\sqrt N}\right)
        \sigma T|\theta_i^*-\tilde\theta_i^*|
        +\widetilde{\mathcal O}\left(
            \frac{\sqrt{LR}\ptail}{\sqrt N}
        \right),
        &\text{(Independent)},\\[3mm]
        \displaystyle
        \widetilde{\mathcal O}\left(
            \frac{\sqrt{LR}U}{\sqrt N}
        \right)
        \sigma T|\theta_i^*-\tilde\theta_i^*|
        +\widetilde{\mathcal O}\left(
            \frac{LR\ptail}{\sqrt N}
        \right),
        &\text{(Shared)}.
    \end{cases}
\end{aligned}
\end{equation}
Although the dimension of $\E(\theta)$ increases with $L$ and $R$, independent sampling reduces the cross-term fluctuation by a factor of $\sqrt{LR}$ relative to the shared-sampling bound.

For the squared-error term $Z$, the identity
\begin{equation}
    \left|
        \|\E(\theta)\|_F^2-\|\E(\theta')\|_F^2
    \right|
    \le
    \left(\|\E(\theta)\|_F+\|\E(\theta')\|_F\right)
    \|\E(\theta)-\E(\theta')\|_F
\end{equation}
together with Eqs.~\eqref{eq:lemma_frob_error_bound} and~\eqref{eq:lemma_frob_error_bound_diff} gives
\begin{equation}\label{eq:intuitive_eigenvalue_Z}
    |Z|
    \le
    \widetilde{\mathcal O}\left(\frac{LR}{N}\right)
    \sigma T|\theta_i^*-\tilde\theta_i^*|.
\end{equation}
However, under the \textit{Independent Time Sampling} scheme, a more careful analysis improves this bound. Indeed,
\begin{equation}
    \|\E(\theta)\|_F^2
    =
    \Ex[\|\E(\theta)\|_F^2]+
    \sum_{l,r}
    \left(
        |\E_{l,r}(\theta)|^2
        -\Ex[|\E_{l,r}(\theta)|^2]
    \right),
\end{equation}
where
\begin{equation}
    \Ex[\|\E(\theta)\|_F^2]
    =
    \frac{2LR-\|\F^{\mathrm{trunc}}(\theta)\|_F^2}{N}
    +\|\E^{\mathrm{bias}}(\theta)\|_F^2.
\end{equation}

The centered terms are independent across $(l,r)$, which reduces their fluctuation from order $LR/N$ to order $\sqrt{LR}/N$. The local version of this argument in Lemma~\ref{lemma:error_term}, including the truncation bias, gives Eq.~\eqref{eq:Z_bound}:
\begin{equation}
    Z
    \le
    \begin{cases}
        \displaystyle
        -\frac XN
        +\widetilde{\mathcal O}\left(\frac{\sqrt{LR}}N\right)
        \sigma T|\theta_i^*-\tilde\theta_i^*|,
        &\text{(Independent)},\\[3mm]
        \displaystyle
        \widetilde{\mathcal O}\left(\frac{LR}N\right)
        \sigma T|\theta_i^*-\tilde\theta_i^*|,
        &\text{(Shared)}.
    \end{cases}
\end{equation}
Under independent sampling, combining $X$ with $-X/N$ gives $(1-1/N)X$. Since $N\ge2$, this changes only absolute constants. Under shared sampling, Eq.~\eqref{eq:thm_eigenvalue_cond_shared} allows the $Z$ term to be absorbed into the linear statistical term in Eq.~\eqref{eq:intuitive_eigenvalue_Y}, i.e. the term $Y$.

Substituting these bounds into Eq.~\eqref{eq:intuitive_eigenvalue_GFE}, using $T|\theta_i^*-\tilde\theta_i^*|\le\delta_\theta+q/2$, and applying Eq.~\eqref{eq:quadratic} gives a quadratic inequality for $\delta_\theta$. Under independent sampling, its coefficients satisfy
\begin{equation}
\begin{aligned}
    CU^2\delta_\theta^2
    &\le c_1\delta_\theta+c_2,\\
    c_1
    &=
    \mathcal O(\sigma\sqrt{LR}\ptail U)
    +\widetilde{\mathcal O}\left(
        \frac{U}{\sqrt N}+\frac{\sqrt{LR}}N
    \right)\sigma,\\
    c_2
    &=
    \mathcal O(LR\ptail^2)
    +\mathcal O(\sigma\sqrt{LR}\ptail Uq)
    +\mathcal O(q^2U^2)\\
    &\quad+
    \widetilde{\mathcal O}\left(
        \frac{U}{\sqrt N}+\frac{\sqrt{LR}}N
    \right)\sigma q
    +\widetilde{\mathcal O}\left(
        \frac{\sqrt{LR}\ptail}{\sqrt N}
    \right),
\end{aligned}
\end{equation}
as in Eq.~\eqref{eq:quadratic_ineq_indep}. The shared-sampling counterpart is Eq.~\eqref{eq:quadratic_ineq_shared}. Using $q\le c_q\ptail/\sigma$, the inequality $ax^2\le bx+c\implies x\le b/a+\sqrt{c/a}$, and the AM--GM estimates following Eq.~\eqref{eq:c2_simplify_indep}, we obtain
\begin{equation}
    \delta_\theta
    \le
    \frac{\sigma(1+\chi)}{\pmin}
    \begin{cases}
        \displaystyle
        \widetilde{\mathcal O}\left(
                \left(1+\frac{1+\chi}{\pmin\sqrt N}\right)
                \frac1{\sqrt{LRN}}
            \right)
        +\mathcal O(\ptail),
        &\text{(Independent)},\\[3mm]
        \displaystyle
        \widetilde{\mathcal O}\left(\frac1{\sqrt N}\right)
        +\mathcal O(\ptail),
        &\text{(Shared)}.
    \end{cases}
\end{equation}
Dividing by $T$ recovers Eqs.~\eqref{eq:thm_eigenvalue_result_indep} and~\eqref{eq:thm_eigenvalue_result_shared}.

To achieve an $\epsilon$-accuracy, both the statistical error and the tail contribution must be controlled. Since $\chi=\mathcal O(1)$ and $\sigma$ contributes only logarithmic factors, the independent-sampling bound becomes
\begin{equation}
    |\theta_i^*-\lambda_{\pi_i}^*|
    \le
    \widetilde{\mathcal O}\left(
        \frac1{T\pmin\sqrt{LRN}}
        +\frac1{T\pmin^2\sqrt{LR}\,N}
        +\frac{\ptail}{T\pmin}
    \right).
\end{equation}
Choose
\begin{equation}
    T=\widetilde{\Theta}\left(
        \frac{\ptail}{\pmin\epsilon}
    \right),
    \qquad \text{so that}
    T_{\max}=\widetilde{\mathcal O}\left(
        \frac{\ptail}{\pmin\epsilon}
    \right),
\end{equation}
provided that this choice satisfies the filtering, cluster-width, and blocking conditions in Eqs.~\eqref{eq:thm_eigenvalue_cond1} and~\eqref{eq:thm_eigenvalue_cond2}. The statistical terms are then controlled by
\begin{equation}
    LRN=\widetilde{\Omega}(\ptail^{-2}),
    \qquad
    \sqrt{LR}\,N
    =\widetilde{\Omega}\left(
        \pmin^{-1}\ptail^{-1}
    \right).
\end{equation}
The second condition also implies Eq.~\eqref{eq:thm_eigenvalue_cond_indep}, since $\ptail<\pmin$. Given $L$ and $R$, these sufficient bounds, together with $N\ge2$, can be met by choosing
\begin{equation}
    N=\widetilde{\Theta}\!\left(
        \max\left\{2,\frac{1}{LR\ptail^2},
        \frac{1}{\sqrt{LR}\pmin\ptail}\right\}\right),
\end{equation}
with sufficiently large implicit constants. This sample count gives
\begin{equation}
    T_{\rm total}
    =
    \begin{cases}
        \displaystyle
        \widetilde{\mathcal O}\left(
            \frac1{\ptail\pmin\epsilon}
            +\frac{\sqrt{LR}}{\pmin^2\epsilon}
            +\frac{LR\ptail}{\pmin\epsilon}
        \right),
        &\text{(Independent)},\\[3mm]
        \displaystyle
        \widetilde{\mathcal O}\left(
            \frac{LR}{\ptail\pmin\epsilon}
        \right),
        &\text{(Shared)}.
    \end{cases}
\end{equation}
For shared sampling, we used $N=\widetilde{\Theta}(\ptail^{-2})$, which also satisfies Eq.~\eqref{eq:thm_eigenvalue_cond_shared}.

\subsection{Analysis of Theorem~\ref{thm:multiplicity_informal}}
\label{subsec:intuitive_multiplicity}

To recover the exact multiplicity of a target dominant cluster $\mathcal D_{\pi_i}$ for a candidate peak $\theta_i^*$, the algorithm relies on thresholding the singular values of $G(\theta_i^*)$. Success requires a clear spectral gap between the weakest target signal and the remaining singular values. In particular, Eq.~\eqref{eq:thm_multiplicity_dimension} ensures that $\min(L,R)\ge|\mathcal D_{\pi_i}|$ and that the target matrix has the required rank.

Under the \textit{Independent Time Sampling} scheme, the nonzero singular values of $A(\theta_i^*)$ satisfy
\begin{equation}
    s_{|\mathcal D_{\pi_i}|}(A(\theta_i^*))
    \ge
    \frac{\sqrt{LR}\pmin}{(1+\chi)^2}
    e^{-(r_\chi+\delta T)^2}.
\end{equation}
The remaining deterministic terms contribute at most order $\sqrt{LR}\ptail$. Thus, the available normalized gap is $\Delta_{\mathrm{phys}}$ from Eq.~\eqref{eq:defn_physical_gap}:
\begin{equation}
    \Delta_{\mathrm{phys}}
    =
    \frac{\pmin}{(1+\chi)^2}
    e^{-(r_\chi+\delta T)^2}
    -4\ptail
    >0.
\end{equation}
The filter-width condition Eq.~\eqref{eq:thm_multiplicity_T} ensures its positivity.

By Lemma~\ref{lemma:spectral_norm_bound}, specifically Eq.~\eqref{eq:lemma_error_spectral} with $P=I_L$ and $Q=I_R$, the total perturbation, including the truncation bias, satisfies
\begin{equation}
    \|\E(\theta_i^*)\|_2
    \le
    \widetilde{\mathcal O}\left(
        \frac{\sqrt L+\sqrt R}{\sqrt N}
    \right).
\end{equation}
Weyl's inequality then allows a threshold inside the interval specified in Eq.~\eqref{eq:thm_multiplicity_tau_interval}, provided that
\begin{equation}
    N
    =\widetilde{\Omega}\left(
        \frac1{\min(L,R)\Delta_{\mathrm{phys}}^2}
    \right).
\end{equation}
The additional blocking condition Eq.~\eqref{eq:thm_multiplicity_alpha} ensures that candidates selected after the $I$ valid clusters have all singular values below the same threshold. Hence, the true multiplicities are recovered and spurious candidates are rejected.

Using $\Delta_{\mathrm{phys}}=\Theta(\pmin)$ from Eq.~\eqref{eq:defn_physical_gap}, the noise-to-gap ratio is bounded by
\[
    \widetilde{\mathcal O}\left(\frac{1}{\pmin\sqrt{\min(L,R)N}}\right).
\]
Choosing $N=\widetilde{\Theta}(\max\{2,1/(\min(L,R)\pmin^2)\})$ also satisfies Eq.~\eqref{eq:thm_eigenvalue_cond_indep}. With an admissible choice $T=\widetilde{\Theta}(1/\Delta)$, this gives
\begin{equation}
    T_{\rm total}
    =
    \widetilde{\mathcal O}\left(
        \frac{\max(L,R)}{\pmin^2\Delta}
        +\frac{LR}{\Delta}
    \right).
\end{equation}
When $L,R\lesssim\pmin^{-2}$, up to logarithmic factors, the first term dominates and we recover
\begin{equation}
    T_{\rm total}
    =
    \widetilde{\mathcal O}\left(
        \frac{\max(L,R)}{\pmin^2\Delta}
    \right).
\end{equation}
The additional $LR/\Delta$ term accounts for the minimum sample count $N\ge2$.

However, under the \textit{Shared Time Sampling} scheme, the bound on the noise-to-gap ratio does not improve as $L$ and $R$ increase. Indeed, Lemmas~\ref{lemma:truncation_bias} and~\ref{lemma:error_term}, under Eq.~\eqref{eq:truncation_precision}, give
\begin{equation}\label{eq:E_spectral_shared_intuitive}
    \|\E(\theta_i^*)\|_2
    \le\|\E(\theta_i^*)\|_F
    \le
    \widetilde{\mathcal O}\left(
        \sqrt{\frac{LR}{N}}
    \right).
\end{equation}
This dependence on $L$ and $R$ is sharp in general\footnote{If all initial states are identical and measurement noise is omitted, the centered sampling error has the form $\E^{\mathrm{stat}}(\theta)=\xi_N(\theta)\mathbf{1}_L\mathbf{1}_R^\dagger$. For a fixed $\theta$ with nonzero single-sample variance $v_\theta$, independent time draws give $\Ex[|\xi_N(\theta)|^2]=v_\theta/N$, and hence $\sqrt{\Ex[\|\E^{\mathrm{stat}}(\theta)\|_2^2]} =\sqrt{v_\theta LR/N}$.}.

Since both the perturbation bound and the available signal gap scale with $\sqrt{LR}$, their ratio is bounded by $\widetilde{\mathcal O}(1/(\Delta_{\mathrm{phys}}\sqrt N))$. Thus, this thresholding argument uses $N=\widetilde{\Omega}(\Delta_{\mathrm{phys}}^{-2})$. Using the same condition $\Delta_{\mathrm{phys}}=\Theta(\pmin)$, choosing the sample count at this scale gives
\begin{equation}
    T_{\rm total}
    =
    \widetilde{\mathcal O}\left(
        \frac{LR}{\pmin^2\Delta}
    \right).
\end{equation}
Thus, this perturbation-based guarantee retains an $\mathcal O(LR)$ overhead under the \textit{Shared Time Sampling} scheme, consistent with the numerical experiments in Section~\ref{subsec:numerics_multiplicity}.

\subsection{Analysis of Theorem~\ref{thm:observable_informal}}\label{subsec:intuitive_observable}

We work under the setup in Section~\ref{subsec:formal_observable} and the conditions of Theorem~\ref{thm:observable}. As in the informal statement, we assume $\delta=0$, $\Delta>0$, $\pmin=\Omega(1)$, $\chi=\mathcal O(1)$, and $\Phi_{:,\mathcal D^c}=\Psi_{:,\mathcal D^c}=0$. These support conditions are stronger than $\ptail=0$ and control the cross terms in the observable signal for a general $O$. Under the \textit{Independent Time Sampling} scheme, we further assume that the observable sample blocks $(t_{l,r,n},t'_{l,r,n},Z^O_{l,r,n})$ are independent across $(l,r,n)$ and satisfy $|Z^O_{l,r,n}|\le C_O\|O\|_2$ for an absolute constant $C_O$; the stated $\|O\|_2$ dependence relies on this bounded-outcome measurement model. The remaining technical conditions are given in Eqs.~\eqref{eq:thm_observable_cond}, \eqref{eq:thm_eigenvalue_cond2}, and~\eqref{eq:truncation_precision}.

To intuitively establish the error bound, we first introduce the exact noiseless matrices to compare against $\tilde G(\theta_i^*)$ and $\tilde G^O(\theta_i^*)$. For a target cluster $\mathcal D_{\pi_i}$, let
\begin{equation}\label{eq:defn_G_ex}
    G_{\text{exact}}
    =
    \Phi_{:,\mathcal D_{\pi_i}}
    (\Psi_{:,\mathcal D_{\pi_i}})\cj,
\end{equation}
\begin{equation}\label{eq:defn_G_ex_O}
    G^O_{\text{exact}}
    =
    \Phi_{:,\mathcal D_{\pi_i}}
    O_{\mathcal D_{\pi_i}}
    (\Psi_{:,\mathcal D_{\pi_i}})\cj.
\end{equation}
We define the exact reduced singular value decomposition as
\begin{equation}\label{eq:defn_G_ex_SVD}
    G_{\text{exact}}
    =
    U_{\text{ex}}\Sigma_{\text{ex}}V_{\text{ex}}\cj,
\end{equation}
where $\Sigma_{\text{ex}}\in\mathbb R^{m_i\times m_i}$ contains the positive singular values and $m_i=|\mathcal D_{\pi_i}|$. The exact projected matrices are
\begin{equation}\label{eq:defn_G_ex_tilde}
    \tilde G_{\text{exact}}
    =
    U_{\text{ex}}\cj G_{\text{exact}}V_{\text{ex}}
    =\Sigma_{\text{ex}},
\end{equation}
\begin{equation}\label{eq:defn_G_ex_O_tilde}
    \tilde G^O_{\text{exact}}
    =
    U_{\text{ex}}\cj G^O_{\text{exact}}V_{\text{ex}}.
\end{equation}
The eigenvalues of $O_{\mathcal D_{\pi_i}}$ are precisely the solutions to
\begin{equation}\label{eq:gen_eigenvalue_exact}
    \tilde G^O_{\text{exact}}w
    =
    \lambda_{\text{exact}}\tilde G_{\text{exact}}w.
\end{equation}
We compare these eigenvalues with the empirical generalized eigenvalues from Eq.~\eqref{eq:gen_eigenvalue_empirical}.

The localization argument in Step~1 of Appendix~\ref{app:proofs_observable}, specifically Eq.~\eqref{eq:loc_error_obs_proof}, gives
\begin{equation}
    \delta_\theta
    :=
    T|\theta_i^*-\lambda_{\pi_i}^*|
    \le
    \widetilde{\mathcal O}\left(\frac1{\sqrt{LRN}}\right)
    +\mathcal O\left(e^{-\Delta^2T^2/4}\right).
\end{equation}
Here we apply the derivation of Theorem~\ref{thm:eigenvalue_error} to this idealized setting.

Since $\delta=0$, the target signal satisfies $A(\theta_i^*)=e^{-\delta_\theta^2}G_{\text{exact}}$. Combining this identity with Eq.~\eqref{eq:C_bound_observable} gives Eq.~\eqref{eq:det_G_bound}:
\begin{equation}
    \|\F(\theta_i^*)-G_{\text{exact}}\|_2
    \le
    \widetilde{\mathcal O}\left(\frac1{\sqrt{LR}\,N}\right)
    +\mathcal O\left(
        \sqrt{LR}e^{-\Delta^2T^2/4}
    \right).
\end{equation}
Furthermore, Eq.~\eqref{eq:lemma_error_spectral} gives
\begin{equation}
    \|\E(\theta_i^*)\|_2
    \le
    \widetilde{\mathcal O}\left(
        \frac{\sqrt L+\sqrt R}{\sqrt N}
    \right).
\end{equation}
Thus,
\begin{equation}\label{eq:Delta_G_intuitive}
\begin{aligned}
    \|G(\theta_i^*)-G_{\text{exact}}\|_2
    &\le
    \|\F(\theta_i^*)-G_{\text{exact}}\|_2
    +\|\E(\theta_i^*)\|_2\\
    &\le
    \widetilde{\mathcal O}\left(
        \frac{\sqrt L+\sqrt R}{\sqrt N}
    \right)
    +\mathcal O\left(
        \sqrt{LR}e^{-\Delta^2T^2/4}
    \right).
\end{aligned}
\end{equation}
Similarly, the observable filtering argument in Eqs.~\eqref{eq:F_O_proj}--\eqref{eq:Delta_GO_bound} and the observable version of Eq.~\eqref{eq:lemma_error_spectral} give
\begin{equation}\label{eq:Delta_G_O_intuitive}
    \|G^O(\theta_i^*)-G^O_{\text{exact}}\|_2
    \le
    \widetilde{\mathcal O}\left(
        \|O\|_2\frac{\sqrt L+\sqrt R}{\sqrt N}
    \right)+
    \mathcal O\left(
        \|O\|_2\sqrt{LR}e^{-\Delta^2T^2/4}
    \right).
\end{equation}
The observable truncation bias is included through Eqs.~\eqref{eq:lemma_truncation_bias_observable} and~\eqref{eq:lemma_truncation_bias_observable_derivative}.

The matrix $G_{\text{exact}}$ has rank $m_i$, with $s_{m_i}(G_{\text{exact}})=\Theta(\sqrt{LR})$. Under Eq.~\eqref{eq:thm_observable_cond}, the perturbation is smaller than this positive singular-value gap. Wedin's $\sin\Theta$ theorem~\cite[Chapter V, Theorem 4.4]{stewart1993} therefore controls the subspace misalignment.

As in Eq.~\eqref{eq:obs_UV}, choose unitary matrices $W_U,W_V$ to align the empirical singular bases with the exact ones, and define
\begin{equation}
    \Delta U:=U_i'W_U-U_{\text{ex}},
    \qquad
    \Delta V:=V_i'W_V-V_{\text{ex}}.
\end{equation}
The projected deviations must use these aligned bases:
\begin{equation}
\begin{aligned}
    \Delta\tilde G
    &:=
    W_U^\dagger\tilde G(\theta_i^*)W_V
    -\tilde G_{\text{exact}},\\
    \Delta\tilde G^O
    &:=
    W_U^\dagger\tilde G^O(\theta_i^*)W_V
    -\tilde G^O_{\text{exact}}.
\end{aligned}
\end{equation}

A direct use of the full matrix perturbation gives
\begin{equation}\label{eq:intuitive_observable_Delta_G}
    \|\Delta\tilde G\|_2
    \le
    \widetilde{\mathcal O}\left(
        \frac{\sqrt L+\sqrt R}{\sqrt N}
    \right)
    +\mathcal O\left(
        \sqrt{LR}e^{-\Delta^2T^2/4}
    \right),
\end{equation}
and similarly,
\begin{equation}\label{eq:intuitive_observable_Delta_G_O}
    \|\Delta\tilde G^O\|_2
    \le
    \widetilde{\mathcal O}\left(
        \|O\|_2\frac{\sqrt L+\sqrt R}{\sqrt N}
    \right)
    +\mathcal O\left(
        \|O\|_2\sqrt{LR}e^{-\Delta^2T^2/4}
    \right).
\end{equation}
After division by the signal scale $\sqrt{LR}$, these bounds give a statistical observable error of order $\widetilde{\mathcal O}(\|O\|_2/\sqrt{\min(L,R)N})$. With $T=\widetilde{\Theta}(1/\Delta)$, the corresponding statistical cost is $\widetilde{\mathcal O}(\max(L,R)\|O\|_2^2/(\epsilon_O^2\Delta))$. Even when $L=R$, this bound contains an additional factor of $L$ compared with the scaling in Theorem~\ref{thm:observable_informal}.

To obtain a total simulation time independent of $L$ and $R$ in the appropriate regime, we use a more refined expansion of the projected deviations. Recall $\Delta G_{\mathrm{det}}=\F(\theta_i^*)-G_{\text{exact}}$ from Eq.~\eqref{eq:defn_det_G}. Equation~\eqref{eq:Delta_G_tilde_expand} gives
\begin{equation}\label{eq:intuitive_observable_Delta_G_O_expand}
\begin{aligned}
    \Delta\tilde G
    &=
    (U_i'W_U)^\dagger\E(\theta_i^*)(V_i'W_V)\\
    &\quad+
    (U_i'W_U)^\dagger
    \Delta G_{\mathrm{det}}(V_i'W_V)\\
    &\quad+
    (U_i'W_U)^\dagger
    G_{\text{exact}}(V_i'W_V)
    -U_{\text{ex}}^\dagger G_{\text{exact}}V_{\text{ex}}.
\end{aligned}
\end{equation}
The leading projected-noise term is $U_{\text{ex}}^\dagger\E(\theta_i^*)V_{\text{ex}}$. By Eq.~\eqref{eq:lemma_error_spectral} with $P=U_{\text{ex}}^\dagger$ and $Q=V_{\text{ex}}$, and $m_i=\mathcal O(1)$ as in Eq.~\eqref{eq:m_i_bound},
\begin{equation}
    \|U_{\text{ex}}^\dagger
      \E(\theta_i^*)V_{\text{ex}}\|_2
    \le
    \widetilde{\mathcal O}\left(\frac1{\sqrt N}\right).
\end{equation}
Thus, projection onto the exact singular subspaces removes the leading dependence on $L$ and $R$.

The remaining terms cannot be omitted, but the aligned signal contribution is quadratic in the subspace errors. Indeed, Eq.~\eqref{eq:Delta_U_squared} gives
\begin{equation}
    \Delta U^\dagger U_{\text{ex}}
    =-\frac12\Delta U^\dagger\Delta U,
    \qquad
    V_{\text{ex}}^\dagger\Delta V
    =-\frac12\Delta V^\dagger\Delta V.
\end{equation}
Consequently, the last line of Eq.~\eqref{eq:intuitive_observable_Delta_G_O_expand} is bounded by
\begin{equation*}
    \mathcal O(\sqrt{LR})\max\{\|\Delta U\|_2,\|\Delta V\|_2\}^2,
\end{equation*}
as shown in Eq.~\eqref{eq:signal_cancellation}.
Combining this cancellation with Eq.~\eqref{eq:max_UV_norm} and the projected-noise estimates gives Eq.~\eqref{eq:obs_tilde_G_bound}:
\begin{equation}
    \|\Delta\tilde G\|_2
    \le
    \widetilde{\mathcal O}\left(
        \frac1{\sqrt N}
        +\frac{\sqrt{LR}}{\min(L,R)N}
    \right)
    +\mathcal O\left(
        \sqrt{LR}e^{-\Delta^2T^2/4}
    \right).
\end{equation}
The column and row spaces of $G^O_{\text{exact}}$ lie in the same exact singular subspaces. Hence the same cancellation applies, and Eq.~\eqref{eq:obs_tilde_GO_bound} gives
\begin{equation}
    \|\Delta\tilde G^O\|_2
    \le
    \widetilde{\mathcal O}\left(
        \|O\|_2\left[
            \frac1{\sqrt N}
            +\frac{\sqrt{LR}}{\min(L,R)N}
        \right]
    \right)
    +\mathcal O\left(
        \|O\|_2\sqrt{LR}e^{-\Delta^2T^2/4}
    \right).
\end{equation}

Finally, combining the projected-matrix error bounds with Lemma~\ref{lemma:gen_eig_perturbation} and using $m_i=\mathcal O(1)$, we can prove that there exists a permutation $\rho\in S_{m_i}$ such that
\begin{equation}
\begin{aligned}
    &\max_{1\le k\le m_i}
    |\lambda_k^O-\lambda_{\rho(k),\mathrm{exact}}|\\
    &\quad\le
    \widetilde{\mathcal O}\left(
        \|O\|_2\left[
            \frac1{\sqrt{LRN}}
            +\frac1{\min(L,R)N}
        \right]
    \right)
    +\mathcal O\left(
        \|O\|_2e^{-\Delta^2T^2/4}
    \right).
\end{aligned}
\end{equation}

To recover the scaling in Theorem~\ref{thm:observable_informal}, we now specialize to $L=R$. To achieve an $\epsilon_O$-accuracy, choose $T=\widetilde{\Theta}(1/\Delta)$ with logarithmic factors large enough that $\|O\|_2e^{-\Delta^2T^2/4}=\mathcal O(\epsilon_O)$. Both statistical terms are controlled by choosing
\begin{equation}
    N=\widetilde{\Theta}\left(
        \frac{\|O\|_2^2}{L^2\epsilon_O^2}+1
    \right),
\end{equation}
with the constant chosen to ensure $N\ge2$. Consequently,
\begin{equation}
    T_{\rm total}
    =\widetilde{\mathcal O}\left(
        \frac{\|O\|_2^2}{\epsilon_O^2\Delta}
        +\frac{L^2}{\Delta}
    \right).
\end{equation}

For sufficiently small $\epsilon_O$, namely, $\epsilon_O\lesssim\|O\|_2/L$, the first term dominates, giving
\begin{equation}
    T_{\rm total}
    =\widetilde{\mathcal O}\left(
        \frac{\|O\|_2^2}{\epsilon_O^2\Delta}
    \right),
\end{equation}
as stated in Theorem~\ref{thm:observable_informal}.

Finally, we briefly discuss the \textit{Shared Time Sampling} scheme under the same structural assumptions. Eq.~\eqref{eq:E_spectral_shared_intuitive} and its observable counterpart give projected-noise bounds of order $\widetilde{\mathcal O}(\sqrt{LR/N})$ and $\widetilde{\mathcal O}(\|O\|_2\sqrt{LR/N})$, respectively. The independent-entry concentration used for $U_{\text{ex}}^\dagger\E(\theta_i^*)V_{\text{ex}}$ is unavailable because the time samples are shared. For sufficiently large $N$ and $\Delta T$, Lemma~\ref{lemma:gen_eig_perturbation} gives the following matching bound after division by the signal scale $\Theta(\sqrt{LR})$, with $2m_i-1=\mathcal O(1)$ absorbed into the implicit constant:
\begin{equation}
    \min_{\rho\in S_{m_i}}\max_{1\le k\le m_i}
    |\lambda_k^O-\lambda_{\rho(k),\mathrm{exact}}|
    \le
    \widetilde{\mathcal O}\left(
        \frac{\|O\|_2}{\sqrt N}
    \right)
    +\mathcal O\left(
        \|O\|_2e^{-\Delta^2T^2/4}
    \right).
\end{equation}
For $0<\epsilon_O\le\|O\|_2$, choosing $N=\widetilde{\Theta}(\|O\|_2^2/\epsilon_O^2)$ and suppressing the exponential term gives
\begin{equation}
    T_{\rm total}
    =
    \widetilde{\mathcal O}\left(
        \frac{LR\|O\|_2^2}{\epsilon_O^2\Delta}
    \right),
\end{equation}
which, with $L=R$, recovers the shared-sampling scaling in Theorem~\ref{thm:observable_informal} and retains an $LR$ factor in the cost bound, consistent with the numerical results in Section~\ref{subsec:numerics_observable}.


\section{Numerical Results}\label{sec:numerical_results}

We compare the two time-sampling schemes for eigenvalue estimation, multiplicity recovery, and observable estimation using controlled toy models. The physical-model applications of QFAMES~\cite[Sec.~VIII]{QFAMES} carry over to the same filtering framework; here we isolate the effect of time sampling on scaling with the number of initial states. The code and numerical results are available online\footnote{\url{https://github.com/lingwei-z/qfames-error-scaling}}. Throughout, $L=R$ and the dominant cluster widths are zero.

For spectral estimation, evolution times are sampled from the following normalized truncated Gaussian density as an approximation to Eq.~\eqref{eq:truncated_gaussian_density}:
\begin{equation}\label{eq:numerics_time_density}
    a_{T,\sigma}(t)
    =
    \frac{
        e^{-t^2/(4T^2)}
        \mathbf{1}_{\{|t|\le\sigma T\}}
    }{
        \displaystyle
        \int_{-\sigma T}^{\sigma T}e^{-s^2/(4T^2)}\,ds
    }.
\end{equation}
Independent sampling uses separate times for each initial-state pair, whereas shared sampling reuses the same $N$ times across all pairs.

Shared-sampling statistical error comprises time-sampling error, from approximating the filter integral with $N$ random times, and measurement noise from finite-shot correlation estimates. Under shared sampling, time-sampling error scales as $1/\sqrt{N}$ and does not decrease with $L$; independent sampling has no such bottleneck. For eigenvalue and observable estimation, the measurement contribution scales as $1/\sqrt{LRN}$ under both schemes, explaining the brief initial error decrease with $L$ before time-sampling error dominates. In the multiplicity experiment, if time-sampling error dominates measurement noise, the difference between the two schemes will be clearly visible in the singular values.

For simplicity, we model the measurement error by adding $\nu(\xi+i\zeta)$ to each exact correlation $\langle\phi_l|e^{-iHt}|\psi_r\rangle$, where $\xi$ and $\zeta$ are independent standard real Gaussian random variables, sampled independently across entries and measurements under both schemes. For the observable correlations in Eq.~\eqref{eq:Ex_Z_O}, we instead use noise amplitude $\nu_O$. This Gaussian noise serves as a proxy for Hadamard-test sampling errors, whose single-shot measurement outcomes are bounded in absolute value by $1$ and have variance at most $1$ for each real and imaginary component. For eigenvalue and observable estimation, we set $\nu=\nu_O=1$. For multiplicity recovery, we instead choose $\nu=0.1$, which corresponds to approximately $100$ Hadamard-test shots per time sample. Since $\nu$ is a fixed constant, this choice does not affect the asymptotic scaling behavior.

Sections~\ref{subsec:numerics_eigenvalue} and~\ref{subsec:numerics_observable} plot errors against (i) $L=R$ and (ii) total evolution time; Section~\ref{subsec:numerics_multiplicity} plots mean normalized singular values against $L$.
For both eigenvalue and observable estimation, we report the $90$th percentile over $100$ trials of the matching distance between estimated and exact spectra. Here the spectra consist of the target Hamiltonian eigenvalues for eigenvalue estimation, and the eigenvalues of the observable restricted to the target eigenspace for observable estimation. In both cases, the matching distance is defined by:
\begin{equation}\label{eq:numerics_matching_error}
    d_{\mathrm{match}}(
        \widehat{\boldsymbol{\lambda}},
        \boldsymbol{\lambda})
    :=
    \min_{\pi\in S_m}
    \max_{1\le k\le m}
    |\widehat{\lambda}_k-\lambda_{\pi(k)}|.
\end{equation}
Here $S_m$ contains all permutations of the $m$ target indices. Costs $T_{\rm total}$ are averaged over trials using Eq.~\eqref{eq:defn_T_total} for spectral estimation and Eq.~\eqref{eq:numerics_observable_cost} for observable estimation. Both definitions include a factor of two for separate real-part and imaginary-part measurements, and shared times are counted for every state pair. At fixed $T,\sigma$, these costs scale as $L^2N$.

\subsection{Eigenvalue Error Scaling}
\label{subsec:numerics_eigenvalue}

We consider dominant eigenvalues $\{0,0.25\}$ and a tail at $0.01$ of weight $\ptail=0.05$. All initial states are identical, with equal positive amplitudes on the dominant eigenstates, each of weight $\pmin=0.475$, and the remaining weight in the tail. The parameters are
\begin{equation}
\begin{aligned}
    &T=200,\qquad \sigma=6,\qquad \nu=1,\\
    &N\in\{50,200,800\},\\
    &L=R\in\{2,4,8,12,20,32,50,70,100\}.
\end{aligned}
\end{equation}
Here $\nu$ is the noise standard deviation in each real and imaginary part. We search for two peaks over $[-0.025,0.275]$ using a grid spacing at most $0.4/T$, followed by continuous local refinement. The blocking radius is $2/T$, and the acceptance threshold is $\|G(\theta)\|_F/L\ge0.1$, following the search-and-block procedure in Section~\ref{subsec:QFAMES_framework}. Total evolution time is defined by Eq.~\eqref{eq:defn_T_total}.

\begin{figure}[htbp]
    \centering
    \begin{subfigure}[t]{0.49\linewidth}
        \centering
        \includegraphics[width=\linewidth]
        {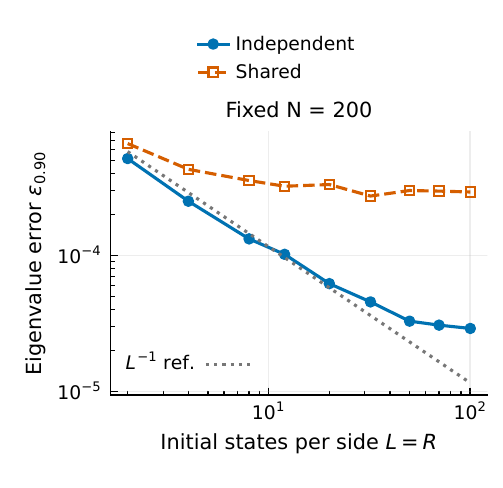}
        \caption{Dependence on $L=R$ at $N=200$.}
        \label{fig:numerics_eigenvalue_states}
    \end{subfigure}
    \hfill
    \begin{subfigure}[t]{0.49\linewidth}
        \centering
        \includegraphics[width=\linewidth]
        {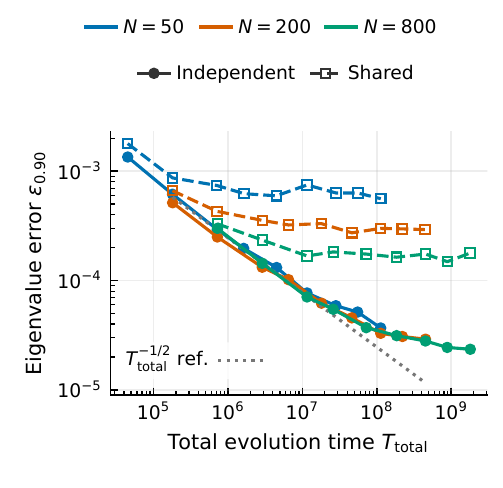}
        \caption{Dependence on total evolution time
        for $N\in\{50,200,800\}$.}
        \label{fig:numerics_eigenvalue_cost}
    \end{subfigure}
    \caption{
        $90$th-percentile eigenvalue matching error over $100$ trials. In the right panel, $L$ varies along each curve, and total evolution time includes both real-part and imaginary-part measurements as in Eq.~\eqref{eq:defn_T_total}. Dotted lines show prescribed $L^{-1}$ and $T_{\rm total}^{-1/2}$ slopes, respectively. }
    \label{fig:thm1_eigenvalue_scaling}
\end{figure}

Figure~\ref{fig:thm1_eigenvalue_scaling} shows approximately $L^{-1}$ error decay under independent sampling before saturation, consistent with the leading $1/\sqrt{LRN}$ term in Eq.~\eqref{eq:thm_eigenvalue_result_indep}. In contrast, shared sampling approaches an $L$-independent error level at fixed $N$. The independent curves approximately collapse onto a $T_{\rm total}^{-1/2}$ trend before saturation. At fixed $T$, the nonzero tail induces a bias for dominant peak locations, producing the large-$L$ plateau, consistent with the $\mathcal O(\ptail/(T\pmin))$ term in Eq.~\eqref{eq:thm_eigenvalue_result_indep}. In this test, we fix $T$ to test the statistical scaling with respect to $L,R,N$, rather than the accuracy-dependent choice in Theorem~\ref{thm:eigenvalue_error_informal}. In addition, $T$ is not large enough to fully suppress the tail bias, which results in the observed saturation.

\subsection{Singular-Value Structure for Multiplicity Recovery}
\label{subsec:numerics_multiplicity}

We consider the dominant spectrum $\{0,0,0.02,0.10\}$, with a tail at $0.01$ of weight $\ptail=0.05$. In the experiment, we mainly focus on characterizing the singular-value structure at $0,0.02$, with multiplicities two and one, respectively, while the dominant eigenvalue at $0.1$ is compressed by the Gaussian filter with $T=150$. The normalized dominant part of each initial state is a four-dimensional complex Gaussian vector conditioned on every coefficient having magnitude greater than $0.4$. The left and right states are drawn independently and reused across the two sampling schemes within each trial. The parameters are
\begin{equation}
\begin{aligned}
    &T=150,\qquad \sigma=6,\qquad N=30,\qquad \nu=0.1,\\
    &L=R\in\{5,8,12,16,24,32,50,70,100\}.
\end{aligned}
\end{equation}
Here $\nu$ again denotes the noise standard deviation per real component. We first estimate the two peak locations near $0$ and $0.02$ using the search-and-block procedure over $[-0.01,0.03]$, then evaluate $G(\theta)$ at these estimated peaks and average the normalized singular values over $50$ trials.

\begin{figure}[tbp]
    \centering
    \begin{subfigure}[t]{\linewidth}
        \centering
        \includegraphics[width=\linewidth]
        {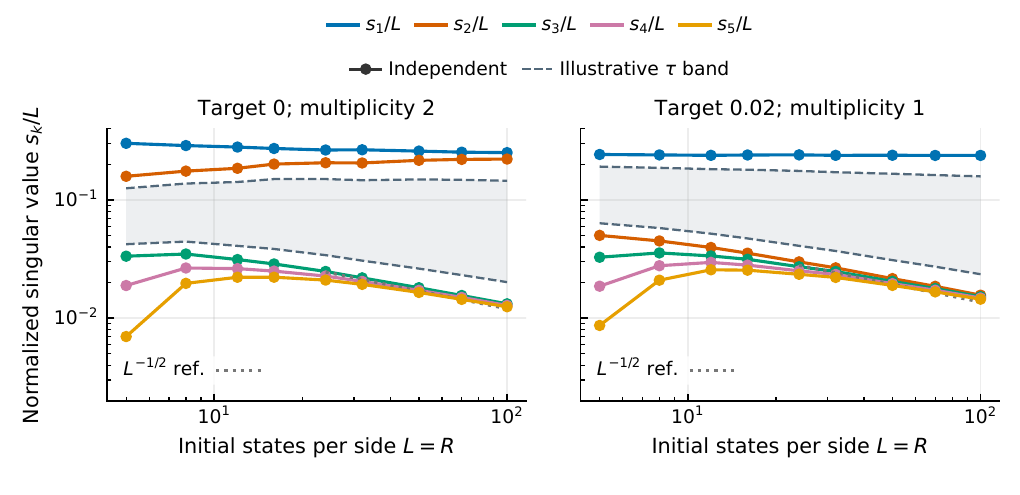}
        \caption{Independent time sampling.}
        \label{fig:numerics_multiplicity_independent}
    \end{subfigure}

    \medskip

    \begin{subfigure}[t]{\linewidth}
        \centering
        \includegraphics[width=\linewidth]
        {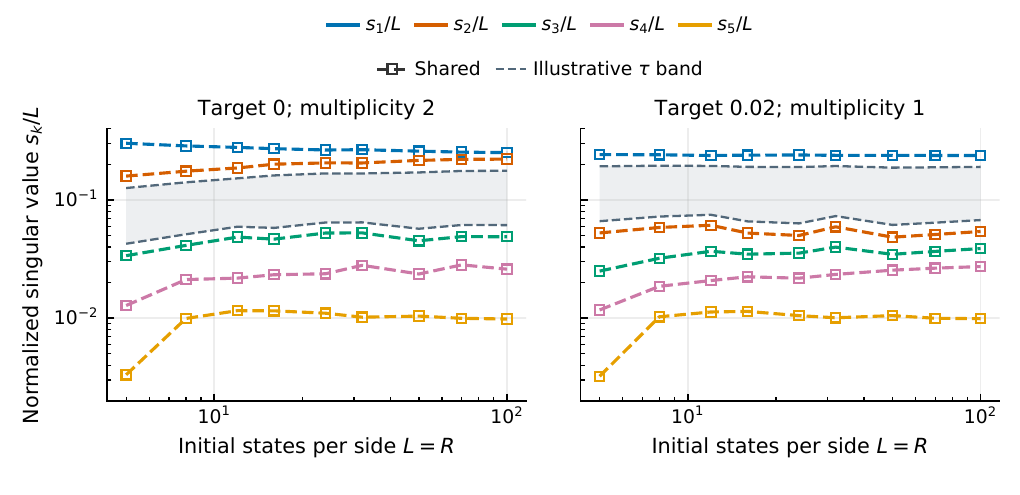}
        \caption{Shared time sampling.}
        \label{fig:numerics_multiplicity_shared}
    \end{subfigure}
    \caption{
        Mean normalized singular values over $50$ trials, with $T=150$, $\sigma=6$, and $N=30$. Left panels: estimated peaks near $0$, with target multiplicity two. Right panels: estimated peaks near $0.02$, with target multiplicity one. Independent-panel dotted lines show prescribed $L^{-1/2}$ slopes, positioned using the independent-sampling $s_4/L$ data for $L\ge24$. }
    \label{fig:thm2_singular_values}
\end{figure}

Figure~\ref{fig:thm2_singular_values} shows the five largest singular values, normalized by $L$. Noise can make $G(\theta)$ full rank, and these values do not correspond one-to-one to eigenstates. The target singular values remain of constant order: two for the cluster at $0$ and one for the cluster at $0.02$. Gaussian filtering suppresses other-cluster contributions by $e^{-T^2(\theta-\lambda)^2}$ [Eq.~\eqref{eq:defn_C}], while truncation bias decays as $e^{-\sigma^2/4}$ [Eq.~\eqref{eq:truncation_bias_bound}]. Both are below the observed statistical fluctuations here. Under independent sampling, $s_3/L$ through $s_5/L$ near $0$, and $s_2/L$ through $s_5/L$ near $0.02$, decrease together at larger $L$. Their behavior is consistent with the $(LN)^{-1/2}$ normalized spectral-noise scale in Lemma~\ref{lemma:spectral_norm_bound}.

Under shared sampling, each non-target spectral component instead carries a common empirical filter coefficient
\begin{equation}
    \widehat{\rho}_N(\omega)
    :=
    \frac1N\sum_{n=1}^N e^{-i\omega t_n}.
\end{equation}
Its $N^{-1/2}$ fluctuations are common to all initial-state pairs and persist as $L$ grows, consistent with Eq.~\eqref{eq:E_spectral_shared_intuitive}. The leading non-target normalized singular values remain near $L$-independent levels.

The shaded regions illustrate selectable ranges of the normalized threshold $\tau/L$ within the gaps between the mean target and non-target singular values. Under independent sampling, these ranges widen as $L$ increases, making threshold selection more flexible. This supports the separation mechanism in Theorem~\ref{thm:multiplicity}. 

\subsection{Observable Estimation Error}
\label{subsec:numerics_observable}

We consider the spectrum $\{0,0,0.1\}$ with $\ptail=0$ and estimate the eigenvalues of the observable restricted to the two-dimensional eigenspace at zero. In the Hamiltonian eigenbasis, we choose
\begin{equation}
    O=
    \begin{pmatrix}
        0.75 & 0.25 & 0\\
        0.25 & 0.75 & 0\\
        0 & 0 & 1
    \end{pmatrix},
    \qquad
    \|O\|_2=1,
\end{equation}
with target observable eigenvalues $0.5$ and $1$. The overlap matrices are
\begin{equation}\label{eq:numerics_observable_frames}
\Phi_{l,:}=\Psi_{l,:}=
\left(
\sqrt{0.175},\;
\sqrt{0.175}\,e^{2\pi i(l-1)/L},\;
\sqrt{0.65}
\right),
\qquad l=1,\ldots,L.
\end{equation}
This gives $\bar p_{\min}=0.175$ and $
\Phi_{:,1:2}^{\dagger}\Phi_{:,1:2}
=\Psi_{:,1:2}^{\dagger}\Psi_{:,1:2}
=0.175L I_2
$. The target columns are perfectly conditioned.

The parameters are
\begin{equation}
\begin{aligned}
    &T=50,\qquad \sigma=6,\qquad \nu=\nu_O=1,\\
    &N\in\{1000,2000,4000\},\\
    &L=R\in\{8,12,16,24,32,50\}.
\end{aligned}
\end{equation}
Here $\nu$ and $\nu_O$ are the spectral and observable noise standard deviations per real component, respectively. Each observable sample uses two independent draws from $a_{T/\sqrt{2},\sigma}$ in Eq.~\eqref{eq:numerics_time_density}; shared sampling reuses each pair across initial-state pairs. We estimate $\theta^*$ by maximizing $\|G(\theta)\|_F$ within the known neighborhood $[-0.04,0.04]$. With multiplicity two supplied, we use the two leading singular vectors to form Eqs.~\eqref{eq:defn_G_tilde}--\eqref{eq:defn_G_O_tilde} and solve Eq.~\eqref{eq:gen_eigenvalue_empirical}. Spectral and observable data are generated independently. The matching error in Eq.~\eqref{eq:numerics_matching_error} tests the full spectrum of the observable restricted to the target eigenspace, as in Theorem~\ref{thm:observable}. Total cost includes spectral acquisition and both observable evolution segments:
\begin{equation}\label{eq:numerics_observable_cost}
    T_{\rm total}
    =2\sum_{l,r,n}|t^{\mathrm{spec}}_{l,r,n}|
    +2\sum_{l,r,n}\left(|t^O_{l,r,n}|+|t^{O\prime}_{l,r,n}|\right).
\end{equation}

\begin{figure}[htbp]
    \centering
    \begin{subfigure}[t]{0.49\linewidth}
        \centering
        \includegraphics[width=\linewidth]
        {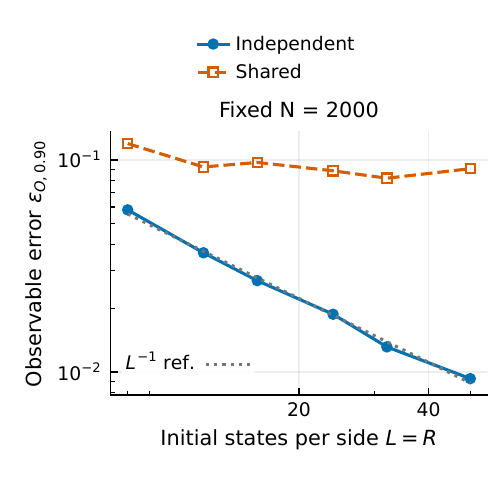}
        \caption{Dependence on $L=R$ at $N=2000$.}
        \label{fig:numerics_observable_states}
    \end{subfigure}
    \hfill
    \begin{subfigure}[t]{0.49\linewidth}
        \centering
        \includegraphics[width=\linewidth]
        {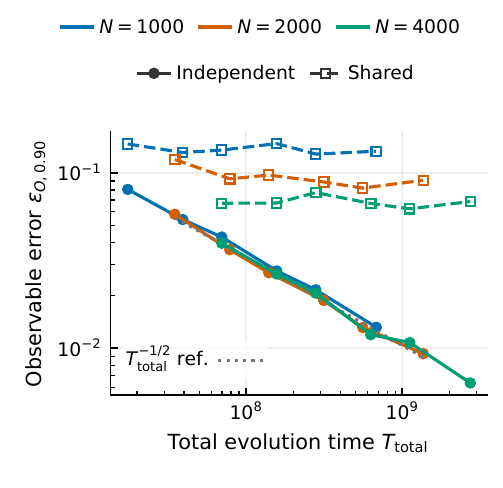}
        \caption{Dependence on total evolution time
        for $N\in\{1000,2000,4000\}$.}
        \label{fig:numerics_observable_cost}
    \end{subfigure}
    \caption{
        90th-percentile matching error over 100 trials for the observable restricted to the zero-energy eigenspace, with unit noise variance in each real and imaginary component. Left: dependence on $L=R$ at $N=2000$. Right: dependence on total evolution time for $N\in\{1000,2000,4000\}$, with $L$ varying along each curve. Cost includes separate real-part and imaginary-part measurements and both observable evolution segments, as in Eq.~\eqref{eq:numerics_observable_cost}. Dotted lines indicate prescribed $L^{-1}$ and $T_{\mathrm{total}}^{-1/2}$ reference slopes.}
    \label{fig:thm3_observable_scaling}
\end{figure}

Figure~\ref{fig:thm3_observable_scaling} shows approximately $L^{-1}$ error decay under independent sampling, consistent with Eq.~\eqref{eq:thm_observable_result}. The independent curves also approximately collapse onto a common $T_{\rm total}^{-1/2}$ trend. In contrast, shared sampling retains an approximately $L$-independent error level at fixed $N$.


\section{Discussion}\label{sec:discussion}

QFAMES uses cross-correlations between multiple initial states to estimate dominant eigenvalues, their multiplicities, and observables within the corresponding eigenspaces~\cite{QFAMES}. Access to these quantities requires sufficient coverage of the target subspaces, which may call for many initial states. Under \textit{Shared Time Sampling}, however, the resulting matrix entries have correlated sampling fluctuations. Increasing the number of states therefore need not improve the signal-to-noise ratio, and the total evolution-time bounds retain an overhead proportional to the number of initial-state pairs.

In this work, we reduce this overhead by sampling evolution times independently for each matrix entry. This change allows statistical fluctuations to average out across entries while preserving the QFAMES filtering and estimation procedures. Under the stated assumptions and within the corresponding parameter ranges, the total evolution-time bounds for eigenvalue and observable estimation become independent of the numbers of initial states, while the multiplicity-recovery bound improves from a dependence on $LR$ to $\max(L,R)$. The bounds retain the $\widetilde{\mathcal O}(\epsilon^{-1})$ dependence for eigenvalue estimation and the $\widetilde{\mathcal O}(\epsilon_O^{-2})$ dependence for observable estimation. Thus, a larger collection of initial states can improve subspace coverage while allowing fewer samples per state pair, without a proportional increase in total evolution time in these regimes.

These improvements are relevant to applications that rely on several initial states to explore a low-energy subspace. For example, the topological and symmetry-breaking models studied in~\cite{QFAMES} require resolving ground-state multiplicities and observables within degenerate or nearly degenerate manifolds. Reducing the sampling overhead makes it more affordable to use additional trial states for this purpose. Quantum chemistry provides another possible application: multireference and nonorthogonal subspace methods use collections of trial states to describe correlated low-energy states~\cite{StairHuangEvangelista2020Krylov,BaekEtAl2023NOQE}. Combining such state constructions with independent time sampling may offer a useful route to spectral characterization when the required overlap conditions hold.

Two directions remain for future work. First, our analysis assigns the same sample count $N$ to every matrix entry. Allowing entry-dependent counts $N_{l,r}$, possibly chosen using preliminary estimates of signal strength and noise, may further reduce the total sampling cost and evolution time. Second, this work mainly focuses on theoretical guarantees for the independent time-sampling scheme to demonstrate the improvement over the shared time-sampling scheme. A more comprehensive numerical study of the independent time-sampling scheme, including its performance on physical models, would be valuable to assess its practical utility.

\section*{Author contributions}

Both authors contributed to the research and the preparation of the manuscript.
AI tools were used to assist with drafting and revising text, reviewing mathematical derivations, and generating and revising numerical code. All AI-assisted material was reviewed and verified by the authors, who take full responsibility for the content.


\appendix
\numberwithin{equation}{section}
\numberwithin{theorem}{section}
\numberwithin{remark}{section}

\section{Collection of Notation}
\label{app:notation}

\begin{longtable}{@{}p{0.25\textwidth} p{0.7\textwidth}@{}}
$[N]$ & The index set $\{1,2,\ldots,N\}$ \\
\multicolumn{2}{l}{\textbf{Hamiltonian notation}} \\
$H=\sum_m\lambda_m |E_m\rangle \langle E_m|$ & Hamiltonian and its exact spectral decomposition \\
$\mathcal{D} \subset [M]$ & Index set of all dominant eigenstates \\
$K = |\mathcal{D}|$ & The total number of dominant eigenstates (counting multiplicity) \\
$I \in \mathbb{N}_+$ & The number of distinct dominant eigenvalue clusters \\
$\{\lambda_i^*\}_{i\in[I]} \subset \mathbb{R}$ & The dominant cluster centers (midpoints of $\mathcal{I}_i$) \\
$\mathcal{D}_i \subset \mathcal{D}$ & The index set of dominant eigenvectors within the $i$-th cluster \\
$\mathcal{I}_i \subset \mathbb{R}$ & The spectral interval spanned by the eigenvalues in cluster $\mathcal{D}_i$ \\
$\delta \in \mathbb{R}$ & The maximum spectral width among all individual dominant clusters \\
$\Delta \in \mathbb{R}_+$ & The minimum spectral separation between distinct dominant clusters \\
\\
\multicolumn{2}{l}{\textbf{Initial-state notation}} \\
$L, R \in \mathbb{N}_+$ & Number of left and right initial states \\
$\{U_l\}_{l\in[L]}, \{V_r\}_{r\in[R]}$ & Initial-state preparation unitaries \\
$\{|\phi_l\rangle\}_{l\in[L]}, \{|\psi_r\rangle\}_{r\in[R]}$ & Left and right initial states prepared by $U_l$ and $V_r$ \\
$\Phi \in \mathbb{C}^{L \times M}$ & Overlaps between left initial states $\{|\phi_l\rangle\}$ and exact eigenstates $\{|E_m\rangle\}$ \\
$\Psi \in \mathbb{C}^{R \times M}$ & Overlaps between right initial states $\{|\psi_r\rangle\}$ and exact eigenstates $\{|E_m\rangle\}$ \\
$p_m \in \mathbb{R}$ & Overlap amplitude of eigenstate $|E_m\rangle$ \\
$\bar{p}_m \in \mathbb{R}$ & Normalized overlap amplitude of eigenstate $|E_m\rangle$ \\
$\bar{p}_{\text{tail}} \in \mathbb{R}$ & Total normalized overlap with the non-dominant tail eigenvectors \\
$\bar{p}_{\min} \in \mathbb{R}_+$ & Minimum normalized overlap among all dominant eigenvectors \\
$\chi \in \mathbb{R}_+$ & Conditioning constant for the uniform overlap assumption \\
\\
\multicolumn{2}{l}{\textbf{Hamiltonian evolution time notation}} \\
$N \in \mathbb{N}_+$ & Number of time samples per entry \\
$T \in \mathbb{R}_+$ & Filter time scale; the spectral filter width is of
order $T^{-1}$ \\
$\sigma \in \mathbb{R}_+$ & Gaussian truncation parameter such that $a_T^{\text{trunc}}(t)$ is supported over $[-\sigma T, \sigma T]$ \\
$a_T^{\text{trunc}}(t)$ & Truncated Gaussian probability measure in Eq.~\eqref{eq:truncated_gaussian_density} \\
$\{t_{l,r,n}\}_{n\in[N]} \subset \mathbb{R}$ & Evolution time samples for the $(l,r)$-th entry under the \textit{Independent Time Sampling} scheme \\
$T_{\text{total}} \in \mathbb{R}_+$ & Total Hamiltonian simulation time \\
\\
\multicolumn{2}{l}{\textbf{Data and tensor notation}} \\
$\Z_{l,r}(t) \in \mathbb{C}$ & The exact time-evolved cross-correlation $\langle \phi_l | e^{-i H t} | \psi_r \rangle$ \\
$Z_{l,r,n} \in \mathbb{C}$ & Measurement outcomes from the generalized Hadamard test \\
$\F(\theta) \in \mathbb{C}^{L \times R}$ & The ideal noiseless filtered matrix \\
$A(\theta), B(\theta), C(\theta)$ & Target cluster, tail residuals, and cross-cluster components of $\F(\theta)$ \\
$G(\theta) \in \mathbb{C}^{L \times R}$ & The empirical filtered data matrix (unbiased estimator for $\F^{\mathrm{trunc}}(\theta)$) \\
$\E(\theta) \in \mathbb{C}^{L \times R}$ & The total error matrix $G(\theta) - \F(\theta)$ \\
$\F^O(\theta), G^O(\theta), \E^O(\theta)$ & Observable-weighted counterparts to $\F(\theta), G(\theta), \E(\theta)$ \\
\\
\multicolumn{2}{l}{\textbf{Algorithm notation}} \\
$q \in \mathbb{R}_+$ & The grid-search parameter such that the grid width is at most $q/T$ \\
$J \in \mathbb{N}_+$ & The number of grid points \\
$\Theta = \{\theta_j\} \subset [-\pi, \pi]$ & The evaluation grid \\
$\alpha \in \mathbb{R}_+$ & Width parameter of the block interval \\
$\eta \in (0, 1)$ & The failure probability of the algorithm \\
$\tilde{I} \in \mathbb{N}_+$ & Prescribed upper bound on the number of dominant clusters before rejection of spurious candidates \\
$\{\theta_j^*\}_{j\in[\tilde{I}]} \subset [-\pi, \pi]$ & Estimated dominant eigenvalues (candidate peaks) \\
$\tau \in \mathbb{R}_+$ & The threshold for singular value truncation \\
$\{m_i\} \subset \mathbb{N}$ & Estimated multiplicities \\
$\lambda^O, \lambda_{\text{exact}}$ & Estimated and exact eigenvalues of the projected observable $O_{\D_i}$ \\
\end{longtable}

\section{Review of QFAMES Framework}
\label{app:review}

In this appendix, we provide a review of the data generation circuits and algorithmic procedures for the QFAMES framework. Although the algorithm has been thoroughly described in~\cite{QFAMES}, we summarize the key steps here for completeness and to establish a unified notation for our analysis.

\paragraph{Step 1: Generate Data.}
The first step of the algorithm is to evaluate the time-evolved cross-correlations between the left and right initial states. To do so, we must generate sequences of Hamiltonian evolution times. Recall the two distinct time-sampling architectures introduced in Section~\ref{sec:main_results}: \textit{Shared Time Sampling} scheme and \textit{Independent Time Sampling} scheme. For both architectures, we draw time samples independently from the truncated Gaussian probability measure $a_T^{\text{trunc}}(t)$ defined in Eq.~\eqref{eq:truncated_gaussian_density}. Specifically, to guarantee a finite-time circuit depth for Hamiltonian simulation, this Gaussian distribution is strictly supported over the truncated interval $[-\sigma T, \sigma T]$, where $\sigma$ is the truncation parameter. For a unified notation, we denote the evolution times as $t_{l,r,n}$ for both architectures, noting that under the \textit{Shared Time Sampling} scheme, $t_{l,r,n} = t_n$ for all $l \in [L]$ and $r \in [R]$.

Specifically, we want to measure quantities of the form:
\begin{equation}
    \Z_{l,r}(t_{l,r,n}) := \langle \phi_l | e^{-i H t_{l,r,n}} | \psi_r \rangle = \langle 0 | U_l^\dagger e^{-i H t_{l,r,n}} V_r | 0 \rangle.
\end{equation}
Furthermore, according to Eq.~\eqref{eq:H_decomp} and Eq.~\eqref{eq:defn_Phi_Psi}, we know
\begin{equation}
    \Z_{l,r}(t_{l,r,n}) = \sum_{m \in [M]} \Phi_{l,m} \cdot e^{-i \lambda_m t_{l,r,n}} \cdot  \Psi_{r,m}\cj.
\end{equation}
The real and imaginary parts of $\Z_{l,r}(t_{l,r,n})$ can be obtained with the generalized Hadamard test as drawn below:
\begin{equation}\label{eq:circuit_H}
\vcenter{
\Qcircuit @C=0.8em @R=1.2em {
    \lstick{|0\rangle}
    & \gate{H}
    & \multigate{1}{\mathrm{INIT}_{l,r}}
    & \ctrl{1}
    & \gate{W}
    & \gate{H}
    & \meter \\
    \lstick{|0\rangle}
    & \qw
    & \ghost{\mathrm{INIT}_{l,r}}
    & \gate{e^{-iHt_{l,r,n}}}
    & \qw
    & \qw
    & \qw
}
}
\end{equation}
\begin{equation}\label{eq:circuit_H_init}
\vcenter{
\Qcircuit @C=1.5em @R=1.5em {
    & \multigate{1}{\text{INIT}_{l,r}} & \qw \\
    & \ghost{\text{INIT}_{l,r}} & \qw
}
}
\equiv
\vcenter{
\Qcircuit @C=1.5em @R=1.5em {
    & \ctrlo{1} & \ctrl{1} & \qw \\
    & \gate{U_l} & \gate{V_r} & \qw
}
}
\end{equation}
where $W$ is chosen to be either the identity matrix $I$ (to extract the real part) or the phase gate $S^\dagger=\begin{bmatrix}
    1 & 0 \\
    0 & -i
\end{bmatrix}$ (to extract the imaginary part).

For each $(l,r,n)$, we perform the real-part and imaginary-part Hadamard tests at the same sampled time $t_{l,r,n}$, using one circuit execution for each part. The complex estimator $Z_{l,r,n}$ is formed by taking the corresponding $\pm1$ outcomes as its real and imaginary parts. Thus,
\begin{equation}
    Z_{l,r,n}\in\{1+i,1-i,-1+i,-1-i\},
    \qquad |Z_{l,r,n}|=\sqrt{2},
\end{equation}
and
\begin{equation}
    \Ex[Z_{l,r,n}\mid t_{l,r,n}]
    =\Z_{l,r}(t_{l,r,n}).
\end{equation}
All circuit executions use independent measurement randomness conditional on the sampled times.

\paragraph{Step 2: Compute the Filtered Matrix Estimator.}
Recall the $L$-by-$R$ filtered data matrix $G(\theta)$ introduced in Section~\ref{subsec:QFAMES_framework}. It is constructed from the sampled data as:
\begin{equation}\label{eq:defn_G_app}
    G(\theta)_{l,r} = \frac{1}{N} \sum_{n=1}^N Z_{l,r,n} e^{i \theta t_{l,r,n}}, \quad l \in [L], r \in [R].
\end{equation}
Under truncated time sampling, $G(\theta)$ has expectation $\F^{\mathrm{trunc}}(\theta)$:
\begin{equation}
    \Ex[G(\theta)]=\F^{\mathrm{trunc}}(\theta) \approx \Phi \cdot \diag \left( \left\{ \exp\left(-(\theta - \lambda_m)^2 T^2\right) \right\}_{m \in [M]} \right) \cdot \Psi\cj = \F(\theta)
\end{equation}
In the following algorithms, we assume that the dominant eigenvalues belong to $[-\pi,\pi]$ for simplicity of presentation. To implement this continuous search practically, we discretize the continuous interval $[-\pi,\pi]$ into a finite evaluation grid $\Theta$. The grid points $\theta$ are chosen uniformly with a step size at most $q/T$, where $q$ is the predefined grid-search parameter. This discretization yields a total of $J = \lceil 2\pi T/q\rceil+1$ grid points.

\paragraph{Step 3: Search and Block for Eigenvalue Estimation.}
After computing $G(\theta)$, the dominant eigenvalues are identified as the peaks of $\|G(\theta)\|_F$ and can be located using the search-and-block procedure proposed in~\cite{QMEGS} and adapted in~\cite{QFAMES}. As outlined in Section~\ref{subsec:QFAMES_framework}, this strategy consists of two iterative components—the search step and the block step—which are described as follows:
\begin{enumerate}
    \item \textbf{Search:} In the first step, we search over the discrete grid $\Theta$ and select
    \begin{equation}
        \theta_1^*\in \operatorname*{arg\,max}_{\theta\in\Theta} \|G(\theta)\|_F,
    \end{equation}
    which approximates the center of a dominant eigenvalue cluster.
    \item \textbf{Block:} To approximate the next dominant eigenvalue and avoid repetition, we define a block interval:
    \begin{equation}
        \mathcal{I}_{B,1} = \left[\theta^*_1 - \frac{\alpha}{T}, \theta^*_1 + \frac{\alpha}{T}\right],
    \end{equation}
    where $\alpha$ is a predefined block parameter and $T$ is the filter time scale. Subsequent searches exclude this interval.
    \item \textbf{Iterate:} In the second iteration, we search for the next maximum point $\theta^*_2$ strictly outside the existing blocked region:
    \begin{equation}
        \theta_2^*\in
        \operatorname*{arg\,max}_{\theta\in\Theta\setminus\mathcal{I}_{B,1}}
        \|G(\theta)\|_F.
    \end{equation}
    After locating $\theta^*_2$, we expand the block interval to include this new neighborhood, updating it to $\mathcal{I}_{B,2} = \mathcal{I}_{B,1} \cup [\theta^*_2 - \alpha/T, \theta^*_2 + \alpha/T]$.
\end{enumerate}

This searching and updating process is repeated iteratively until a desired number of $\tilde{I}$ local maxima are discovered, yielding a set of candidate peaks $\{\theta^*_i\}_{i=1}^{\tilde{I}}$ corresponding to the set of dominant eigenvalues $\{\lambda_i^*\}_{i \in [I]}$. For the complete pseudocode and detailed implementation of this procedure, we refer the reader to~\cite[Algorithm 1]{QFAMES}.

\paragraph{Step 4: Estimate Multiplicity via SVD.}
Recall the singular value decomposition $G(\theta_i^*)=U_i \Sigma_i V_i\cj$ performed for each candidate peak $\{\theta^*_i\}_{i=1}^{\tilde{I}}$ from Eq.~\eqref{eq:defn_G_SVD}. Based on this decomposition, we define the multiplicity $m_i=\# \{j \mid \left(\Sigma_i\right)_{j,j} > \tau\}$ for some given threshold $\tau$, which gives the multiplicity associated with each candidate $\theta_i^*$. If $m_i=0$, then $\theta_i^*$ corresponds to a spurious dominant eigenvalue and should be discarded.

\paragraph{Step 5: Observable Estimation.}
Given an observable $O$, it can always be rewritten as the weighted sum of unitary Hermitian operators such as Pauli matrices. For a general observable $O$, we assume that the measurement estimator satisfies $|Z^O_{l,r,n}|\le C_O\|O\|_2$ for an absolute constant $C_O$, as in the setup in Section~\ref{subsec:formal_observable}. The measurement outcomes $Z_{l,r,n}^O$ of the modified circuit in~\cite[Eq.~(29)]{QFAMES} provide an unbiased estimator of
\begin{equation}\label{eq:Ex_Z_O}
    \Ex[Z_{l,r,n}^O\mid t_{l,r,n},t'_{l,r,n}] = \Z^O_{l,r}(t_{l,r,n}, t'_{l,r,n}) := \langle \phi_l | e^{i H t'_{l,r,n}} O e^{-i H t_{l,r,n}} | \psi_r \rangle,
\end{equation}
where $t_{l,r,n}$ and $t'_{l,r,n}$ are sampled independently from $a_{T/\sqrt{2}}^{\mathrm{trunc}}$.

Recall the observable-weighted data matrix $G^O(\theta)$ introduced in Section~\ref{subsec:QFAMES_framework}. It is computed as:
\begin{equation}\label{eq:defn_G_O_app}
    G^O(\theta)_{l,r} = \frac{1}{N} \sum_{n=1}^N Z^O_{l,r,n} e^{i\theta t_{l,r,n}} e^{-i\theta t'_{l,r,n}}
\end{equation}
for all $l \in [L], r \in [R]$. According to~\cite[Eq.~(33)]{QFAMES}, the expectation values are:
\begin{equation}
\begin{aligned}
    \Ex[G^O(\theta)]&=\F^{O,\mathrm{trunc}}(\theta)\\ &\approx \Phi \cdot \diag\big(\{e^{-(\theta-\lambda_k)^2 T^2 / 2}\}_{k \in [M]}\big) \cdot O_{[M]} \\
    &\quad \cdot \diag\big(\{e^{-(\theta-\lambda_{k'})^2 T^2 / 2}\}_{k' \in [M]}\big) \cdot \Psi^\dagger = \F^O(\theta),
\end{aligned}
\end{equation}
where $O_{[M]}:=\left( \langle E_k | O | E_{k'} \rangle \right)_{k,k' \in [M]}$.

Now we focus on degenerate dominant eigenstates located at $\theta_i^*$ given by~\cite[Algorithm 1]{QFAMES} with multiplicity $m_i$. Recall the projected matrices $\tilde{G}(\theta_i^*)$ and $\tilde{G}^O(\theta_i^*)$ from Eq.~\eqref{eq:defn_G_tilde} and Eq.~\eqref{eq:defn_G_O_tilde}, which restrict $G(\theta_i^*)$ and $G^O(\theta_i^*)$ to the large singular value subspace with cutoff threshold $\tau$.

From Eq.~\eqref{eq:gen_eigenvalue_empirical}, the last step is to solve the generalized eigenvalue problem $\tilde{G}^O(\theta^*_i) w = \lambda^O \tilde{G}(\theta^*_i) w$, where the resulting eigenvalues $\lambda_1^O, \dots, \lambda_{m_i}^O$ will be approximately those of the projected observable on the corresponding dominant eigenspace (defined in Eq.~\eqref{eq:defn_proj_observable}), as shown in~\cite[Section V]{QFAMES}.

\section{Rigorous Theoretical Guarantees}
\label{app:thms}

In this section, we present the rigorous formal statements for the theorems introduced informally in Section~\ref{sec:main_results}. These theorems detail the precise parametric dependencies, thresholding regimes, and explicit probability bounds required for our theoretical guarantees.

\subsection{Eigenvalue estimation}
\label{subsec:formal_eigenvalue}

\paragraph{Setup and parameter conditions.}
Suppose that Assumptions~\ref{assum:cluster_sep}, \ref{assum:dom_threshold}, and~\ref{assum:uniform_overlap} hold, with
\begin{equation}\label{eq:thm_eigenvalue_tail}
    0<\ptail<1, \qquad C_p=\frac{\pmin}{\ptail}>8(1+2\chi).
\end{equation}
Let $\eta\in(0,1)$, assume $N\ge2$ and $\tilde I\ge I$, and write
\begin{equation}\label{eq:defn_r_chi}
    r_\chi:=
    \sqrt{\log\left(\frac{2(1+2\chi)}{1+\chi}\right)}.
\end{equation}
The cluster centers satisfy $\lambda_i^*\in\mathcal I_i$, and the evaluation grid $\Theta$ contains a point within distance $q/(2T)$ of each cluster center. Assume that the algorithmic parameters satisfy the following conditions.

\noindent\textit{Filtering and grid resolution:}
\begin{equation}\label{eq:thm_eigenvalue_cond1}
\begin{aligned}
    &T\ge
    \frac{C_T}{\Delta}\log\left(\frac1{\ptail}\right),
    \qquad
    \delta T\le\frac14\ptail^2,\\
    &0<q\le c_q\frac{\ptail}{\sigma},
    \qquad
    q<\sqrt{2\log\left(\frac{1+2\chi}{1+\chi}\right)}.
\end{aligned}
\end{equation}

\noindent\textit{Blocking geometry:}
\begin{equation}\label{eq:thm_eigenvalue_cond2}
    r_\chi+
    \sqrt{\log\left(\frac{16(1+\chi)}{\pmin}\right)}
    <\alpha<
    \Delta T-r_\chi-\frac q2.
\end{equation}

\noindent\textit{Truncation:}
\begin{equation}\label{eq:truncation_precision}
    \sigma\ge
    \max\left\{2\sqrt{\log(LRN)},1\right\}.
\end{equation}
Here $C_T>0$ is a sufficiently large absolute constant, and $c_q\in(0,1]$ is a fixed absolute constant.

The notation $\widetilde{\mathcal O}$ and $\widetilde{\Omega}$ hides logarithmic factors in $|\Theta|$, $L$, $R$, and $\eta^{-1}$. The sample-size conditions use sufficiently large implicit constants, and all dependence on $\chi$ is displayed explicitly.

\begin{theorem}[Rigorous Eigenvalue Error Scaling]
\label{thm:eigenvalue_error}
Under the setup in Section~\ref{subsec:formal_eigenvalue}, suppose that Eqs.~\eqref{eq:thm_eigenvalue_tail}, \eqref{eq:thm_eigenvalue_cond1}, \eqref{eq:thm_eigenvalue_cond2}, and~\eqref{eq:truncation_precision} hold. Under either of the following sample-size conditions, with probability at least $1-\eta$, the search-and-block procedure returns at least $I$ points, and there exists a permutation $\pi:[I]\to[I]$ such that its first $I$ outputs satisfy the corresponding error bound.

\begin{enumerate}
    \item \textbf{Independent Time Sampling:} If
    \begin{equation}\label{eq:thm_eigenvalue_cond_indep}
        \sqrt{LR}\,N
        =\widetilde{\Omega}\left(
            \frac{(1+2\chi)^2}{\pmin^2}
        \right),
    \end{equation}
    then
    \begin{equation}\label{eq:thm_eigenvalue_result_indep}
        \sup_{1\le j\le I}
        |\theta_j^*-\lambda_{\pi_j}^*|
        \le
        \frac{\sigma(1+\chi)}{T\pmin}
        \left[
            \widetilde{\mathcal O}\left(
                \left(1+\frac{1+\chi}{\pmin\sqrt N}\right)
                \frac1{\sqrt{LRN}}
            \right)
            +\mathcal O(\ptail)
        \right].
    \end{equation}

    \item \textbf{Shared Time Sampling:} If
    \begin{equation}\label{eq:thm_eigenvalue_cond_shared}
        N=\widetilde{\Omega}\left(
            \frac{(1+2\chi)^2}{\pmin^2}
        \right),
    \end{equation}
    then
    \begin{equation}\label{eq:thm_eigenvalue_result_shared}
        \sup_{1\le j\le I}
        |\theta_j^*-\lambda_{\pi_j}^*|
        \le
        \frac{\sigma(1+\chi)}{T\pmin}
        \left[
            \widetilde{\mathcal O}\left(\frac1{\sqrt N}\right)
            +\mathcal O(\ptail)
        \right].
    \end{equation}
\end{enumerate}
\end{theorem}

See the proof in Appendix~\ref{app:proofs_eigenvalue}.

\begin{remark}
    Under the \textit{Independent Time Sampling} scheme, if sufficiently many time samples are available, namely,
    \begin{equation}
        N=\widetilde{\Omega}\left(\frac{(1+\chi)^2}{\pmin^2}\right),
    \end{equation}
    then the second statistical-error term in Eq.~\eqref{eq:thm_eigenvalue_result_indep} is dominated by the first. Consequently, up to logarithmic factors, the statistical error depends on $L$, $R$, and $N$ only through the total number of samples $LRN$:
    \begin{equation}
        \sup_{1\le j\le I}
        \left|\theta_j^*-\lambda_{\pi_j}^*\right|
        \le
        \frac{\sigma(1+\chi)}{T\pmin}
        \left(
            \widetilde{\mathcal{O}}\left(\frac{1}{\sqrt{LRN}}\right)
            +\mathcal{O}\left(\ptail\right)
        \right).
    \end{equation}
\end{remark}

\begin{remark}
Even under the \textit{Shared Time Sampling} scheme, Theorem~\ref{thm:eigenvalue_error} refines the eigenvalue estimation guarantee in~\cite[Theorem~2]{QFAMES}. In our normalized notation, the previous result requires $N=\widetilde{\Omega}(\ptail^{-2})$ and gives an error of order $\widetilde{\mathcal{O}}(\ptail/(T\pmin))$. For $\chi=\mathcal{O}(1)$, our result only requires $N=\widetilde{\Omega}(\pmin^{-2})$ and explicitly separates the statistical error from the tail contribution, recovering the previous error scaling when $N=\widetilde{\Omega}(\ptail^{-2})$.
\end{remark}

\subsection{Multiplicity recovery}
\label{subsec:formal_multiplicity}

\paragraph{Setup and parameter conditions.}
Suppose that all the conditions of Theorem~\ref{thm:eigenvalue_error} hold under the \textit{Independent Time Sampling} scheme. In particular, Assumption~\ref{assum:uniform_overlap} requires
\begin{equation}\label{eq:thm_multiplicity_dimension}
    \min(L,R)\ge\max_{1\le j\le I}|\D_j|,
\end{equation}
and both $\Phi_{:,\D_j}$ and $\Psi_{:,\D_j}$ have full column rank for every $j\in[I]$. Let $r_\chi$ be defined by Eq.~\eqref{eq:defn_r_chi}. Assume further that the filter width satisfies
\begin{equation}\label{eq:thm_multiplicity_T}
    r_\chi+\delta T
    <
    \sqrt{\log\left(
        \frac{\pmin}{4\ptail(1+\chi)^2}
    \right)}.
\end{equation}
This guarantees a positive normalized singular-value gap, which we further require to be of the same order as $\pmin$:
\begin{equation}\label{eq:defn_physical_gap}
    \Delta_{\text{phys}}
    :=
    \frac{\pmin}{(1+\chi)^2}
    \exp\left(-(r_\chi+\delta T)^2\right)
    -4\ptail
    =\Theta(\pmin).
\end{equation}
To reject additional candidates after all dominant clusters have been selected, assume also that
\begin{equation}\label{eq:thm_multiplicity_alpha}
    \alpha>
    r_\chi+\delta T+
    \sqrt{\log\left(\frac1{\ptail}\right)}.
\end{equation}

Then we choose a cutoff threshold
\begin{equation}\label{eq:thm_multiplicity_tau}
    \tau
    =
    \Theta\left(
        \sqrt{LR}\left(\ptail+\Delta_{\text{phys}}\right)
    \right),
\end{equation}
where the threshold lies inside the interval
\begin{equation}\label{eq:thm_multiplicity_tau_interval}
    2\sqrt{LR}\ptail
    <
    \tau
    <
    \sqrt{LR}\left(2\ptail+\Delta_{\text{phys}}\right),
\end{equation}
with both margins at least $c_\tau\sqrt{LR}\Delta_{\text{phys}}$ for some fixed absolute constant $c_\tau\in(0,1/2)$. The implicit sample-size constant is chosen sufficiently large depending on $c_\tau$.

The notation $\widetilde{\Omega}$ hides logarithmic factors in $|\Theta|$, $L$, $R$, and $\eta^{-1}$, and the implicit sample-size constant is sufficiently large.

\begin{theorem}[Rigorous Multiplicity Recovery]
\label{thm:multiplicity}
Under the setup in Section~\ref{subsec:formal_multiplicity} and the conditions of Theorem~\ref{thm:eigenvalue_error} for the \textit{Independent Time Sampling} scheme, suppose that Eqs.~\eqref{eq:thm_multiplicity_dimension}, \eqref{eq:thm_multiplicity_T}, \eqref{eq:defn_physical_gap}, and~\eqref{eq:thm_multiplicity_alpha} hold. Choose $\tau$ as in Eqs.~\eqref{eq:thm_multiplicity_tau} and~\eqref{eq:thm_multiplicity_tau_interval}, with the stated margins. If, in addition to Eq.~\eqref{eq:thm_eigenvalue_cond_indep}, the sample size satisfies
\begin{equation}\label{eq:thm_multiplicity_N}
    N=\widetilde{\Omega}\left(
        \frac{1}{\min(L,R)\Delta_{\text{phys}}^2}
    \right),
\end{equation}
then, with probability at least $1-\eta$, the search-and-block procedure returns at least $I$ points, and recovers the true multiplicities with the cutoff threshold $\tau$. Specifically,
\begin{enumerate}
    \item \textbf{Valid Clusters ($1\le i\le I$):} There exists a permutation $\pi:[I]\to[I]$ such that the estimated multiplicities satisfy $m_i=|\D_{\pi_i}|$ for all $i\in[I]$.

    \item \textbf{Spurious Candidates ($i>I$):} Every additional candidate returned by the procedure satisfies $m_i=0$.
\end{enumerate}
\end{theorem}

See the proof in Appendix~\ref{app:proofs_multiplicity}.

\begin{remark}
    After discarding all candidates with $m_i=0$, exactly $I$ peaks remain, so the number of dominant clusters is determined and we may set $\tilde I=I$ for the retained candidates.
\end{remark}

\begin{remark}
The upper bound on $T$ reflects the spectral resolution needed to recover the multiplicity of a prescribed cluster. The Gaussian filter has width of order $T^{-1}$. If $T$ is too large relative to the inverse cluster width, it begins to resolve finer spectral structure within the cluster and may strongly suppress some of its eigencomponents. Their associated singular values may then fall below the threshold, so the estimated multiplicity can correspond to only part of the intended cluster. Thus, $T$ should separate distinct clusters while retaining all eigencomponents within each target cluster. This restriction does not arise from internal spectral splitting for an exactly degenerate cluster ($\delta=0$).
\end{remark}

\subsection{Observable estimation}
\label{subsec:formal_observable}

\paragraph{Setup and parameter conditions.}
Suppose that Assumption~\ref{assum:uniform_overlap} holds and that all left and right initial states are supported on the dominant spectral subspace:
\begin{equation}\label{eq:thm_observable_no_tail}
    \Phi_{:,\mathcal D^c}=0,
    \qquad
    \Psi_{:,\mathcal D^c}=0.
\end{equation}
Let $\eta\in(0,1)$, $N\ge2$, and $\tilde I\ge I$. Assume that
\begin{equation}\label{eq:thm_observable_cond}
\begin{aligned}
    &\delta=0,\qquad \Delta>0,\qquad
    \pmin=\Omega(1),\qquad \chi=\mathcal O(1),\\
    &\Delta T\ge C_T,\qquad
    \sigma=\Theta\left(\sqrt{\log(LRN)}\right),\\
    &0<q\le\frac{c_q}{\sqrt{LRN}},\qquad
    \min(L,R)N=\widetilde{\Omega}(1).
\end{aligned}
\end{equation}
Here $C_T$ and the implicit sample-size constant are sufficiently large, and $c_q>0$ is sufficiently small. The grid covers every cluster center within distance $q/(2T)$, and the blocking and truncation conditions in Eqs.~\eqref{eq:thm_eigenvalue_cond2} and~\eqref{eq:truncation_precision} hold.

Under the \textit{Independent Time Sampling} scheme, assume also that the observable sample blocks $(t_{l,r,n},t'_{l,r,n},Z^O_{l,r,n})$ are independent across $(l,r,n)$ and satisfy $|Z^O_{l,r,n}|\le C_O\|O\|_2$ for an absolute constant $C_O$. Moreover, use the cutoff
\begin{equation}\label{eq:thm_observable_tau}
    \tau=\sqrt{LR}\pmin e^{-r_\chi^2}/(2(1+\chi)^2),
\end{equation}
where $r_\chi$ is defined in Eq.~\eqref{eq:defn_r_chi}.

The notation $\widetilde{\mathcal O}$ and $\widetilde{\Omega}$ hides logarithmic factors in $|\Theta|$, $L$, $R$, $N$, and $\eta^{-1}$.

\begin{theorem}[Observable Estimation Error]
\label{thm:observable}
Under the setup in Section~\ref{subsec:formal_observable} and the \textit{Independent Time Sampling} scheme, suppose that Eqs.~\eqref{eq:thm_observable_no_tail}, \eqref{eq:thm_observable_cond}, \eqref{eq:thm_eigenvalue_cond2}, and~\eqref{eq:truncation_precision} hold, and use the cutoff in Eq.~\eqref{eq:thm_observable_tau}. Then, with probability at least $1-\eta$, the first $I$ outputs recover the dominant clusters up to a permutation $\pi:[I]\to[I]$ and their multiplicities $m_i=|\mathcal D_{\pi_i}|$. On the same event, for each $i\in[I]$, there exists a permutation $\rho\in S_{m_i}$ such that the generalized eigenvalues from Eq.~\eqref{eq:gen_eigenvalue_empirical} satisfy
\begin{equation}\label{eq:thm_observable_result}
    \max_{1\le k\le m_i}
    |\lambda_k^O-\lambda_{\rho(k),\mathrm{exact}}|\le
    \widetilde{\mathcal O}\left(
        \|O\|_2\left[
            \frac1{\sqrt{LRN}}
            +\frac1{\min(L,R)N}
        \right]
    \right)
    +\mathcal O\left(
        \|O\|_2e^{-\Delta^2T^2/4}
    \right),
\end{equation}
where $\{\lambda_{j,\mathrm{exact}}\}_{j=1}^{m_i}$ are the eigenvalues of $O_{\mathcal D_{\pi_i}}$. Here $S_{m_i}$ denotes the permutations of $[m_i]$, and both spectra are counted with algebraic multiplicity.
\end{theorem}

See the proof in Appendix~\ref{app:proofs_observable}.

\section{Proofs of Main Theorems}
\label{app:proofs}

\subsection{Proof of Theorem~\ref{thm:eigenvalue_error}}
\label{app:proofs_eigenvalue}

\begin{proof}[Proof of Theorem~\ref{thm:eigenvalue_error}]

\textbf{Step 0: Uniform error bounds.} The truncation condition in Eq.~\eqref{eq:truncation_precision} ensures that Lemmas~\ref{lemma:error_term} and~\ref{lemma:cross_terms} apply to the total error $\E=G-\F$. Applying each of the five required inequalities with failure probability $\eta/5$ and taking a union bound, we obtain the following bounds simultaneously for all $\theta,\theta'\in\Theta$, with probability at least $1-\eta$.

\noindent \textbf{Independent Time Sampling:}
\begin{equation}\label{eq:prob_eqs_indep}
\begin{aligned}
    &\left|
        \|\E(\theta)\|_F^2-\|\E(\theta')\|_F^2
        +\frac{\|\F(\theta)\|_F^2-\|\F(\theta')\|_F^2}{N}
    \right|\le
    \widetilde{\mathcal O}\left(\frac{\sqrt{LR}}{N}\right)
    \min\{1,\sigma T|\theta-\theta'|\},\\
    &|2\RE\langle\F(\theta),\E(\theta)\rangle|
    \le
    \widetilde{\mathcal O}\left(\frac1{\sqrt N}\right)
    \|\F(\theta)\|_F,\\
    &|2\RE\langle\F(\theta)-\F(\theta'),\E(\theta)\rangle|
    \le
    \widetilde{\mathcal O}\left(\frac1{\sqrt N}\right)
    \|\F(\theta)-\F(\theta')\|_F,\\
    &|2\RE\langle\F(\theta'),\E(\theta)-\E(\theta')\rangle|
    \le
    \widetilde{\mathcal O}\left(\frac1{\sqrt N}\right)
    \sigma T|\theta-\theta'|\|\F(\theta')\|_F.
\end{aligned}
\end{equation}

\noindent \textbf{Shared Time Sampling:}
\begin{equation}\label{eq:prob_eqs_shared}
\begin{aligned}
    &\left|
        \|\E(\theta)\|_F^2-\|\E(\theta')\|_F^2
    \right|
    \le
    \widetilde{\mathcal O}\left(\frac{LR}{N}\right)
    \min\{1,\sigma T|\theta-\theta'|\},\\
    &|2\RE\langle\F(\theta),\E(\theta)\rangle|
    \le
    \widetilde{\mathcal O}\left(\sqrt{\frac{LR}{N}}\right)
    \|\F(\theta)\|_F,\\
    &|2\RE\langle\F(\theta)-\F(\theta'),\E(\theta)\rangle|
    \le
    \widetilde{\mathcal O}\left(\sqrt{\frac{LR}{N}}\right)
    \|\F(\theta)-\F(\theta')\|_F,\\
    &|2\RE\langle\F(\theta'),\E(\theta)-\E(\theta')\rangle|
    \le
    \widetilde{\mathcal O}\left(\sqrt{\frac{LR}{N}}\right)
    \sigma T|\theta-\theta'|\|\F(\theta')\|_F.
\end{aligned}
\end{equation}
The first line in each case combines the absolute and Lipschitz bounds. We work on this event throughout the proof.

\medskip
\textbf{Step 1: Coarse localization and distinct clusters.} For each selected point $\theta_j^*$, let $\pi_j$ be an index of a nearest spectral interval $\mathcal I_i$, with ties resolved arbitrarily. We show that $\pi_1,\ldots,\pi_I$ are distinct and
\begin{equation}\label{eq:step1_error_bound}
    \sup_{1\le j\le I}
    |\theta_j^*-\lambda_{\pi_j}^*|
    \le\frac{r_\chi}{T}.
\end{equation}

We first record a comparison bound that will be used both for the first iteration and for the induction. Since $G=\F+\E$,
\begin{equation}\label{eq:squared_norm_decomp}
    \|G(\theta)\|_F^2
    =
    \|\F(\theta)\|_F^2
    +2\RE\langle\F(\theta),\E(\theta)\rangle
    +\|\E(\theta)\|_F^2.
\end{equation}
Suppose that $\|G(\theta)\|_F\ge\|G(\theta')\|_F$. Under independent sampling, substituting Eq.~\eqref{eq:prob_eqs_indep} into Eq.~\eqref{eq:squared_norm_decomp} gives
\begin{equation}\label{eq:squared_norm_diff}
    0\le
    \left(1-\frac1N\right)
    \left(\|\F(\theta)\|_F^2-\|\F(\theta')\|_F^2\right)+
    \widetilde{\mathcal O}\left(
        \frac{\|\F(\theta)\|_F+\|\F(\theta')\|_F}{\sqrt N}
    \right)
    +\widetilde{\mathcal O}\left(\frac{\sqrt{LR}}{N}\right).
\end{equation}
Using $N\ge2$, dividing by $\|\F(\theta)\|_F+\|\F(\theta')\|_F$, and applying the same argument under shared sampling, we obtain
\begin{equation}\label{eq:signal_comparison_general}
\begin{aligned}
    &\|\F(\theta')\|_F-\|\F(\theta)\|_F\\
    &\le
    \begin{cases}
        \displaystyle
        \widetilde{\mathcal O}\left(\frac1{\sqrt N}\right)
        +
        \widetilde{\mathcal O}\left(
            \frac{\sqrt{LR}}
            {N(\|\F(\theta)\|_F+\|\F(\theta')\|_F)}
        \right),
        &\text{(Independent)},\\[3mm]
        \displaystyle
        \widetilde{\mathcal O}\left(\sqrt{\frac{LR}{N}}\right)
        +
        \widetilde{\mathcal O}\left(
            \frac{LR}
            {N(\|\F(\theta)\|_F+\|\F(\theta')\|_F)}
        \right),
        &\text{(Shared)}.
    \end{cases}
\end{aligned}
\end{equation}
This bound applies whenever the denominator is positive.

Let $U_{\min}:=\sqrt{LR}\pmin/(1+\chi)$. By Lemma~\ref{lemma:bound_U}, every cluster amplitude is at least $U_{\min}$. Suppose that $\|G(\theta)\|_F\ge\|G(\theta')\|_F$ and $\|\F(\theta)\|_F+\|\F(\theta')\|_F\ge U_{\min}/2$. Dividing Eq.~\eqref{eq:signal_comparison_general} by $U_{\min}$ gives
\begin{equation}\label{eq:signal_comparison_general_ratio}
\begin{aligned}
    &\frac{\|\F(\theta')\|_F-\|\F(\theta)\|_F}{U_{\min}}\\
    &\le
    \begin{cases}
        \displaystyle
        \widetilde{\mathcal O}\left(
            \frac{1+\chi}{\pmin\sqrt{LRN}}
            +\frac{(1+\chi)^2}{\pmin^2\sqrt{LR}\,N}
        \right),
        &\text{(Independent)},\\[3mm]
        \displaystyle
        \widetilde{\mathcal O}\left(
            \frac{1+\chi}{\pmin\sqrt N}
            +\frac{(1+\chi)^2}{\pmin^2N}
        \right),
        &\text{(Shared)}.
    \end{cases}
\end{aligned}
\end{equation}
Substituting the sample-size conditions in Eqs.~\eqref{eq:thm_eigenvalue_cond_indep} and~\eqref{eq:thm_eigenvalue_cond_shared} into Eq.~\eqref{eq:signal_comparison_general_ratio}, we obtain
\begin{equation}\label{eq:small_signal_comparison}
    \|\F(\theta')\|_F-\|\F(\theta)\|_F
    \le\frac{U_{\min}}{16},
\end{equation}
where the implicit constants in the sample-size conditions are chosen sufficiently large.

We next bound the deterministic signal $\F(\theta)=A(\theta)+B(\theta)+C(\theta)$. The tail term satisfies
\begin{equation}\label{eq:B_bound}
    \|B(\theta)\|_F
    \le
    \sum_{m\in\mathcal D^c}\sqrt{LR}\bar p_m
    =\sqrt{LR}\ptail.
\end{equation}
If $\mathcal I_{\pi_\theta}$ is a nearest spectral interval to $\theta$, then every eigenvalue in another cluster is at least $\Delta/2$ away from $\theta$. Indeed, this is immediate when $\theta\in\mathcal I_{\pi_\theta}$. Otherwise, let $\lambda_*$ be a nearest endpoint of $\mathcal I_{\pi_\theta}$ to $\theta$. For any $m\in\mathcal D\setminus\mathcal D_{\pi_\theta}$, the nearest-interval property gives $|\theta-\lambda_*|\le|\theta-\lambda_m|$. Since $\lambda_*$ is an eigenvalue in the target cluster,
\begin{equation}
    \Delta
    \le|\lambda_m-\lambda_*|
    \le|\lambda_m-\theta|+|\theta-\lambda_*|
    \le2|\theta-\lambda_m|.
\end{equation}

By Eq.~\eqref{eq:defn_p_m_avg}, $\|\Phi_{:,m}\Psi_{:,m}^\dagger\|_F =\sqrt{LR}\bar p_m$. Moreover, Cauchy--Schwarz and normalization of the initial states give
\begin{equation}\label{eq:normalized_overlap_sum}
    \sum_{m=1}^M\bar p_m
    =
    \frac1{\sqrt{LR}}
    \sum_{m=1}^M
        \|\Phi_{:,m}\|_2\|\Psi_{:,m}\|_2
    \le
    \frac{\|\Phi\|_F\|\Psi\|_F}{\sqrt{LR}}
    =1.
\end{equation}
Thus, expanding $C(\theta)$ using Eq.~\eqref{eq:defn_C}, we obtain
\begin{equation}\label{eq:C_bound}
\begin{aligned}
    \|C(\theta)\|_F
    \le
    \sqrt{LR}
    \sum_{m\in\mathcal D\setminus\mathcal D_{\pi_\theta}}
        \bar p_m e^{-T^2(\theta-\lambda_m)^2}\le
    \sqrt{LR}e^{-\Delta^2T^2/4}\le\frac14\sqrt{LR}\ptail.
\end{aligned}
\end{equation}
The last inequality follows from Eq.~\eqref{eq:thm_eigenvalue_cond1} with sufficiently large $C_T$.

For the target cluster, write
\begin{equation}\label{eq:A_decomp}
    \begin{aligned}
        A(\theta) &= \underbrace{\Phi_{:,\mathcal{D}_{\pi_\theta}} \cdot \underbrace{\diag\left(\left\{\exp\left(-(\theta - \lambda_{\pi_\theta}^*)^2T^2\right)\right\}_{m \in \mathcal{D}_{\pi_\theta}}\right)}_{\text{constant diagonal}} \cdot \left(\Psi_{:,\mathcal{D}_{\pi_\theta}}\right)\cj}_{A_{\mathrm{center}}(\theta)} \\
        &+ \underbrace{\Phi_{:,\mathcal{D}_{\pi_\theta}} \cdot \diag\left(\left\{\exp\left(-(\theta - \lambda_m)^2T^2\right)-\exp\left(-(\theta - \lambda_{\pi_\theta}^*)^2T^2\right)\right\}_{m \in \mathcal{D}_{\pi_\theta}}\right) \cdot \left(\Psi_{:,\mathcal{D}_{\pi_\theta}}\right)\cj}_{\Delta A_{\mathrm{width}}(\theta)}
    \end{aligned}
\end{equation}

Since $|\lambda_m-\lambda_{\pi_\theta}^*|\le\delta$ for $m\in\mathcal D_{\pi_\theta}$ and $|\frac{d}{dx}e^{-T^2x^2}|\le T$, Eq.~\eqref{eq:normalized_overlap_sum} gives
\begin{equation}\label{eq:A2_bound}
\begin{aligned}
    \|\Delta A_{\mathrm{width}}(\theta)\|_F
    &\le
    \delta T
    \sum_{m\in\mathcal D_{\pi_\theta}}
        \|\Phi_{:,m}\Psi_{:,m}^\dagger\|_F\\
    &=
    \delta T\sqrt{LR}
    \sum_{m\in\mathcal D_{\pi_\theta}}\bar p_m
    \le\delta T\sqrt{LR}
    \le\frac14\sqrt{LR}\ptail^2.
\end{aligned}
\end{equation}
The last inequality follows from the cluster-width condition in Eq.~\eqref{eq:thm_eigenvalue_cond1}. Also,
\begin{equation}\label{eq:A1_bound}
    \|A_{\mathrm{center}}(\theta)\|_F
    =
    e^{-(\theta-\lambda_{\pi_\theta}^*)^2T^2}
    \left\|
        \Phi_{:,\mathcal D_{\pi_\theta}}
        \left(\Psi_{:,\mathcal D_{\pi_\theta}}\right)^\dagger
    \right\|_F.
\end{equation}

Consider the first selected point $\theta_1^*$. Let $\tilde\theta_1^*$ be a grid point nearest to $\lambda_{\pi_1}^*$, and set $U:=\|\Phi_{:,\mathcal D_{\pi_1}} (\Psi_{:,\mathcal D_{\pi_1}})^\dagger\|_F$. Then $|\tilde\theta_1^*-\lambda_{\pi_1}^*|\le q/(2T)$ and $U_{\min}\le U\le\sqrt{LR}$, where $U_{\min}=\sqrt{LR}\pmin/(1+\chi)$. The blocking condition given by Eq.~\eqref{eq:thm_eigenvalue_cond2} ensures that $\mathcal I_{\pi_1}$ is also the nearest cluster interval to $\tilde\theta_1^*$. By Eqs.~\eqref{eq:B_bound}--\eqref{eq:A1_bound},
\begin{equation}\label{eq:F_tilde_theta1_unsquared}
    \|\F(\tilde\theta_1^*)\|_F
    \ge e^{-q^2/4}U-\frac32\sqrt{LR}\ptail,
\end{equation}
and
\begin{equation}\label{eq:F_theta1_unsquared}
    \|\F(\theta_1^*)\|_F
    \le
    e^{-(\theta_1^*-\lambda_{\pi_1}^*)^2T^2}U
    +\frac32\sqrt{LR}\ptail.
\end{equation}
The conditions on $q$ and $C_p$ imply $e^{-q^2/4}>1/\sqrt2$ and $\sqrt{LR}\ptail<U_{\min}/8$. Hence Eq.~\eqref{eq:F_tilde_theta1_unsquared} gives $\|\F(\tilde\theta_1^*)\|_F>U/2$.

Since $\|G(\theta_1^*)\|_F\ge
\|G(\tilde\theta_1^*)\|_F$ and $\|\F(\theta_1^*)\|_F +
\|\F(\tilde\theta_1^*)\|_F \ge U_{\min}/2$, we may apply
Eq.~\eqref{eq:small_signal_comparison}. Combining it with Eqs.~\eqref{eq:F_tilde_theta1_unsquared} and~\eqref{eq:F_theta1_unsquared}, we obtain
\begin{equation}\label{eq:squared_norm_diff_4}
\begin{aligned}
    e^{-(\theta_1^*-\lambda_{\pi_1}^*)^2T^2}
    &\ge
    e^{-q^2/4}
    -\frac{3\sqrt{LR}\ptail}{U}
    -\frac{U_{\min}}{16U}\\
    &\ge
    \frac{1+\chi}{1+2\chi}
    -\frac{3(1+\chi)}{8(1+2\chi)}
    -\frac{1+\chi}{8(1+2\chi)}\\
    &=\frac{1+\chi}{2(1+2\chi)}.
\end{aligned}
\end{equation}
Therefore,
\begin{equation}\label{eq:theta1_diff}
    |\theta_1^*-\lambda_{\pi_1}^*| \le \frac{1}{T}\sqrt{\log\left(\frac{2(1+2\chi)}{1+\chi}\right)}
    =\frac{r_\chi}{T}.
\end{equation}

We now extend this result by induction. Suppose that the first $j-1$ selected points correspond to distinct clusters and satisfy $|\theta_k^*-\lambda_{\pi_k}^*|\le r_\chi/T$, for all $1 \le k \le j-1$. For any undiscovered cluster $i$, let $\tilde\theta_i$ be a grid point within $q/(2T)$ of $\lambda_i^*$. By the triangle inequality and Eq.~\eqref{eq:thm_eigenvalue_cond2}, for every $k<j$,
\begin{equation}\label{eq:unblocked_center}
    |\tilde\theta_i-\theta_k^*|
    \ge |\lambda_i^*-\lambda_{\pi_k}^*| - |\theta_k^*-\lambda_{\pi_k}^*| - |\tilde\theta_i-\lambda_i^*| \ge
    \Delta-\frac{r_\chi}{T}-\frac{q}{2T}
    >\frac{\alpha}{T}.
\end{equation}
Thus $\tilde\theta_i$ remains available, and the same bound as in Eq.~\eqref{eq:F_tilde_theta1_unsquared} gives $\|\F(\tilde\theta_i)\|_F>U_{\min}/2$.

If $\pi_j=\pi_k$ for some $k<j$, then $T|\theta_j^*-\lambda_{\pi_k}^*|>\alpha-r_\chi$, because $\theta_j^*$ is outside the blocked interval centered at $\theta_k^*$. Using the same upper bound as in Eq.~\eqref{eq:F_theta1_unsquared} and the blocking condition Eq.~\eqref{eq:thm_eigenvalue_cond2}, we have
\begin{equation}\label{eq:discovered_signal_upper}
    \|\F(\theta_j^*)\|_F
    \le
    e^{-(\alpha-r_\chi)^2}\sqrt{LR}
    +\frac32\sqrt{LR}\ptail <\frac1{16}U_{\min}+\frac3{16}U_{\min}
    =\frac14U_{\min}.
\end{equation}
However, $\theta_j^*$ maximizes the empirical norm over the remaining grid, so $\|G(\theta_j^*)\|_F\ge\|G(\tilde\theta_i)\|_F$. Equation~\eqref{eq:small_signal_comparison} would then give $\|\F(\tilde\theta_i)\|_F-\|\F(\theta_j^*)\|_F
\le U_{\min}/16$, contradicting the gap
greater than $U_{\min}/4$ established above.

Therefore $\pi_j$ is an undiscovered cluster. Its center grid point is available by Eq.~\eqref{eq:unblocked_center}, so the first-iteration argument applies and gives $|\theta_j^*-\lambda_{\pi_j}^*|\le r_\chi/T$. This completes the induction and proves Eq.~\eqref{eq:step1_error_bound}. In particular, $\pi_1,\ldots,\pi_I$ define a permutation of $[I]$.

\medskip
\textbf{Step 2: Refined error bounds.} We now use the local variation of the error terms to refine Eq.~\eqref{eq:step1_error_bound}. We present the argument for $j=1$. For every $j\le I$, Step~1 guarantees that the corresponding center grid point remains available at iteration $j$, so the same comparison and estimates apply.

Recall $U=\|\Phi_{:,\mathcal D_{\pi_1}} (\Psi_{:,\mathcal D_{\pi_1}})^\dagger\|_F$ and $\tilde\theta_1^*$ from Step~1, and let $\delta_\theta:=T|\theta_1^*-\lambda_{\pi_1}^*|$. By Eq.~\eqref{eq:squared_norm_decomp},
\begin{equation}\label{eq:squared_norm_diff_s2}
\begin{aligned}
    0
    &\le
    \|G(\theta_1^*)\|_F^2
    -\|G(\tilde\theta_1^*)\|_F^2
    =X+Y+Z,\\
    X&:=
    \|\F(\theta_1^*)\|_F^2
    -\|\F(\tilde\theta_1^*)\|_F^2,\\
    Y&:=
    2\RE\langle\F(\theta_1^*),\E(\theta_1^*)\rangle
    -2\RE\langle\F(\tilde\theta_1^*),
                  \E(\tilde\theta_1^*)\rangle,\\
    Z&:=
    \|\E(\theta_1^*)\|_F^2
    -\|\E(\tilde\theta_1^*)\|_F^2.
\end{aligned}
\end{equation}

\noindent\textit{Evaluation of $X$.} Using $\F=A+B+C$, write $X=X_1+X_2+X_3$, where
\begin{equation}\label{eq:F_diff_decomp}
\begin{aligned}
    X_1&:=
    \|A(\theta_1^*)\|_F^2
    -\|A(\tilde\theta_1^*)\|_F^2,\\
    X_2&:=
    \|B(\theta_1^*)+C(\theta_1^*)\|_F^2
    -\|B(\tilde\theta_1^*)+C(\tilde\theta_1^*)\|_F^2,\\
    X_3&:=
    2\RE\langle A(\theta_1^*),
                   B(\theta_1^*)+C(\theta_1^*)\rangle-
    2\RE\langle A(\tilde\theta_1^*),
                   B(\tilde\theta_1^*)+C(\tilde\theta_1^*)\rangle.
\end{aligned}
\end{equation}
By Eqs.~\eqref{eq:A1_bound} and~\eqref{eq:thm_eigenvalue_cond1},
\begin{equation}
    \|A_{\mathrm{center}}(\tilde\theta_1^*)\|_F
    \ge e^{-q^2/4}U
    >\frac{U}{\sqrt2}.
\end{equation}
On the other hand, Eq.~\eqref{eq:A2_bound}, Assumption~\ref{assum:dom_threshold}, and Lemma~\ref{lemma:bound_U} give
\begin{equation}
    \|\Delta A_{\mathrm{width}}(\tilde\theta_1^*)\|_F
    \le\frac14\sqrt{LR}\ptail^2
    \le\frac{(1+\chi)\ptail}{4C_p}U
    <\frac{U}{32},
\end{equation}
where we used $\ptail<1$ and $C_p>8(1+2\chi)$. Thus $\|A_{\mathrm{center}}(\tilde\theta_1^*)\|_F -\|\Delta A_{\mathrm{width}}(\tilde\theta_1^*)\|_F>0$, so we may square the reverse triangle inequality to obtain
\begin{equation}\label{eq:X1_bound}
\begin{aligned}
    X_1
    &\le
    \left(
        \|A_{\mathrm{center}}(\theta_1^*)\|_F
        +\|\Delta A_{\mathrm{width}}(\theta_1^*)\|_F
    \right)^2\\
    &\quad-
    \left(
        \|A_{\mathrm{center}}(\tilde\theta_1^*)\|_F
        -\|\Delta A_{\mathrm{width}}(\tilde\theta_1^*)\|_F
    \right)^2\\
    &\le
    -\left(e^{-q^2/2}-e^{-2\delta_\theta^2}\right)U^2
    +\sqrt{LR}\ptail^2U+\frac1{16}LR\ptail^4.
\end{aligned}
\end{equation}
Also, Eqs.~\eqref{eq:B_bound} and~\eqref{eq:C_bound} give
\begin{equation}\label{eq:X2_bound}
    X_2
    \le
    \left(\|B(\theta_1^*)\|_F+\|C(\theta_1^*)\|_F\right)^2
    \le\frac{25}{16}LR\ptail^2.
\end{equation}

For $X_3$, Cauchy--Schwarz gives
\begin{equation}\label{eq:X3_bound_0}
\begin{aligned}
    X_3
    &\le
    2\|A(\theta_1^*)-A(\tilde\theta_1^*)\|_F
      \|B(\theta_1^*)+C(\theta_1^*)\|_F\\
    &\quad+
    2\|A(\tilde\theta_1^*)\|_F
      \|B(\theta_1^*)-B(\tilde\theta_1^*)
        +C(\theta_1^*)-C(\tilde\theta_1^*)\|_F.
\end{aligned}
\end{equation}
For the Gaussian filter, we have
\begin{equation}\label{eq:gaussian_derivative_bound}
    \left|\frac{d}{dx}e^{-T^2x^2}\right|
    =2T^2|x|e^{-T^2x^2}
    \le\sqrt{\frac2e}\,T
    \le T.
\end{equation}
Since $\theta_1^*$ and $\tilde\theta_1^*$ correspond to the same cluster $\mathcal D_{\pi_1}$, Eq.~\eqref{eq:A_decomp} gives
\begin{equation}\label{eq:lipschitz_A}
\begin{aligned}
    \|A(\theta_1^*)-A(\tilde\theta_1^*)\|_F
    &\le
    \left|
        e^{-T^2(\theta_1^*-\lambda_{\pi_1}^*)^2}
        -e^{-T^2(\tilde\theta_1^*-\lambda_{\pi_1}^*)^2}
    \right|U\\
    &\quad+
    \|\Delta A_{\mathrm{width}}(\theta_1^*)\|_F
    +\|\Delta A_{\mathrm{width}}(\tilde\theta_1^*)\|_F\\
    &\le
    TU|\theta_1^*-\tilde\theta_1^*|
    +2\delta T\sqrt{LR},
\end{aligned}
\end{equation}
where the last inequality follows from Eq.~\eqref{eq:gaussian_derivative_bound}, Eq.~\eqref{eq:A2_bound} and Lipschitz continuity.

Similarly, expanding $B$ using Eq.~\eqref{eq:defn_B} and noting that $\|\Phi_{:,m}\Psi_{:,m}^\dagger\|_F =\sqrt{LR}\bar p_m$, we obtain
\begin{equation}\label{eq:lipschitz_B}
\begin{aligned}
    \|B(\theta_1^*)-B(\tilde\theta_1^*)\|_F
    &\le
    \sqrt{LR}\sum_{m\in\mathcal D^c}\bar p_m
    \left|
        e^{-T^2(\theta_1^*-\lambda_m)^2}
        -e^{-T^2(\tilde\theta_1^*-\lambda_m)^2}
    \right|\\
    &\le
    \sqrt{LR}\sum_{m\in\mathcal D^c}\bar p_m
    T|\theta_1^*-\tilde\theta_1^*|\\
    &=\sqrt{LR}\ptail T|\theta_1^*-\tilde\theta_1^*|.
\end{aligned}
\end{equation}

For $C$, Step~1 gives $|\theta_1^*-\lambda_{\pi_1}^*|\le r_\chi/T$, while $|\tilde\theta_1^*-\lambda_{\pi_1}^*|
\le q/(2T)<r_\chi/T$.
Hence every point $s$ on the segment joining $\theta_1^*$ and $\tilde\theta_1^*$ satisfies $|s-\lambda_{\pi_1}^*|\le r_\chi/T$. Since $\lambda_{\pi_1}^*\in\mathcal I_{\pi_1}$, cluster separation and Eq.~\eqref{eq:thm_eigenvalue_cond2} imply
\begin{equation}
    |s-\lambda_m|
    \ge
    |\lambda_m-\lambda_{\pi_1}^*|
    -|s-\lambda_{\pi_1}^*|
    \ge\Delta-\frac{r_\chi}{T}
    >\frac{\Delta}{2},
    \qquad m\in\mathcal D\setminus\mathcal D_{\pi_1}.
\end{equation}
Here we used $\Delta T>2r_\chi$.

The function $x\mapsto 2T^2x e^{-T^2x^2}$ is decreasing for $x\ge1/(\sqrt2T)$. Thus, when $\Delta T\ge\sqrt2$,
\begin{equation}
    \sup_{|x|\ge\Delta/2}
    \left|\frac{d}{dx}e^{-T^2x^2}\right|
    =\Delta T^2e^{-\Delta^2T^2/4}.
\end{equation}
Apply the mean value theorem to each Gaussian term in Eq.~\eqref{eq:defn_C}. Using $\|\Phi_{:,m}\Psi_{:,m}^\dagger\|_F=\sqrt{LR}\bar p_m$ and Eq.~\eqref{eq:normalized_overlap_sum}, we obtain
\begin{equation}\label{eq:lipschitz_C}
\begin{aligned}
    \|C(\theta_1^*)-C(\tilde\theta_1^*)\|_F
    &\le
    \sqrt{LR}
    \sum_{m\in\mathcal D\setminus\mathcal D_{\pi_1}}
    \bar p_m\,
    \Delta T^2e^{-\Delta^2T^2/4}
    |\theta_1^*-\tilde\theta_1^*|\\
    &\le
    \sqrt{LR}\,\Delta T e^{-\Delta^2T^2/4}
    T|\theta_1^*-\tilde\theta_1^*|\\
    &\le
    \sqrt{LR}\ptail T|\theta_1^*-\tilde\theta_1^*|.
\end{aligned}
\end{equation}
The last inequality uses $\Delta T e^{-\Delta^2T^2/4}\le\ptail$, which follows from Eq.~\eqref{eq:thm_eigenvalue_cond1} with sufficiently large $C_T$.

By Eqs.~\eqref{eq:A_decomp}, \eqref{eq:A2_bound}, and~\eqref{eq:A1_bound}, $\|A(\tilde\theta_1^*)\|_F
\le U+\sqrt{LR}\ptail^2/4$.
Substituting Eqs.~\eqref{eq:lipschitz_A}, \eqref{eq:lipschitz_B}, and~\eqref{eq:lipschitz_C} into Eq.~\eqref{eq:X3_bound_0}, and using Eqs.~\eqref{eq:B_bound} and~\eqref{eq:C_bound}, we establish
\begin{equation}\label{eq:X3_bound}
\begin{aligned}
    X_3
    &\le
    2\left(
        TU|\theta_1^*-\tilde\theta_1^*|
        +\frac12\sqrt{LR}\ptail^2
    \right)\frac54\sqrt{LR}\ptail\\
    &\quad+
    4\left(U+\frac14\sqrt{LR}\ptail^2\right)
    \sqrt{LR}\ptail T|\theta_1^*-\tilde\theta_1^*|\\
    &=
    \left(
        \frac{13}{2}\sqrt{LR}\ptail U
        +LR\ptail^3
    \right)T|\theta_1^*-\tilde\theta_1^*|
    +\frac54LR\ptail^3\\
    &\le
    \mathcal O(\sqrt{LR}\ptail U)
    T|\theta_1^*-\tilde\theta_1^*|
    +\mathcal O(LR\ptail^3).
\end{aligned}
\end{equation}
For the last inequality, Lemma~\ref{lemma:bound_U} and Assumption~\ref{assum:dom_threshold} imply
\begin{equation}
    \frac{\sqrt{LR}\ptail}{U}
    \le\frac{(1+\chi)\ptail}{\pmin}
    \le\frac{1+\chi}{C_p}
    <\frac18.
\end{equation}
Hence $LR\ptail^3
\le(\ptail/8)\sqrt{LR}\ptail U$,
which controls the second term in the coefficient of $T|\theta_1^*-\tilde\theta_1^*|$.

Combining Eqs.~\eqref{eq:X1_bound}, \eqref{eq:X2_bound}, and~\eqref{eq:X3_bound}, we obtain
\begin{equation}\label{eq:X_bound}
    X
    \le
    -\left(e^{-q^2/2}-e^{-2\delta_\theta^2}\right)U^2+
    \mathcal O(\sqrt{LR}\ptail U)
    T|\theta_1^*-\tilde\theta_1^*|
    +\mathcal O(LR\ptail^2).
\end{equation}
Here $U\le\sqrt{LR}$ and $\ptail<1$ allow $\sqrt{LR}\ptail^2U$, $LR\ptail^3$, and $LR\ptail^4$ to be absorbed into $\mathcal O(LR\ptail^2)$.

\noindent\textit{Evaluation of $Y$.} Write
\begin{equation}
    Y=
    2\RE\langle
        \F(\theta_1^*)-\F(\tilde\theta_1^*),
        \E(\theta_1^*)
    \rangle+
    2\RE\langle
        \F(\tilde\theta_1^*),
        \E(\theta_1^*)-\E(\tilde\theta_1^*)
    \rangle.
\end{equation}
Equations~\eqref{eq:lipschitz_A}--\eqref{eq:lipschitz_C} give $\|\F(\theta_1^*)-\F(\tilde\theta_1^*)\|_F
\le\mathcal O(U)T|\theta_1^*-\tilde\theta_1^*|
+\mathcal O(\sqrt{LR}\ptail^2)$, while $\|\F(\tilde\theta_1^*)\|_F=\mathcal O(U)$. Applying Eq.~\eqref{eq:prob_eqs_indep} or Eq.~\eqref{eq:prob_eqs_shared}, and using $\ptail^2\le\ptail$, we obtain
\begin{equation}\label{eq:Y_bound}
    Y\le
    \begin{cases}
        \displaystyle
        \widetilde{\mathcal O}\left(\frac{U}{\sqrt N}\right)
        \sigma T|\theta_1^*-\tilde\theta_1^*|
        +
        \widetilde{\mathcal O}\left(
            \frac{\sqrt{LR}\ptail}{\sqrt N}
        \right),
        &\text{(Independent)},\\[3mm]
        \displaystyle
        \widetilde{\mathcal O}\left(
            \frac{\sqrt{LR}U}{\sqrt N}
        \right)
        \sigma T|\theta_1^*-\tilde\theta_1^*|
        +
        \widetilde{\mathcal O}\left(
            \frac{LR\ptail}{\sqrt N}
        \right),
        &\text{(Shared)}.
    \end{cases}
\end{equation}

\noindent\textit{Evaluation of $Z$.} The local squared-error bounds in Eqs.~\eqref{eq:prob_eqs_indep} and~\eqref{eq:prob_eqs_shared} give
\begin{equation}\label{eq:Z_bound}
    Z\le
    \begin{cases}
        \displaystyle
        -\frac XN+
        \widetilde{\mathcal O}\left(\frac{\sqrt{LR}}N\right)
        \sigma T|\theta_1^*-\tilde\theta_1^*|,
        &\text{(Independent)},\\[3mm]
        \displaystyle
        \widetilde{\mathcal O}\left(\frac{LR}N\right)
        \sigma T|\theta_1^*-\tilde\theta_1^*|,
        &\text{(Shared)}.
    \end{cases}
\end{equation}
Under independent sampling, combining $X$ with $-X/N$ gives $(1-1/N)X$. Since $1/2\le1-1/N<1$, division by this factor changes only absolute constants. Under shared sampling, the coefficient in Eq.~\eqref{eq:Z_bound} is absorbed into the first coefficient in Eq.~\eqref{eq:Y_bound}, using Eq.~\eqref{eq:thm_eigenvalue_cond_shared} and $U\ge\sqrt{LR}\pmin/(1+\chi)$.

Substituting Eqs.~\eqref{eq:X_bound}, \eqref{eq:Y_bound}, and~\eqref{eq:Z_bound} into Eq.~\eqref{eq:squared_norm_diff_s2}, and applying the preceding absorption of the $Z$ term, we have
\begin{equation}\label{eq:master_ineq}
\begin{aligned}
    &\left(e^{-q^2/2}-e^{-2\delta_\theta^2}\right)U^2\\
    &\le
    \mathcal O(\sqrt{LR}\ptail U)
    T|\theta_1^*-\tilde\theta_1^*|
    +\mathcal O(LR\ptail^2)\\
    &\quad+
    \begin{cases}
        \displaystyle
        \widetilde{\mathcal O}\left(
            \frac{U}{\sqrt N}+\frac{\sqrt{LR}}N
        \right)
        \sigma T|\theta_1^*-\tilde\theta_1^*|
        +
        \widetilde{\mathcal O}\left(
            \frac{\sqrt{LR}\ptail}{\sqrt N}
        \right),
        &\text{(Independent)},\\[3mm]
        \displaystyle
        \widetilde{\mathcal O}\left(
            \frac{\sqrt{LR}U}{\sqrt N}
        \right)
        \sigma T|\theta_1^*-\tilde\theta_1^*|
        +
        \widetilde{\mathcal O}\left(
            \frac{LR\ptail}{\sqrt N}
        \right),
        &\text{(Shared)}.
    \end{cases}
\end{aligned}
\end{equation}

The triangle inequality gives
\begin{equation}\label{eq:triangle_ineq}
    T|\theta_1^*-\tilde\theta_1^*|
    \le\delta_\theta+\frac q2.
\end{equation}
By Eq.~\eqref{eq:step1_error_bound}, $\delta_\theta\le r_\chi<\sqrt{\log4}$. Convexity of $x\mapsto e^{-2x}$ on $[0,\log4]$ therefore gives
\begin{equation}\label{eq:quadratic}
    e^{-q^2/2}-e^{-2\delta_\theta^2}
    \ge C\delta_\theta^2-\frac{q^2}{2},
    \qquad C:=\frac{15}{16\log4}.
\end{equation}
Substituting Eqs.~\eqref{eq:triangle_ineq} and~\eqref{eq:quadratic} into Eq.~\eqref{eq:master_ineq} gives a quadratic inequality. We finish the two sampling cases separately.

\medskip
\noindent\textit{Independent Time Sampling.} There are nonnegative coefficients $c_1,c_2$ such that
\begin{equation}\label{eq:quadratic_ineq_indep}
\begin{aligned}
    CU^2\delta_\theta^2
    &\le c_1\delta_\theta+c_2,\\
    c_1
    &=
    \mathcal O(\sigma\sqrt{LR}\ptail U)
    +
    \widetilde{\mathcal O}\left(
        \frac{U}{\sqrt N}+\frac{\sqrt{LR}}N
    \right)\sigma,\\
    c_2
    &=
    \mathcal O(LR\ptail^2)
    +\mathcal O(\sigma\sqrt{LR}\ptail Uq)
    +\mathcal O(q^2U^2)\\
    &\quad+
    \widetilde{\mathcal O}\left(
        \frac{U}{\sqrt N}+\frac{\sqrt{LR}}N
    \right)\sigma q
    +
    \widetilde{\mathcal O}\left(
        \frac{\sqrt{LR}\ptail}{\sqrt N}
    \right).
\end{aligned}
\end{equation}
For $a>0$ and $b,c,x\ge0$, the inequality $ax^2\le bx+c$ implies $x\le b/a+\sqrt{c/a}$. The linear coefficient satisfies
\begin{equation}\label{eq:linear_ratio_indep}
    \frac{c_1}{U^2}
    =
    \mathcal O\left(\frac{\sigma\sqrt{LR}\ptail}{U}\right)
    +
    \widetilde{\mathcal O}\left(
        \frac{\sigma}{U\sqrt N}
        +\frac{\sigma\sqrt{LR}}{U^2N}
    \right).
\end{equation}

To bound $c_2$, use $q\le c_q\ptail/\sigma$ from Eq.~\eqref{eq:thm_eigenvalue_cond1} and $U\le\sqrt{LR}$. They give
\begin{equation}\label{eq:c2_simplify_indep}
    \sigma\sqrt{LR}\ptail Uq+q^2U^2
    \le
    c_qLR\ptail^2+c_q^2LR\ptail^2
    =\mathcal O(LR\ptail^2).
\end{equation}
For the remaining terms, the AM--GM inequality $\sqrt{ab}\le(a+b)/2$ yields
\begin{equation}
\begin{aligned}
    &\sqrt{
        \widetilde{\mathcal O}\left(
            \frac{\sigma q}{U\sqrt N}
            +\frac{\sigma q\sqrt{LR}}{U^2N}
        \right)
    }\le
    \mathcal O(q)
    +
    \widetilde{\mathcal O}\left(
        \frac{\sigma}{U\sqrt N}
        +\frac{\sigma\sqrt{LR}}{U^2N}
    \right),\\
    &\sqrt{
        \widetilde{\mathcal O}\left(
            \frac{\sqrt{LR}\ptail}{U^2\sqrt N}
        \right)
    }
    \le
    \mathcal O\left(\frac{\sqrt{LR}\ptail}{U}\right)
    +
    \widetilde{\mathcal O}\left(\frac1{U\sqrt N}\right).
\end{aligned}
\end{equation}

Since $q\le\ptail\le\sqrt{LR}\ptail/U$ and $\sigma\ge1$, we conclude that
\begin{equation}\label{eq:constant_ratio_indep}
    \sqrt{\frac{c_2}{U^2}}
    \le
    \mathcal O\left(\frac{\sqrt{LR}\ptail}{U}\right)
    +
    \widetilde{\mathcal O}\left(
        \frac{\sigma}{U\sqrt N}
        +\frac{\sigma\sqrt{LR}}{U^2N}
    \right).
\end{equation}
Combining Eqs.~\eqref{eq:quadratic_ineq_indep}, \eqref{eq:linear_ratio_indep}, and~\eqref{eq:constant_ratio_indep}, and using $U\ge\sqrt{LR}\pmin/(1+\chi)$, gives
\begin{equation}\label{eq:step2_error_bound}
\begin{aligned}
    \delta_\theta
    &\le
    \mathcal O\left(\frac{\sigma\sqrt{LR}\ptail}{U}\right)
    +
    \widetilde{\mathcal O}\left(
        \frac{\sigma}{U\sqrt N}
        +\frac{\sigma\sqrt{LR}}{U^2N}
    \right)\\
    &\le
    \frac{\sigma(1+\chi)}{\pmin}
    \left[
        \widetilde{\mathcal O}\left(
            \left(1+\frac{1+\chi}{\pmin\sqrt N}\right)
            \frac1{\sqrt{LRN}}
        \right)
        +\mathcal O(\ptail)
    \right].
\end{aligned}
\end{equation}
Dividing by $T$ proves Eq.~\eqref{eq:thm_eigenvalue_result_indep}.

\medskip
\noindent\textit{Shared Time Sampling.} In this case the quadratic inequality has coefficients
\begin{equation}\label{eq:quadratic_ineq_shared}
\begin{aligned}
    CU^2\delta_\theta^2
    &\le c_1\delta_\theta+c_2,\\
    c_1
    &=
    \mathcal O(\sigma\sqrt{LR}\ptail U)
    +
    \widetilde{\mathcal O}\left(
        \frac{\sigma\sqrt{LR}U}{\sqrt N}
    \right),\\
    c_2
    &=
    \mathcal O(LR\ptail^2)
    +\mathcal O(\sigma\sqrt{LR}\ptail Uq)
    +\mathcal O(q^2U^2)\\
    &\quad+
    \widetilde{\mathcal O}\left(
        \frac{\sigma q\sqrt{LR}U}{\sqrt N}
    \right)
    +
    \widetilde{\mathcal O}\left(
        \frac{LR\ptail}{\sqrt N}
    \right).
\end{aligned}
\end{equation}
Thus
\begin{equation}\label{eq:linear_ratio_shared}
    \frac{c_1}{U^2}
    =
    \mathcal O\left(\frac{\sigma\sqrt{LR}\ptail}{U}\right)
    +
    \widetilde{\mathcal O}\left(
        \frac{\sigma\sqrt{LR}}{U\sqrt N}
    \right).
\end{equation}
The deterministic terms in $c_2$ are handled by Eq.~\eqref{eq:c2_simplify_indep}. For the statistical terms, the same AM--GM argument gives
\begin{equation}
\begin{aligned}
    \sqrt{
        \widetilde{\mathcal O}\left(
            \frac{\sigma q\sqrt{LR}}{U\sqrt N}
        \right)
    }
    &\le
    \mathcal O(q)
    +
    \widetilde{\mathcal O}\left(
        \frac{\sigma\sqrt{LR}}{U\sqrt N}
    \right),\\
    \sqrt{
        \widetilde{\mathcal O}\left(
            \frac{LR\ptail}{U^2\sqrt N}
        \right)
    }
    &\le
    \mathcal O\left(\frac{\sqrt{LR}\ptail}{U}\right)
    +
    \widetilde{\mathcal O}\left(
        \frac{\sqrt{LR}}{U\sqrt N}
    \right).
\end{aligned}
\end{equation}
Consequently,
\begin{equation}\label{eq:constant_ratio_shared}
    \sqrt{\frac{c_2}{U^2}}
    \le
    \mathcal O\left(\frac{\sqrt{LR}\ptail}{U}\right)
    +
    \widetilde{\mathcal O}\left(
        \frac{\sigma\sqrt{LR}}{U\sqrt N}
    \right).
\end{equation}
Combining Eqs.~\eqref{eq:quadratic_ineq_shared}, \eqref{eq:linear_ratio_shared}, and~\eqref{eq:constant_ratio_shared}, we obtain
\begin{equation}\label{eq:step2_error_bound_shared}
    \delta_\theta \le
    \mathcal O\left(\frac{\sigma\sqrt{LR}\ptail}{U}\right)
    +
    \widetilde{\mathcal O}\left(
        \frac{\sigma\sqrt{LR}}{U\sqrt N}
    \right)\le
    \frac{\sigma(1+\chi)}{\pmin}
    \left[
        \widetilde{\mathcal O}\left(\frac1{\sqrt N}\right)
        +\mathcal O(\ptail)
    \right].
\end{equation}
Dividing by $T$ proves Eq.~\eqref{eq:thm_eigenvalue_result_shared}.

The estimates hold for every $1\le j\le I$ on the same event from Step~0, with the corresponding cluster amplitude in place of $U$. Taking the supremum over these iterations completes the proof.
\end{proof}

\subsection{Proof of Theorem~\ref{thm:multiplicity}}
\label{app:proofs_multiplicity}

\begin{proof}[Proof of Theorem~\ref{thm:multiplicity}]

\textbf{Step 0: Localization and uniform error bounds.} Applying Eq.~\eqref{eq:step1_error_bound} in Step~1 of the proof of Theorem~\ref{thm:eigenvalue_error} with failure probability $\eta/2$, the search-and-block procedure returns at least $I$ points, and there exists a permutation $\pi:[I]\to[I]$ such that
\begin{equation}
    \sup_{1\le i\le I}
    |\theta_i^*-\lambda_{\pi_i}^*|
    \le\frac{r_\chi}{T}.
\end{equation}
Applying Lemma~\ref{lemma:spectral_norm_bound} with $P=I_L$, $Q=I_R$, and failure probability $\eta/2$, the sample-size condition in Eq.~\eqref{eq:thm_multiplicity_N} gives
\begin{equation}\label{eq:E_spectral_bound}
    \sup_{\theta\in\Theta}\|\E(\theta)\|_2
    \le
    \widetilde{\mathcal O}\left(
        \frac{\sqrt L+\sqrt R}{\sqrt N}
    \right)
    \le
    \frac{c_\tau}{2}\sqrt{LR}\Delta_{\text{phys}},
\end{equation}
where the implicit sample-size constant is chosen sufficiently large depending on $c_\tau$. By a union bound, these conclusions hold simultaneously with probability at least $1-\eta$. We work on this event throughout the proof.

\medskip
\textbf{Step 1: Valid clusters ($1\le i\le I$).} We present the argument for $i=1$. The same estimates apply to every $1\le i\le I$.

Recall the decomposition $G=A+B+C+\E$ from Eqs.~\eqref{eq:defn_E} and~\eqref{eq:defn_FABC}, with $A$ corresponding to the target cluster $\D_{\pi_1}$. By Eq.~\eqref{eq:step1_error_bound} and the triangle inequality, for every $m\in\D_{\pi_1}$,
\begin{equation}
    |\theta_1^*-\lambda_m|
    \le
    |\theta_1^*-\lambda_{\pi_1}^*|
    +|\lambda_{\pi_1}^*-\lambda_m|
    \le\frac{r_\chi}{T}+\delta.
\end{equation}
To simplify the notation for the minimum signal attenuation factor, let
\begin{equation}
    C_{\text{filter}}
    :=
    \exp\left(-(r_\chi+\delta T)^2\right).
\end{equation}
By Assumption~\ref{assum:uniform_overlap}, the $|\D_{\pi_1}|$-th singular value of the target signal matrix satisfies
\begin{equation}
\begin{aligned}
    s_{|\D_{\pi_1}|}(A(\theta_1^*))
    &\ge
    C_{\text{filter}}\,
    s_{\min}(\Phi_{:,\D_{\pi_1}})
    s_{\min}(\Psi_{:,\D_{\pi_1}})\\
    &\ge
    \frac{C_{\text{filter}}}{(1+\chi)^2}
    \frac{
        \|\Phi_{:,\D_{\pi_1}}\|_F
        \|\Psi_{:,\D_{\pi_1}}\|_F
    }{|\D_{\pi_1}|}\\
    &=
    \frac{C_{\text{filter}}}
         {(1+\chi)^2|\D_{\pi_1}|}
    \sqrt{
        \sum_{m\in\D_{\pi_1}}\|\Phi_{:,m}\|_2^2
    }
    \sqrt{
        \sum_{m\in\D_{\pi_1}}\|\Psi_{:,m}\|_2^2
    }\\
    &\ge
    \frac{C_{\text{filter}}}
         {(1+\chi)^2|\D_{\pi_1}|}
    \sum_{m\in\D_{\pi_1}}
        \|\Phi_{:,m}\|_2\|\Psi_{:,m}\|_2\\
    &\ge
    \sqrt{LR}\,
    \frac{C_{\text{filter}}\pmin}{(1+\chi)^2}\\
    &=
    \sqrt{LR}\left(
        4\ptail+\Delta_{\text{phys}}
    \right).
\end{aligned}
\end{equation}
Here the first inequality follows from the full column rank of both overlap matrices and the positive diagonal filter in Eq.~\eqref{eq:defn_A}; the fourth line follows from the Cauchy--Schwarz inequality; and the last equality is given by Eq.~\eqref{eq:defn_physical_gap}.

According to Eqs.~\eqref{eq:B_bound} and~\eqref{eq:C_bound}, the tail and cross-cluster contributions satisfy
\begin{equation}
    \|B(\theta_1^*)+C(\theta_1^*)\|_2
    \le2\sqrt{LR}\ptail.
\end{equation}
By Weyl's inequality for singular values~\cite[Theorem~3.3.16]{horn1991topics}, $|s_k(X+Y)-s_k(X)|\le\|Y\|_2$. Applying this inequality with $X=A(\theta_1^*)$ and $Y=B(\theta_1^*)+C(\theta_1^*)+\E(\theta_1^*)$, we obtain
\begin{equation}
\begin{aligned}
    s_{|\D_{\pi_1}|}(G(\theta_1^*))
    &\ge
    s_{|\D_{\pi_1}|}(A(\theta_1^*))
    -\|B(\theta_1^*)+C(\theta_1^*)\|_2
    -\|\E(\theta_1^*)\|_2\\
    &\ge
    \sqrt{LR}\left(
        2\ptail+\Delta_{\text{phys}}
    \right)
    -\frac{c_\tau}{2}\sqrt{LR}\Delta_{\text{phys}}\\
    &\ge
    \tau+\frac{c_\tau}{2}\sqrt{LR}\Delta_{\text{phys}}
    >\tau.
\end{aligned}
\end{equation}
By Eq.~\eqref{eq:thm_multiplicity_dimension}, $|\D_{\pi_1}|\le\min(L,R)$. Moreover, the full column rank of both overlap matrices and the positive diagonal filter imply $\operatorname{rank}(A(\theta_1^*))=|\D_{\pi_1}|$. Thus, for every $|\D_{\pi_1}|<k\le\min(L,R)$, we have $s_k(A(\theta_1^*))=0$, and Weyl's inequality gives
\begin{equation}
\begin{aligned}
    s_k(G(\theta_1^*))
    &\le
    \|B(\theta_1^*)+C(\theta_1^*)\|_2
    +\|\E(\theta_1^*)\|_2\\
    &\le
    2\sqrt{LR}\ptail
    +\frac{c_\tau}{2}\sqrt{LR}\Delta_{\text{phys}}\\
    &\le
    \tau-\frac{c_\tau}{2}\sqrt{LR}\Delta_{\text{phys}}
    <\tau.
\end{aligned}
\end{equation}
Thus exactly $|\D_{\pi_1}|$ singular values exceed $\tau$, giving $m_1=|\D_{\pi_1}|$. The same argument proves $m_i=|\D_{\pi_i}|$ for all $1\le i\le I$.

\medskip
\textbf{Step 2: Spurious candidates ($i>I$).} Let $\theta_i^*$ be any additional candidate returned after the first $I$ iterations. Since it lies outside every previously blocked interval, for each $1\le k\le I$,
\begin{equation}
    |\theta_i^*-\theta_k^*|>\frac{\alpha}{T}.
\end{equation}
For every $m\in\D_{\pi_k}$, the triangle inequality therefore gives
\begin{equation}
\begin{aligned}
    T|\theta_i^*-\lambda_m|
    &\ge
    T|\theta_i^*-\theta_k^*|
    -T|\theta_k^*-\lambda_{\pi_k}^*|
    -T|\lambda_{\pi_k}^*-\lambda_m|\\
    &>
    \alpha-r_\chi-\delta T\\
    &>
    \sqrt{\log\left(\frac1{\ptail}\right)},
\end{aligned}
\end{equation}
where the last inequality follows from Eq.~\eqref{eq:thm_multiplicity_alpha}. Because $\pi$ is a permutation of $[I]$, this bound holds for every dominant eigenvalue. Using Eq.~\eqref{eq:normalized_overlap_sum}, we obtain
\begin{equation}
    \|A(\theta_i^*)+C(\theta_i^*)\|_2
    \le
    \sqrt{LR}
    \sum_{m\in\D}
        \bar p_m
        e^{-T^2|\theta_i^*-\lambda_m|^2}\le\sqrt{LR}\ptail.
\end{equation}
Together with Eq.~\eqref{eq:B_bound} and Eq.~\eqref{eq:E_spectral_bound}, this gives
\begin{equation}
\begin{aligned}
    s_1(G(\theta_i^*))
    &=\|G(\theta_i^*)\|_2\\
    &\le
    \|A(\theta_i^*)+C(\theta_i^*)\|_2
    +\|B(\theta_i^*)\|_2
    +\|\E(\theta_i^*)\|_2\\
    &\le
    2\sqrt{LR}\ptail
    +\frac{c_\tau}{2}\sqrt{LR}\Delta_{\text{phys}}\\
    &\le
    \tau-\frac{c_\tau}{2}\sqrt{LR}\Delta_{\text{phys}}
    <\tau.
\end{aligned}
\end{equation}
Therefore no singular value exceeds $\tau$, and $m_i=0$ for every additional candidate. This completes the proof.
\end{proof}

\subsection{Proof of Theorem~\ref{thm:observable}}
\label{app:proofs_observable}

\begin{proof}[Proof of Theorem~\ref{thm:observable}]

\textbf{Step 0: Uniform error bounds.} The truncation condition in $\sigma=\Theta\left(\sqrt{\log(LRN)}\right)$ ensures that Lemmas~\ref{lemma:error_term}, \ref{lemma:cross_terms}, and~\ref{lemma:spectral_norm_bound} apply to the total errors $\E$ and $\E^O$.

For a target cluster $\mathcal D_{\pi_i}$, let $U_{\mathrm{ex}}$ and $V_{\mathrm{ex}}$ be the exact singular-vector bases defined in Eq.~\eqref{eq:defn_G_ex_SVD}. These matrices depend only on the exact signal. Moreover, normalization of the initial states gives
\begin{equation}\label{eq:m_i_bound}
    m_i\pmin
    \le
    \sum_{m\in\mathcal D_{\pi_i}}\bar p_m
    \le
    \frac{
        \|\Phi_{:,\mathcal D_{\pi_i}}\|_F
        \|\Psi_{:,\mathcal D_{\pi_i}}\|_F
    }{\sqrt{LR}}
    \le1.
\end{equation}
Thus $m_i=\mathcal O(1)$. The same argument gives $|\mathcal D|=\mathcal O(1)$, and hence $I=\mathcal O(1)$.

Applying Lemma~\ref{lemma:spectral_norm_bound} and taking a union bound, we obtain Eq.~\eqref{eq:prob_eqs_indep} together with the following bounds, simultaneously for every target cluster and every $\theta\in\Theta$, with probability at least $1-\eta$:
\begin{equation}\label{eq:prob_eqs_S}
\begin{aligned}
    &\|\E(\theta)\|_2
    \le
    \widetilde{\mathcal O}\left(
        \frac{\sqrt L+\sqrt R}{\sqrt N}
    \right),
    \qquad
    \|\E^O(\theta)\|_2
    \le
    \widetilde{\mathcal O}\left(
        \|O\|_2\frac{\sqrt L+\sqrt R}{\sqrt N}
    \right),\\
    &\|\E(\theta)V_{\mathrm{ex}}\|_2
    \le
    \widetilde{\mathcal O}\left(\frac{\sqrt L}{\sqrt N}\right),
    \qquad
    \|\E^O(\theta)V_{\mathrm{ex}}\|_2
    \le
    \widetilde{\mathcal O}\left(
        \|O\|_2\frac{\sqrt L}{\sqrt N}
    \right),\\
    &\|U_{\mathrm{ex}}^\dagger\E(\theta)\|_2
    \le
    \widetilde{\mathcal O}\left(\frac{\sqrt R}{\sqrt N}\right),
    \qquad
    \|U_{\mathrm{ex}}^\dagger\E^O(\theta)\|_2
    \le
    \widetilde{\mathcal O}\left(
        \|O\|_2\frac{\sqrt R}{\sqrt N}
    \right),\\
    &\|U_{\mathrm{ex}}^\dagger\E(\theta)V_{\mathrm{ex}}\|_2
    \le
    \widetilde{\mathcal O}\left(\frac1{\sqrt N}\right),
    \qquad
    \|U_{\mathrm{ex}}^\dagger\E^O(\theta)V_{\mathrm{ex}}\|_2
    \le
    \widetilde{\mathcal O}\left(\frac{\|O\|_2}{\sqrt N}\right).
\end{aligned}
\end{equation}
Here we used $m_i=\mathcal O(1)$ in the projected bounds. Since these estimates hold uniformly over $\Theta$, they also hold at the selected points $\theta_i^*$. We work on this event throughout the proof.

\medskip
\textbf{Step 1: Localization and multiplicity recovery.} We first prove that
\begin{equation}\label{eq:loc_error_obs_proof}
    \sup_{1\le j\le I}
    |\theta_j^*-\lambda_{\pi_j}^*|
    \le
    \frac1T\left[
        \mathcal O\left(e^{-\Delta^2T^2/4}\right)
        +\widetilde{\mathcal O}\left(\frac1{\sqrt{LRN}}\right)
    \right].
\end{equation}

Since $\delta=0$ and the initial states satisfy Eq.~\eqref{eq:thm_observable_no_tail}, both $B(\theta)$ and $\Delta A_{\mathrm{width}}(\theta)$ vanish. The remaining deterministic interference is $C(\theta)$. For a nearest cluster to $\theta$, every eigenvalue in another cluster is at least $\Delta/2$ away. Consequently,
\begin{equation}\label{eq:C_bound_observable}
    \|C(\theta)\|_F
    \le
    \mathcal O\left(
        \sqrt{LR}e^{-\Delta^2T^2/4}
    \right).
\end{equation}
This is the same cross-cluster estimate as in Eq.~\eqref{eq:C_bound}.

We follow the coarse-localization argument in Step~1 of the proof of Theorem~\ref{thm:eigenvalue_error}. Our assumption $\min(L,R)N=\widetilde{\Omega}(1)$ implies $\sqrt{LR}N=\widetilde{\Omega}(1)$. Since $\pmin=\Omega(1)$ and $\chi=\mathcal O(1)$, this satisfies the sample-size requirement in Eq.~\eqref{eq:thm_eigenvalue_cond_indep}. The same argument therefore gives Eq.~\eqref{eq:small_signal_comparison}: whenever $\|G(\theta)\|_F\ge\|G(\theta')\|_F$ and $\|\F(\theta)\|_F+\|\F(\theta')\|_F\ge U_{\min}/2$,
\begin{equation}\label{eq:F_comparison_in_obs}
    \|\F(\theta')\|_F-\|\F(\theta)\|_F
    \le\frac{U_{\min}}{16},
    \qquad
    U_{\min}:=\frac{\sqrt{LR}\pmin}{1+\chi}.
\end{equation}

Recall that $\F(\theta)=A(\theta)+C(\theta)$ in the present setting. Since $\Delta T\ge C_T$, Eq.~\eqref{eq:C_bound_observable} implies, for sufficiently large $C_T$,
\begin{equation}
    \sup_{\theta\in\Theta}\|C(\theta)\|_F
    \le\frac{U_{\min}}{16}.
\end{equation}
For the first selected point $\theta_1^*$, let $\pi_1$ index a nearest cluster and let $\tilde\theta_1^*$ be a grid point within $q/(2T)$ of $\lambda_{\pi_1}^*$. Writing $U=\|\Phi_{:,\mathcal D_{\pi_1}}
\Psi_{:,\mathcal D_{\pi_1}}^\dagger\|_F$,
Lemma~\ref{lemma:bound_U} gives $U\ge U_{\min}$. Recall the decomposition $A_{\text{center}}$ and $\Delta A_{\text{width}}$ defined in Eq.~\eqref{eq:A_decomp}, while $\Delta A_{\text{width}}=0$ since $\delta=0$. According to Eq.~\eqref{eq:A1_bound}, we obtain
\begin{equation}
\begin{aligned}
    \|\F(\tilde\theta_1^*)\|_F
    &\ge \|A_{\text{center}}(\tilde\theta_1^*)\|_F - \|C(\tilde\theta_1^*)\|_F \ge e^{-q^2/4}U-\frac{U_{\min}}{16},\\
    \|\F(\theta_1^*)\|_F
    &\le \|A_{\text{center}}(\theta_1^*)\|_F + \|C(\theta_1^*)\|_F \le e^{-T^2|\theta_1^*-\lambda_{\pi_1}^*|^2}U
         +\frac{U_{\min}}{16}.
\end{aligned}
\end{equation}
For sufficiently small $c_q$, $e^{-q^2/4}\ge3/4$, so $\|\F(\tilde\theta_1^*)\|_F>U_{\min}/2$. According to Eq.~\eqref{eq:F_comparison_in_obs}, we have
\begin{equation}
    e^{-T^2|\theta_1^*-\lambda_{\pi_1}^*|^2}
    \ge e^{-q^2/4}-\frac{3U_{\min}}{16U} \ge\frac9{16}
    >\frac{1+\chi}{2(1+2\chi)}
    =e^{-r_\chi^2},
\end{equation}
where $r_\chi$ is defined in Eq.~\eqref{eq:defn_r_chi}. Hence $|\theta_1^*-\lambda_{\pi_1}^*|\le r_\chi/T$.

We extend this conclusion by induction. Suppose that the first $j-1$ outputs correspond to distinct clusters and satisfy the same bound. By the blocking condition Eq.~\eqref{eq:thm_eigenvalue_cond2}, assumed here, a center grid point $\tilde\theta_i$ of every undiscovered cluster remains available, since
\begin{equation}
    |\tilde\theta_i-\theta_k^*|
    \ge\Delta-\frac{r_\chi}{T}-\frac q{2T}
    >\frac{\alpha}{T},
    \qquad k<j,
\end{equation}
as in Eq.~\eqref{eq:unblocked_center}. Moreover, $\|\F(\tilde\theta_i)\|_F>U_{\min}/2$.

If the next selected point $\theta_j^*$ had a previously selected cluster $\pi_k$ as its nearest cluster, then $T|\theta_j^*-\lambda_{\pi_k}^*|\ge\alpha-r_\chi$. The lower bound on $\alpha$ in Eq.~\eqref{eq:thm_eigenvalue_cond2} would give, as in Eq.~\eqref{eq:discovered_signal_upper},
\begin{equation}
    \|\F(\theta_j^*)\|_F
    \le e^{-(\alpha-r_\chi)^2}\sqrt{LR}
        +\frac{U_{\min}}{16}
    <\frac{U_{\min}}8.
\end{equation}
This contradicts the comparison bound Eq.~\eqref{eq:F_comparison_in_obs} with $\tilde\theta_i$, because $\theta_j^*$ maximizes the empirical norm over the remaining grid.

Thus $\pi_j$ is an undiscovered cluster, and its center grid point is still available. Applying the first-iteration comparison again yields
\begin{equation}\label{eq:obs_coarse_localization}
    |\theta_j^*-\lambda_{\pi_j}^*|
    \le\frac{r_\chi}{T},
    \qquad 1\le j\le I.
\end{equation}
Therefore, the first $I$ outputs correspond to distinct clusters, and $\pi:[I]\to[I]$ is a permutation.

To refine Eq.~\eqref{eq:obs_coarse_localization}, consider the representative cluster $\pi_1$ and write $\delta_\theta=T|\theta_1^*-\lambda_{\pi_1}^*|$. Recall $U=\|\Phi_{:,\mathcal D_{\pi_1}}
\Psi_{:,\mathcal D_{\pi_1}}^\dagger\|_F
=\Theta(\sqrt{LR})$. As in Eq.~\eqref{eq:squared_norm_diff_s2}, we compare $\theta_1^*$ with its center grid point $\tilde\theta_1^*$ and write
\begin{equation}
    0\le
    \|G(\theta_1^*)\|_F^2
    -\|G(\tilde\theta_1^*)\|_F^2
    =X+Y+Z.
\end{equation}

We first estimate the signal difference $X$. Since $\F=A+C$,
\begin{equation}\label{eq:obs_X_expand}
\begin{aligned}
    X
    &=
    \|A(\theta_1^*)\|_F^2
    -\|A(\tilde\theta_1^*)\|_F^2\\
    &\quad+
    2\RE\left(
        \langle A(\theta_1^*),C(\theta_1^*)\rangle
        -
        \langle A(\tilde\theta_1^*),C(\tilde\theta_1^*)\rangle
    \right)\\
    &\quad+
    \|C(\theta_1^*)\|_F^2
    -\|C(\tilde\theta_1^*)\|_F^2.
\end{aligned}
\end{equation}
Since $\delta=0$, the first difference is bounded as in Eq.~\eqref{eq:X1_bound}, with no cluster-width error. For the remaining terms, we apply the same argument using the variation estimates Eqs.~\eqref{eq:lipschitz_A} and~\eqref{eq:lipschitz_C}, together with the bounds on $A$ and $C$ in Eqs.~\eqref{eq:A1_bound} and~\eqref{eq:C_bound_observable}. Every other dominant eigenvalue remains at distance at least $\Delta-r_\chi/T$ along the segment joining $\theta_1^*$ and $\tilde\theta_1^*$. Since $x\mapsto 2xe^{-x^2}$ is decreasing for $x\ge1/\sqrt2$, Eq.~\eqref{eq:normalized_overlap_sum} gives, for every $s$ on this segment,
\begin{equation}
    \|C'(s)\|_F
    \le
    2T\sqrt{LR}(\Delta T-r_\chi)
    e^{-(\Delta T-r_\chi)^2}
    \le
    \mathcal O\left(
        T\sqrt{LR}e^{-\Delta^2T^2/4}
    \right),
\end{equation}
where the last inequality holds for sufficiently large $C_T$, since $r_\chi<\sqrt{\log4}$. Consequently,
\begin{equation}
    \|C(\theta_1^*)-C(\tilde\theta_1^*)\|_F
    \le
    \mathcal O\left(
        \sqrt{LR}e^{-\Delta^2T^2/4}
    \right)
    T|\theta_1^*-\tilde\theta_1^*|.
\end{equation}
Using $U=\Theta(\sqrt{LR})$ and Eq.~\eqref{eq:triangle_ineq}, we therefore obtain
\begin{equation}\label{eq:obs_X_bound}
    X
    \le
    -U^2\left(e^{-q^2/2}-e^{-2\delta_\theta^2}\right)
    +\mathcal O\left(
        U^2e^{-\Delta^2T^2/4}
    \right)(\delta_\theta+q).
\end{equation}

The estimates for $Y$ and $Z$ follow from Eq.~\eqref{eq:prob_eqs_indep} by the same argument as in Step~2 of the proof of Theorem~\ref{thm:eigenvalue_error}. Specifically, $|Y|\le
\widetilde{\mathcal O}(\sigma U/\sqrt N)
(\delta_\theta+q)$, and $Z=-X/N+
\widetilde{\mathcal O}(\sigma\sqrt{LR}/N)
(\delta_\theta+q)$.

Substituting Eq.~\eqref{eq:obs_X_bound} and applying Eq.~\eqref{eq:quadratic}, the same quadratic-inequality argument as in Eq.~\eqref{eq:quadratic_ineq_indep} therefore yields
\begin{equation}
    \delta_\theta
    \le
    \mathcal O\left(
        q+e^{-\Delta^2T^2/4}
    \right)
    +\widetilde{\mathcal O}\left(
        \frac{\sigma}{U\sqrt N}
        +\frac{\sigma\sqrt{LR}}{U^2N}
    \right).
\end{equation}
Using $U=\Theta(\sqrt{LR})$, $q=\mathcal O(1/\sqrt{LRN})$, and $\sigma=\Theta(\sqrt{\log(LRN)})$, and dividing by $T$, proves Eq.~\eqref{eq:loc_error_obs_proof}. The same argument applies to every $j\le I$.

We also verify multiplicity recovery. By Assumption~\ref{assum:uniform_overlap} and Eq.~\eqref{eq:obs_coarse_localization}, the target matrix has rank $|\mathcal D_{\pi_i}|$ and
\begin{equation}
    s_{|\mathcal D_{\pi_i}|}(A(\theta_i^*))
    \ge
    \frac{\sqrt{LR}\pmin}{(1+\chi)^2}e^{-r_\chi^2}
    =2\tau.
\end{equation}
Eqs.~\eqref{eq:prob_eqs_S}, \eqref{eq:C_bound_observable}, and $\min(L,R)N=\widetilde{\Omega}(1)$, with sufficiently large constants, ensure
\begin{equation}
    \|C(\theta_i^*)+\E(\theta_i^*)\|_2<\frac{\tau}{2}.
\end{equation}
Weyl's inequality therefore places the first $|\mathcal D_{\pi_i}|$ singular values of $G(\theta_i^*)$ above $\tau$, and all remaining singular values below $\tau$. Thus $m_i=|\mathcal D_{\pi_i}|$.

\medskip
\textbf{Step 2: Errors in the projected matrices.} Fix $1\le i\le I$. Define the deterministic part of the deviation from $G_{\mathrm{exact}}$, which is defined in Eq.~\eqref{eq:defn_G_ex}, as
\begin{equation}\label{eq:defn_det_G}
    \Delta G_{\mathrm{det}}
    :=\F(\theta_i^*)-G_{\mathrm{exact}}
    =G(\theta_i^*)-\E(\theta_i^*)-G_{\mathrm{exact}}.
\end{equation}
Since $\delta=0$,
\begin{equation}\label{eq:A_exact_diff}
\begin{aligned}
    \|A(\theta_i^*)-G_{\mathrm{exact}}\|_2
    &=
    \left(1-e^{-T^2|\theta_i^*-\lambda_{\pi_i}^*|^2}\right)
    \|G_{\mathrm{exact}}\|_2\\
    &\le
    T^2|\theta_i^*-\lambda_{\pi_i}^*|^2
    \|G_{\mathrm{exact}}\|_2\\
    &\le
    \widetilde{\mathcal O}\left(\frac1{\sqrt{LR}N}\right)
    +\mathcal O\left(
        \sqrt{LR}e^{-\Delta^2T^2/2}
    \right).
\end{aligned}
\end{equation}
Here we used $1-e^{-x^2}\le x^2$, Eq.~\eqref{eq:loc_error_obs_proof}, and $\|G_{\mathrm{exact}}\|_2\le\sqrt{LR}$. Combining this with Eq.~\eqref{eq:C_bound_observable} gives
\begin{equation}\label{eq:det_G_bound}
    \|\Delta G_{\mathrm{det}}\|_2
    \le
    \widetilde{\mathcal O}\left(\frac1{\sqrt{LR}N}\right)
    +\mathcal O\left(
        \sqrt{LR}e^{-\Delta^2T^2/4}
    \right).
\end{equation}

Recall $\F^O$, $G^O$, $\E^O$ and $G^O_{\mathrm{exact}}$ defined in Eqs.~\eqref{eq:defn_F_O},~\eqref{eq:defn_G_O_app},~\eqref{eq:defn_E} and~\eqref{eq:defn_G_ex_O}, respectively. Similar to Eq.~\eqref{eq:defn_det_G}, we define
\begin{equation}\label{eq:defn_det_GO}
    \Delta G^O_{\mathrm{det}}
    :=\F^O(\theta_i^*)-G^O_{\mathrm{exact}}
    =G^O(\theta_i^*)-\E^O(\theta_i^*)-G^O_{\mathrm{exact}}.
\end{equation}
Recall from Eq.~\eqref{eq:defn_F_O} and the definition of $K(\theta)$ in Eq.~\eqref{eq:defn_K} that
\begin{equation}
    \F^O_{l,r}(\theta)
    =
    \langle\phi_l|K(\theta)OK(\theta)|\psi_r\rangle.
\end{equation}
To separate the target-cluster contribution from the remaining terms, we define the spectral projectors
\begin{equation}
    P_{\pi_i}
    :=\sum_{m\in\mathcal D_{\pi_i}}
        |E_m\rangle\langle E_m|,
    \qquad
    P_{\mathcal D}
    :=\sum_{m\in\mathcal D}
        |E_m\rangle\langle E_m|.
\end{equation}
The support condition Eq.~\eqref{eq:thm_observable_no_tail} gives $P_{\mathcal D}|\phi_l\rangle=|\phi_l\rangle$ and $P_{\mathcal D}|\psi_r\rangle=|\psi_r\rangle$. Since spectral projectors commute with $K(\theta)$,
\begin{equation}\label{eq:F_O_proj}
    \F^O_{l,r}(\theta)
    =
    \langle\phi_l|
    [K(\theta)P_{\mathcal D}]
    O
    [K(\theta)P_{\mathcal D}]
    |\psi_r\rangle.
\end{equation}

Since $\delta=0$, the filter acts as a scalar on the target eigenspace, so
\begin{equation}\label{eq:obs_filter_decomp}
    K(\theta_i^*)P_{\mathcal D}
    =
    e^{-T^2|\theta_i^*-\lambda_{\pi_i}^*|^2/2}P_{\pi_i}
    +K(\theta_i^*)(P_{\mathcal D}-P_{\pi_i}).
\end{equation}
The first term selects the target cluster, while the second contains the other dominant clusters. By Eq.~\eqref{eq:obs_coarse_localization},
\begin{equation}
    \|K(\theta_i^*)(P_{\mathcal D}-P_{\pi_i})\|_2
    \le
    e^{-(\Delta T-r_\chi)^2/2}
    \le e^{-\Delta^2T^2/4},
\end{equation}
where the last inequality holds for sufficiently large $C_T$, in particular when $\Delta T\ge4(r_\chi+1)$.

Substituting Eq.~\eqref{eq:obs_filter_decomp} into Eq.~\eqref{eq:F_O_proj} gives
\begin{equation}
\begin{aligned}
    \F^O_{l,r}(\theta_i^*)
    &=
    e^{-T^2|\theta_i^*-\lambda_{\pi_i}^*|^2}
    \langle\phi_l|P_{\pi_i}OP_{\pi_i}|\psi_r\rangle\\
    &\quad+
    e^{-T^2|\theta_i^*-\lambda_{\pi_i}^*|^2/2}
    \langle\phi_l|
    P_{\pi_i}OK(\theta_i^*)(P_{\mathcal D}-P_{\pi_i})
    |\psi_r\rangle\\
    &\quad+
    e^{-T^2|\theta_i^*-\lambda_{\pi_i}^*|^2/2}
    \langle\phi_l|
    K(\theta_i^*)(P_{\mathcal D}-P_{\pi_i})OP_{\pi_i}
    |\psi_r\rangle\\
    &\quad+
    \langle\phi_l|
    K(\theta_i^*)(P_{\mathcal D}-P_{\pi_i})
    OK(\theta_i^*)(P_{\mathcal D}-P_{\pi_i})
    |\psi_r\rangle.
\end{aligned}
\end{equation}
By Eq.~\eqref{eq:defn_G_ex_O}, the first term equals $e^{-T^2|\theta_i^*-\lambda_{\pi_i}^*|^2} (G^O_{\mathrm{exact}})_{l,r}$. The remaining three terms constitute $C^O_{l,r}(\theta_i^*)$ in the decomposition Eq.~\eqref{eq:defn_FABC_O}, since $B^O=0$ under Eq.~\eqref{eq:thm_observable_no_tail}. The second and third terms each contain one remainder factor $K(\theta_i^*)(P_{\mathcal D}-P_{\pi_i})$, while the fourth contains two. Using its norm bound above, $\|P_{\pi_i}\|_2=1$, and the normalization of the initial states, we obtain
\begin{equation}
    |C^O_{l,r}(\theta_i^*)|
    \le
    \|O\|_2\left(
        2e^{-\Delta^2T^2/4}
        +e^{-\Delta^2T^2/2}
    \right).
\end{equation}
Consequently,
\begin{equation}\label{eq:C_exp_bound_O}
    \|C^O(\theta_i^*)\|_2
    \le\|C^O(\theta_i^*)\|_F
    \le
    \mathcal O\left(
        \|O\|_2\sqrt{LR}e^{-\Delta^2T^2/4}
    \right).
\end{equation}

Substituting $\F^O(\theta_i^*) =e^{-T^2|\theta_i^*-\lambda_{\pi_i}^*|^2} G^O_{\mathrm{exact}}+C^O(\theta_i^*)$ into Eq.~\eqref{eq:defn_det_GO}, we obtain
\begin{equation}\label{eq:G_det_O_expand}
    \Delta G^O_{\mathrm{det}}
    =
    \left(e^{-T^2|\theta_i^*-\lambda_{\pi_i}^*|^2}-1\right)
    G^O_{\mathrm{exact}}
    +C^O(\theta_i^*).
\end{equation}
Moreover,
\begin{equation}\label{eq:G_ex_bound}
    \|G^O_{\mathrm{exact}}\|_2
    \le
    \|\Phi_{:,\mathcal D_{\pi_i}}\|_2
    \|O_{\mathcal D_{\pi_i}}\|_2
    \|\Psi_{:,\mathcal D_{\pi_i}}\|_2
    \le\|O\|_2\sqrt{LR}.
\end{equation}
Substituting Eqs.~\eqref{eq:loc_error_obs_proof}, \eqref{eq:C_exp_bound_O}, and~\eqref{eq:G_ex_bound} into Eq.~\eqref{eq:G_det_O_expand}, we establish
\begin{equation}\label{eq:Delta_GO_bound}
    \|\Delta G^O_{\mathrm{det}}\|_2
    \le
    \|O\|_2\left[
        \widetilde{\mathcal O}\left(\frac1{\sqrt{LR}N}\right)
        +\mathcal O\left(
            \sqrt{LR}e^{-\Delta^2T^2/4}
        \right)
    \right].
\end{equation}

Next, we align the empirical singular-vector bases with the exact ones. Recall $U_{\text{ex}}$, $V_\text{ex}$, $U_i'$ and $V_i'$ defined in Eq.~\eqref{eq:defn_G_ex_SVD} and Eq.~\eqref{eq:defn_G_tilde}. Let $\Theta_U$ and $\Theta_V$ denote the canonical angle matrices between the corresponding left and right singular subspaces, following
\cite[Chapter I, Definition 5.3]{stewart1993}\footnote{In the notation of Definition 5.3, we set $\mathcal X=\mathcal R(U_{\mathrm{ex}})$ and $\mathcal Y=\mathcal R(U_i')$ for the left singular subspaces. The right singular subspaces are treated analogously.}. We first construct unitary transformations that align the empirical singular-vector bases with the exact ones.

According to~\cite[Chapter I, Theorem 5.2]{stewart1993}\footnote{Here, the orthonormal bases $X_1$ and $Y_1$ in Theorem 5.2 correspond to our matrices $U_{\mathrm{ex}}$ and $U_i'$, respectively, with dimensions $n=L$ and $l=m_i$.}, the SVD of $U_{\mathrm{ex}}^\dagger U_i'$ can be written as
\begin{equation}\label{eq:UU}
    U_{\mathrm{ex}}^\dagger U_i'
    =
    \begin{cases}
        Y\Gamma Z^\dagger,
        & \text{if }L\ge2m_i,\\[1mm]
        Y
        \begin{pmatrix}
            \Gamma&0\\
            0&I_{2m_i-L}
        \end{pmatrix}
        Z^\dagger,
        & \text{if }L<2m_i,
    \end{cases}
\end{equation}
where $Y,Z\in\mathbb C^{m_i\times m_i}$ are unitary. Let $k=\min(m_i,L-m_i)$. The matrix $\Gamma$ is a $k$-by-$k$ diagonal matrix, and the corresponding diagonal matrix $\Sigma$ satisfies $\Gamma^2+\Sigma^2=I_k$. Thus,
\begin{equation}\label{eq:defn_angles}
\begin{aligned}
    \Gamma
    &=\diag(\cos\theta_1,\ldots,\cos\theta_k),\\
    \Sigma
    &=\diag(\sin\theta_1,\ldots,\sin\theta_k),\\
    \Theta_U
    &=\diag(\theta_1,\ldots,\theta_k),
    \qquad \theta_j\in[0,\pi/2].
\end{aligned}
\end{equation}
The identity block in the second case corresponds to additional zero canonical angles. If $k=0$, the two left singular subspaces are both $\mathbb C^L$; in this case all their canonical angles are zero. An analogous construction applies to $\mathcal R(V_{\mathrm{ex}})$ and $\mathcal R(V_i')$, with $L$ replaced by $R$, yielding $\Theta_V$.

Now consider $W_U:=ZY^\dagger$. By Eq.~\eqref{eq:UU},
\begin{equation}\label{eq:UUW}
    U_{\mathrm{ex}}^\dagger(U_i'W_U)
    =
    \begin{cases}
        Y\Gamma Y^\dagger,
        & \text{if }L\ge2m_i,\\[1mm]
        Y
        \begin{pmatrix}
            \Gamma&0\\
            0&I_{2m_i-L}
        \end{pmatrix}
        Y^\dagger,
        & \text{if }L<2m_i,
    \end{cases}
\end{equation}
which is Hermitian and positive semidefinite in either case. Consequently,
\begin{equation}
\begin{aligned}
    \|U_i'W_U-U_{\mathrm{ex}}\|_2^2
    &=
    \lambda_{\max}\left(
        (U_i'W_U-U_{\mathrm{ex}})^\dagger
        (U_i'W_U-U_{\mathrm{ex}})
    \right)\\
    &=
    \lambda_{\max}\left(
        2I_{m_i}
        -W_U^\dagger(U_i')^\dagger U_{\mathrm{ex}}
        -U_{\mathrm{ex}}^\dagger U_i'W_U
    \right)\\
    &=
    \begin{cases}
        \lambda_{\max}\left(
            2I_{m_i}-2Y\Gamma Y^\dagger
        \right),
        & \text{if }L\ge2m_i,\\[1mm]
        \lambda_{\max}\left(
            2I_{m_i}
            -2Y
            \begin{pmatrix}
                \Gamma&0\\
                0&I_{2m_i-L}
            \end{pmatrix}
            Y^\dagger
        \right),
        & \text{if }L<2m_i,
    \end{cases}\\
    &=2(1-\cos\theta_{\max}),
\end{aligned}
\end{equation}
where $\theta_{\max}:=\max_{1\le j\le k}\theta_j$ when $k>0$, and $\theta_{\max}:=0$ when $k=0$. Since $\theta_{\max}\in[0,\pi/2]$,
\begin{equation}
    \|U_i'W_U-U_{\mathrm{ex}}\|_2
    =2\sin(\theta_{\max}/2)
    =\frac{\sin\theta_{\max}}{\cos(\theta_{\max}/2)}
    \le\sqrt2\sin\theta_{\max}
    =\sqrt2\|\sin\Theta_U\|_2.
\end{equation}
Here the last norm is understood to be zero if $k=0$.

The same argument applies to $V_i'$. Hence there exist unitary matrices $W_U,W_V\in\mathbb C^{m_i\times m_i}$ such that
\begin{equation}\label{eq:obs_UV}
    \|U_i'W_U-U_{\mathrm{ex}}\|_2
    \le\sqrt2\|\sin\Theta_U\|_2, \qquad
    \|V_i'W_V-V_{\mathrm{ex}}\|_2
    \le\sqrt2\|\sin\Theta_V\|_2.
\end{equation}
We denote the deviations of the aligned singular-vector bases by
\begin{equation}\label{eq:defn_Delta_UV}
    \Delta U:=U_i'W_U-U_{\mathrm{ex}},
    \qquad
    \Delta V:=V_i'W_V-V_{\mathrm{ex}}.
\end{equation}

Next, we bound $\|\Delta U\|_2$ and $\|\Delta V\|_2$ using Wedin's $\sin\Theta$ theorem
\cite[Chapter V, Theorem 4.4]{stewart1993}\footnote{To apply Wedin's theorem, we set $A=G_{\mathrm{exact}}$ and $\tilde A=G(\theta_i^*)$. The dominant unperturbed singular subspaces correspond to $U_1=U_{\mathrm{ex}}$ and $V_1=V_{\mathrm{ex}}$, with $\Sigma_1=\Sigma_{\mathrm{ex}}$. The perturbed subspaces correspond to $\tilde U_1=U_i'$ and $\tilde V_1=V_i'$, with $\tilde\Sigma_1=\Sigma_i'$. The canonical angle matrices denoted by $\Phi$ and $\Theta$ in the theorem correspond to our $\Theta_U$ and $\Theta_V$, respectively.}.

We define the right residual matrix $R$ as
\begin{equation}
\begin{aligned}
    R
    &:=G_{\mathrm{exact}}V_i'-U_i'\Sigma_i'\\
    &=(G_{\mathrm{exact}}-G(\theta_i^*))V_i'\\
    &=-(\Delta G_{\mathrm{det}}+\E(\theta_i^*))
      (V_{\mathrm{ex}}+\Delta V)W_V^\dagger.
\end{aligned}
\end{equation}
Here $\Delta G_{\mathrm{det}}$ is defined in Eq.~\eqref{eq:defn_det_G}. Taking the spectral norm of $R$ yields
\begin{equation}\label{eq:R_norm}
    \|R\|_2
    \le
    \|\E(\theta_i^*)V_{\mathrm{ex}}\|_2
    +\|\E(\theta_i^*)\|_2\|\Delta V\|_2
    +\|\Delta G_{\mathrm{det}}\|_2.
\end{equation}
Similarly, the left residual matrix $S$ is
\begin{equation}
\begin{aligned}
    S
    &:=G_{\mathrm{exact}}^\dagger U_i'-V_i'\Sigma_i'\\
    &=(G_{\mathrm{exact}}-G(\theta_i^*))^\dagger U_i'\\
    &=-(\Delta G_{\mathrm{det}}+\E(\theta_i^*))^\dagger
      (U_{\mathrm{ex}}+\Delta U)W_U^\dagger,
\end{aligned}
\end{equation}
and satisfies
\begin{equation}\label{eq:S_norm}
    \|S\|_2
    \le
    \|\E(\theta_i^*)^\dagger U_{\mathrm{ex}}\|_2
    +\|\E(\theta_i^*)\|_2\|\Delta U\|_2
    +\|\Delta G_{\mathrm{det}}\|_2.
\end{equation}

To apply Wedin's theorem, we determine the spectral
gap parameters $(\alpha_W,\delta_W)$\footnote{The spectral gap parameters in \cite[Chapter V, Theorem 4.4]{stewart1993} are denoted by $\alpha$ and $\delta$. We write $\alpha_W$ and $\delta_W$ to distinguish them from the block parameter $\alpha$ and the cluster width $\delta$ used in this paper.}. By Assumption~\ref{assum:uniform_overlap}, both $\Phi_{:,\mathcal D_{\pi_i}}$ and $\Psi_{:,\mathcal D_{\pi_i}}$ have full column rank. Thus $G_{\mathrm{exact}}$ has rank exactly $m_i$, and all its background singular values vanish. We may therefore take $\alpha_W=0$, understood as the limit
$\alpha_W\downarrow0$ in Wedin's theorem.

By Weyl's inequality,
\begin{equation}
    s_{\min}(\Sigma_i')
    =s_{m_i}(G(\theta_i^*))\ge
    s_{m_i}(G_{\mathrm{exact}})
    -\|G(\theta_i^*)-G_{\mathrm{exact}}\|_2
    =:\delta_W.
\end{equation}
Here $\Sigma_i'$ contains the leading $m_i$ singular values from Eq.~\eqref{eq:defn_G_SVD}. Assumption~\ref{assum:uniform_overlap} gives
\begin{equation}
    s_{m_i}(G_{\mathrm{exact}})
    \ge
    \frac{\sqrt{LR}\pmin}{(1+\chi)^2}
    =\Omega(\sqrt{LR}).
\end{equation}
On the other hand, Eqs.~\eqref{eq:prob_eqs_S} and~\eqref{eq:det_G_bound} imply
\begin{equation}
    \frac{
        \|G(\theta_i^*)-G_{\mathrm{exact}}\|_2
    }{\sqrt{LR}}\le
    \widetilde{\mathcal O}\left(
        \frac1{\sqrt{\min(L,R)N}}
    \right)
    +\widetilde{\mathcal O}\left(\frac1{LRN}\right)
    +\mathcal O\left(e^{-\Delta^2T^2/4}\right).
\end{equation}
The assumption $\min(L,R)N=\widetilde{\Omega}(1)$, with a sufficiently large implicit constant, and a sufficiently large $C_T$ therefore ensure $\delta_W=\Omega(\sqrt{LR})$.

Thus, applying Wedin's theorem with $\alpha_W=0$ and $\delta_W=\Omega(\sqrt{LR})$, and substituting Eqs.~\eqref{eq:prob_eqs_S}, \eqref{eq:det_G_bound}, \eqref{eq:obs_UV}, \eqref{eq:R_norm}, and~\eqref{eq:S_norm}, we obtain
\begin{equation}\label{eq:sin_theta_bound}
\begin{aligned}
    \max\{\|\Delta U\|_2,\|\Delta V\|_2\}
    &\le
    \sqrt2
    \max\{\|\sin\Theta_U\|_2,\|\sin\Theta_V\|_2\}\\
    &\le
    \frac{\sqrt2}{\delta_W}
    \max\{\|R\|_2,\|S\|_2\}\\
    &\le
    \widetilde{\mathcal O}\left(
        \frac1{\sqrt{\min(L,R)N}}
    \right)
    \left(
        1+\max\{\|\Delta U\|_2,\|\Delta V\|_2\}
    \right)\\
    &\quad+
    \widetilde{\mathcal O}\left(\frac1{LRN}\right)
    +\mathcal O\left(e^{-\Delta^2T^2/4}\right).
\end{aligned}
\end{equation}
Since $\min(L,R)N=\widetilde{\Omega}\left(1\right)$, choosing a sufficiently large implicit constant finally gives
\begin{equation}\label{eq:max_UV_norm}
    \max\{\|\Delta U\|_2,\|\Delta V\|_2\}
    \le
    \widetilde{\mathcal O}\left(
        \frac1{\sqrt{\min(L,R)N}}
    \right)
    +\mathcal O\left(e^{-\Delta^2T^2/4}\right).
\end{equation}

Recall the exact and empirical projected matrices defined in Eqs.~\eqref{eq:defn_G_ex_tilde}, \eqref{eq:defn_G_ex_O_tilde}, \eqref{eq:defn_G_tilde}, and~\eqref{eq:defn_G_O_tilde}. Let
\begin{equation}\label{eq:defn_Delta_G_tilde}
\begin{aligned}
    \Delta\tilde G
    &:=W_U^\dagger\tilde G(\theta_i^*)W_V
      -\tilde G_{\mathrm{exact}},\\
    \Delta\tilde G^O
    &:=W_U^\dagger\tilde G^O(\theta_i^*)W_V
      -\tilde G^O_{\mathrm{exact}}.
\end{aligned}
\end{equation}
Expanding the first deviation gives
\begin{equation}\label{eq:Delta_G_tilde_expand}
\begin{aligned}
    \Delta\tilde G
    &=(U_i'W_U)^\dagger\E(\theta_i^*)(V_i'W_V)+(U_i'W_U)^\dagger
             \Delta G_{\mathrm{det}}(V_i'W_V)\\
    &\quad+
    \left[
        (U_i'W_U)^\dagger G_{\mathrm{exact}}(V_i'W_V)
        -U_{\mathrm{ex}}^\dagger
         G_{\mathrm{exact}}V_{\mathrm{ex}}
    \right].
\end{aligned}
\end{equation}
By Eq.~\eqref{eq:UUW}, $U_{\mathrm{ex}}^\dagger\Delta U =\Delta U^\dagger U_{\mathrm{ex}}$. Using the orthonormality of $U_i'W_U$, we obtain
\begin{equation}\label{eq:Delta_U_squared}
\begin{aligned}
    I
    &=(U_{\mathrm{ex}}+\Delta U)^\dagger
      (U_{\mathrm{ex}}+\Delta U)\\
    &=I+2\Delta U^\dagger U_{\mathrm{ex}}
      +\Delta U^\dagger\Delta U,
\end{aligned}
\qquad
    \implies \Delta U^\dagger U_{\mathrm{ex}}
    =-\frac12\Delta U^\dagger\Delta U.
\end{equation}
Similarly, $V_{\mathrm{ex}}^\dagger\Delta V =-\frac12\Delta V^\dagger\Delta V$. Substituting $G_{\mathrm{exact}} =U_{\mathrm{ex}}\Sigma_{\mathrm{ex}}V_{\mathrm{ex}}^\dagger$ therefore gives
\begin{equation}\label{eq:signal_cancellation}
\begin{aligned}
    &\|(U_i'W_U)\cj G_{\text{exact}} (V_i'W_V) - U_{\text{ex}}\cj G_{\text{exact}} V_{\text{ex}}\|_2 \\
    &\quad= \|\Delta U\cj U_{\text{ex}} \Sigma_{\text{ex}} + \Sigma_{\text{ex}} V_{\text{ex}}\cj \Delta V + \Delta U\cj G_{\text{exact}} \Delta V\|_2 \\
    &\quad\le \frac{1}{2} \left( \|\Delta U\|_2 + \|\Delta V\|_2 \right)^2 \|G_{\text{exact}}\|_2 \\
    &\quad\le 2\max(\|\Delta U\|_2, \|\Delta V\|_2)^2 \cdot \mathcal{O}(\sqrt{LR}),
\end{aligned}
\end{equation}
For the projected noise, we expand
\begin{equation}
\begin{aligned}
    &(U_i'W_U)^\dagger\E(\theta_i^*)(V_i'W_V) \\
    &\quad= U_{\mathrm{ex}}^\dagger\E(\theta_i^*)V_{\mathrm{ex}}
    +\Delta U^\dagger\E(\theta_i^*)V_{\mathrm{ex}}+
    U_{\mathrm{ex}}^\dagger\E(\theta_i^*)\Delta V
    +\Delta U^\dagger\E(\theta_i^*)\Delta V.
\end{aligned}
\end{equation}
Combining Eqs.~\eqref{eq:prob_eqs_S}, \eqref{eq:det_G_bound}, \eqref{eq:max_UV_norm}, and~\eqref{eq:signal_cancellation}, we obtain
\begin{equation}\label{eq:obs_tilde_G_bound}
\begin{aligned}
    \|\Delta\tilde G\|_2
    &\le
    \|U_{\mathrm{ex}}^\dagger\E(\theta_i^*)V_{\mathrm{ex}}\|_2\\
    &\quad+
    \|\Delta U\|_2\|\E(\theta_i^*)V_{\mathrm{ex}}\|_2
    +\|U_{\mathrm{ex}}^\dagger\E(\theta_i^*)\|_2
       \|\Delta V\|_2\\
    &\quad+
    \|\Delta U\|_2\|\E(\theta_i^*)\|_2\|\Delta V\|_2
    +\|\Delta G_{\mathrm{det}}\|_2\\
    &\quad+
    \mathcal O(\sqrt{LR})
    \max\{\|\Delta U\|_2,\|\Delta V\|_2\}^2\\
    &\le
    \widetilde{\mathcal O}\left(
        \frac1{\sqrt N}
        +\sqrt{\frac{\max(L,R)}{\min(L,R)}}\frac1N
    \right)
    +\mathcal O\left(
        \sqrt{LR}e^{-\Delta^2T^2/4}
    \right).
\end{aligned}
\end{equation}

The same expansion applies to $\Delta\tilde G^O$. The column and row spaces of $G^O_{\mathrm{exact}}$ are contained in those spanned by $U_{\mathrm{ex}}$ and $V_{\mathrm{ex}}$, respectively. Thus
\begin{equation}
    G^O_{\mathrm{exact}}
    =
    U_{\mathrm{ex}}\tilde G^O_{\mathrm{exact}}
    V_{\mathrm{ex}}^\dagger.
\end{equation}
Using Eq.~\eqref{eq:Delta_U_squared} again gives
\begin{equation}\label{eq:signal_cancellation_O}
\begin{aligned}
    &\|(U_i'W_U)\cj G^O_{\text{exact}} (V_i'W_V) - U_{\text{ex}}\cj G^O_{\text{exact}} V_{\text{ex}}\|_2 \\
    &\quad= \|\Delta U\cj U_{\text{ex}} \tilde{G}^O_{\text{exact}} + \tilde{G}^O_{\text{exact}} V_{\text{ex}}\cj \Delta V + \Delta U\cj G^O_{\text{exact}} \Delta V\|_2 \\
    &\quad\le \frac{1}{2}\left( \|\Delta U\|_2^2+\|\Delta V\|_2^2 \right) \|\tilde{G}^O_{\text{exact}}\|_2+ \|\Delta U\|_2 \|\Delta V\|_2 \|G^O_{\text{exact}}\|_2\\
    &\quad\le 2\max(\|\Delta U\|_2, \|\Delta V\|_2)^2 \cdot \mathcal{O}(\|O\|_2\sqrt{LR}),
\end{aligned}
\end{equation}
Therefore, Eqs.~\eqref{eq:prob_eqs_S}, \eqref{eq:Delta_GO_bound}, \eqref{eq:max_UV_norm}, and~\eqref{eq:signal_cancellation_O} imply
\begin{equation}\label{eq:obs_tilde_GO_bound}
    \|\Delta\tilde G^O\|_2
    \le
    \widetilde{\mathcal O}\left(
        \|O\|_2\left[
            \frac1{\sqrt N}
            +\sqrt{\frac{\max(L,R)}{\min(L,R)}}\frac1N
        \right]
    \right)+
    \mathcal O\left(
        \|O\|_2\sqrt{LR}e^{-\Delta^2T^2/4}
    \right).
\end{equation}

\medskip
\textbf{Step 3: Generalized eigenvalue perturbation.} We apply Lemma~\ref{lemma:gen_eig_perturbation} with $m=m_i$, $A=\tilde G^O_{\mathrm{exact}}$, $B=\tilde G_{\mathrm{exact}}=\Sigma_{\mathrm{ex}}$, $\Delta A=\Delta\tilde G^O$, and $\Delta B=\Delta\tilde G$. First, according to Assumption~\ref{assum:uniform_overlap} and $\chi=\mathcal{O}\left(1\right)$,
\begin{equation}
    s_{\min}(B)
    =s_{m_i}(G_{\mathrm{exact}})
    =\Omega(\sqrt{LR}).
\end{equation}
By Eq.~\eqref{eq:obs_tilde_G_bound}, $\min(L,R)N=\widetilde{\Omega}(1)$ and $\Delta T\ge C_T$, with sufficiently large constants, ensure $\|\Delta\tilde G\|_2\le\frac12s_{\min}(B)$. Hence, $B+\Delta\tilde G$ is invertible and $s_{\min}(B)-\|\Delta\tilde G\|_2=\Omega(\sqrt{LR})$.

Next, we verify the diagonalizability and the condition number of $M=B^{-1}A$. By Assumption~\ref{assum:uniform_overlap}, $\tilde{G}_{\text{exact}} = \Sigma_{\text{ex}}$ is strictly invertible. Substituting $G_{\text{exact}} = \Phi_{:,\mathcal{D}_{\pi_i}} (\Psi_{:,\mathcal{D}_{\pi_i}})^\dagger$, we can factorize $\Sigma_{\text{ex}}$ into the product of two $m_i \times m_i$ matrices:
\begin{equation}
    \Sigma_{\mathrm{ex}}
    =
    \left(U_{\mathrm{ex}}^\dagger
          \Phi_{:,\mathcal D_{\pi_i}}\right)
    \left(\Psi_{:,\mathcal D_{\pi_i}}^\dagger
          V_{\mathrm{ex}}\right)
\end{equation}
Since their product $\Sigma_{\text{ex}}$ is invertible, both $U_{\text{ex}}^\dagger \Phi_{:,\mathcal{D}_{\pi_i}}$ and $\Psi_{:,\mathcal{D}_{\pi_i}}^\dagger V_{\text{ex}}$ are also invertible. This yields $\Sigma_{\text{ex}}^{-1} U_{\text{ex}}^\dagger \Phi_{:,\mathcal{D}_{\pi_i}} = (\Psi_{:,\mathcal{D}_{\pi_i}}^\dagger V_{\text{ex}})^{-1}$, allowing us to express $M$ as:
    \begin{equation}
        M = (\Psi_{:,\mathcal{D}_{\pi_i}}^\dagger V_{\text{ex}})^{-1} O_{\mathcal{D}_{\pi_i}} (\Psi_{:,\mathcal{D}_{\pi_i}}^\dagger V_{\text{ex}}).
    \end{equation}
Here we recall $\tilde G^O_{\mathrm{exact}}=U_{\mathrm{ex}}^\dagger\Phi_{:,\mathcal D_{\pi_i}}O_{\mathcal D_{\pi_i}}\Psi_{:,\mathcal D_{\pi_i}}^\dagger V_{\mathrm{ex}}$. Let $O_{\mathcal D_{\pi_i}}=Q\Lambda Q^\dagger$ be its unitary diagonalization. An eigenbasis of $M$ is
\begin{equation}\label{eq:defn_VM}
    V_M=
    \left(\Psi_{:,\mathcal D_{\pi_i}}^\dagger
          V_{\mathrm{ex}}\right)^{-1}Q.
\end{equation}

Because $\Phi_{:,\mathcal D_{\pi_i}}$ has full column rank, $V_{\mathrm{ex}}$ spans the column space of $\Psi_{:,\mathcal D_{\pi_i}}$. Writing its exact SVD as $\Psi_{:,\mathcal D_{\pi_i}} =U_\Psi\Sigma_\Psi V_\Psi^\dagger$, there is a unitary matrix $W$ such that $V_{\mathrm{ex}}=U_\Psi W$. Thus,
\begin{equation}
    \Psi_{:,\mathcal D_{\pi_i}}^\dagger V_{\mathrm{ex}}
    =V_\Psi\Sigma_\Psi W.
\end{equation}
Since both $V_\Psi$ and $W$ are unitary matrices, the singular values of $\Psi_{:,\mathcal{D}_{\pi_i}}^\dagger V_{\text{ex}}$ and $\Psi_{:,\mathcal{D}_{\pi_i}}$ are the same. By Eq.~\eqref{eq:defn_s_avg} and Assumption~\ref{assum:uniform_overlap},
\begin{equation}
\begin{aligned}
    \kappa_2(V_M)
    &=
    \frac{s_{\max}(\Psi_{:,\mathcal D_{\pi_i}})}
         {s_{\min}(\Psi_{:,\mathcal D_{\pi_i}})}\\
    &\le
    \left[
        m_i\left(
            \frac{s_{\mathrm{avg}}(\Psi_{:,\mathcal D_{\pi_i}})}
                 {s_{\min}(\Psi_{:,\mathcal D_{\pi_i}})}
        \right)^2
        -(m_i-1)
    \right]^{1/2}\\
    &\le\sqrt{1+m_i(2\chi+\chi^2)}
    =\mathcal O(1).
\end{aligned}
\end{equation}
Here $m_i=\mathcal O(1)$ was given by Eq.~\eqref{eq:m_i_bound} in Step~0. In particular,
\begin{equation}
    \|M\|_2
    \le\kappa_2(V_M)\|O\|_2
    =\mathcal O(\|O\|_2).
\end{equation}

Finally, the unitary transformations $W_U,W_V$ preserve the generalized eigenvalues. Indeed, setting $w'=W_V^\dagger w$ and left-multiplying the empirical generalized eigenvalue equation by $W_U^\dagger$ gives
\begin{equation}
    \left(\tilde G^O_{\mathrm{exact}}
          +\Delta\tilde G^O\right)w'
    =
    \lambda_k^O
    \left(\tilde G_{\mathrm{exact}}
          +\Delta\tilde G\right)w'.
\end{equation}
Recall that $\lambda_{j,\mathrm{exact}}$, $j\in[m_i]$, are the exact generalized eigenvalues defined by Eq.~\eqref{eq:gen_eigenvalue_exact}. By Eq.~\eqref{eq:lemma_gen_eig_matching} in Lemma~\ref{lemma:gen_eig_perturbation}, there exists a permutation $\rho\in S_{m_i}$ such that, for every $k\in[m_i]$,
\begin{equation}\label{eq:observable_gen_eig_matching}
\begin{aligned}
    |\lambda_k^O-\lambda_{\rho(k),\mathrm{exact}}|
    &\le
    \frac{(2m_i-1)\kappa_2(V_M)}
         {s_{\min}(\tilde G_{\mathrm{exact}})
          -\|\Delta\tilde G\|_2}
    \left(
        \|\Delta\tilde G^O\|_2
        +\|\Delta\tilde G\|_2\|M\|_2
    \right).
\end{aligned}
\end{equation}
Substituting Eqs.~\eqref{eq:obs_tilde_G_bound} and~\eqref{eq:obs_tilde_GO_bound} and using $m_i=\mathcal O(1)$ from Eq.~\eqref{eq:m_i_bound}, we obtain
\begin{equation}
\begin{aligned}
    &|\lambda_k^O-\lambda_{\rho(k),\mathrm{exact}}|\\
    &\quad\le
    \widetilde{\mathcal O}\left(
        \frac{\|O\|_2}{\sqrt{LR}}
        \left[
            \frac1{\sqrt N}
            +\sqrt{\frac{\max(L,R)}{\min(L,R)}}\frac1N
        \right]
    \right)
    +\mathcal O\left(
        \|O\|_2e^{-\Delta^2T^2/4}
    \right)\\
    &\quad=
    \widetilde{\mathcal O}\left(
        \|O\|_2\left[
            \frac1{\sqrt{LRN}}
            +\frac1{\min(L,R)N}
        \right]
    \right)
    +\mathcal O\left(
        \|O\|_2e^{-\Delta^2T^2/4}
    \right).
\end{aligned}
\end{equation}
For each $i\in[I]$, taking the maximum over $k\in[m_i]$ for the permutation $\rho$ supplied by Lemma~\ref{lemma:gen_eig_perturbation} proves Eq.~\eqref{eq:thm_observable_result}. Since all preceding error bounds hold on the same probability-$1-\eta$ event, this conclusion holds simultaneously for all $i\in[I]$.
\end{proof}

\section{Technical Lemmas}
\label{app:lemmas}
In this section, we present several technical lemmas that will be used in the proofs of our main theorems.

\begin{lemma}[Bounds of the Target Cluster Signal]
\label{lemma:bound_U}
Under Assumption~\ref{assum:uniform_overlap}, for every $i\in[I]$,
\begin{equation}
    \frac{\sqrt{LR}\pmin}{1+\chi}
    \le
    \|\Phi_{:,\D_i}\Psi_{:,\D_i}^\dagger\|_F
    \le\sqrt{LR},
\end{equation}
where $\pmin$ is defined in Eq.~\eqref{eq:defn_p_min_tail}.
\end{lemma}

\begin{proof}
For the lower bound, fix a dominant cluster $\D_i$. Following the argument of~\cite[Eq.~(D25)]{QFAMES}, Assumption~\ref{assum:uniform_overlap} and Eq.~\eqref{eq:defn_s_avg} give
\begin{equation}
\begin{aligned}
    \|\Phi_{:,\D_i}\Psi_{:,\D_i}^\dagger\|_F^2
    &=\operatorname{Tr}(\Phi_{:,\D_i}^\dagger \Phi_{:,\D_i}\,\Psi_{:,\D_i}^\dagger \Psi_{:,\D_i})\\
    &\ge s_{\min}(\Phi_{:,\D_i})^2\|\Psi_{:,\D_i}\|_F^2\\
    &\ge\frac{\|\Phi_{:,\D_i}\|_F^2\|\Psi_{:,\D_i}\|_F^2}{|\D_i|(1+\chi)^2}\\
    &\ge\frac{1}{|\D_i|(1+\chi)^2}
        \sum_{m\in\D_i}
        \|\Phi_{:,m}\|_2^2\|\Psi_{:,m}\|_2^2\\
    &\ge\frac{LR\pmin^2}{(1+\chi)^2}.
\end{aligned}
\end{equation}
Taking square roots proves the lower bound uniformly over all dominant clusters.

For the upper bound, let $P_i = \sum_{m \in \D_i} |E_m\rangle \langle E_m|$ denote the orthogonal projection operator onto the $i$-th dominant spectral subspace. Since the eigenvectors $\{|E_m\rangle\}_{m\in\D_i}$ are orthonormal, $P_i$ is an orthogonal projector and satisfies $P_i^2=P_i$ and $0\preceq P_i\preceq I$. Recalling the definition in Eq.~\eqref{eq:defn_Phi_Psi}, the $(l,r)$-th entry of the matrix $\Phi_{:,\D_i}(\Psi_{:,\D_i})\cj$ is given by
\begin{equation}
    \sum_{m \in \D_i} \Phi_{l,m} \overline{\Psi}_{r,m} = \sum_{m \in \D_i} \langle \phi_l | E_m \rangle \langle E_m | \psi_r \rangle = \langle \phi_l | P_i | \psi_r \rangle.
\end{equation}
Applying the Cauchy--Schwarz inequality and the property of $P_i$, we have
\begin{equation}
    |\langle \phi_l | P_i | \psi_r \rangle|^2 = |\langle \phi_l | P_i \cdot P_i | \psi_r \rangle|^2 \le \langle \phi_l | P_i | \phi_l \rangle \langle \psi_r | P_i | \psi_r \rangle \le \langle \phi_l | \phi_l \rangle \langle \psi_r | \psi_r \rangle=1.
\end{equation}
Summing the squared magnitudes of all entries yields:
\begin{equation}
    \|\Phi_{:,\D_i}(\Psi_{:,\D_i})\cj\|_F = \sqrt{\sum_{l=1}^{L}\sum_{r=1}^{R} |\langle \phi_l | P_i | \psi_r \rangle|^2 } \le \sqrt{LR},
\end{equation}
which completes the proof.
\end{proof}

\begin{lemma}[Truncation Bias Bound]\label{lemma:truncation_bias}
For $\sigma\ge1$, the deterministic truncation biases defined in Eq.~\eqref{eq:defn_E_stat_bias} satisfy, uniformly in $\theta$,
\begin{equation}\label{eq:lemma_truncation_bias}
    \|\E^{\mathrm{bias}}(\theta)\|_F\le\mathcal{O}\left(\sqrt{LR}e^{-\sigma^2/4}\right),
\end{equation}
\begin{equation}\label{eq:lemma_truncation_bias_derivative}
    \left\|\frac{d}{d\theta}\E^{\mathrm{bias}}(\theta)\right\|_F\le\mathcal{O}\left(T\sqrt{LR}e^{-\sigma^2/4}\right),
\end{equation}
\begin{equation}\label{eq:lemma_truncation_bias_observable}
    \|\E^{O,\mathrm{bias}}(\theta)\|_F\le\mathcal{O}\left(\|O\|_2\sqrt{LR}e^{-\sigma^2/4}\right),
\end{equation}
\begin{equation}\label{eq:lemma_truncation_bias_observable_derivative}
    \left\|\frac{d}{d\theta}\E^{O,\mathrm{bias}}(\theta)\right\|_F\le\mathcal{O}\left(\|O\|_2T\sqrt{LR}e^{-\sigma^2/4}\right).
\end{equation}
\end{lemma}

\begin{proof}
The Gaussian density and the corresponding truncated probability measure are given by~\cite[Eqs.~(9) and~(16)]{QFAMES}:
\begin{equation}\label{eq:gaussian_density}
    a_T(t)=\frac{1}{2T\sqrt{\pi}}\exp\left(-\frac{t^2}{4T^2}\right),
\end{equation}
\begin{equation}\label{eq:truncated_gaussian_density}
\begin{aligned}
    a_T^{\mathrm{trunc}}(t)
    &=\left(1-\int_{-\sigma T}^{\sigma T}\frac{1}{2T\sqrt{\pi}}\exp\left(-\frac{s^2}{4T^2}\right)\,ds\right)\delta_0(t)\\
    &\quad+\frac{1}{2T\sqrt{\pi}}\exp\left(-\frac{t^2}{4T^2}\right)\mathbf{1}_{[-\sigma T,\sigma T]}(t),
\end{aligned}
\end{equation}
where $\delta_0$ is the Dirac mass at zero. The first term places the omitted Gaussian tail mass at $t=0$, while the second keeps the density on $[-\sigma T,\sigma T]$. Hence, with $h_{l,r}(t,\theta)=\langle\phi_l|e^{i(\theta-H)t}|\psi_r\rangle$,
\begin{equation}\label{eq:F_Gaussian}
\begin{aligned}
    \F^{\mathrm{trunc}}_{l,r}(\theta)&=\int_{\mathbb{R}}h_{l,r}(t,\theta)a_T^{\mathrm{trunc}}(t)\,dt,\\
    \F_{l,r}(\theta)&=\int_{\mathbb{R}}h_{l,r}(t,\theta)a_T(t)\,dt.
\end{aligned}
\end{equation}
Subtracting these identities cancels the integral over $[-\sigma T,\sigma T]$. Using $|h_{l,r}|\le1$ and the tail estimate $\int_x^\infty e^{-s^2}\,ds\le e^{-x^2}/(2x)$, as in the argument of~\cite[Eq.~(22), Lemma~A.1]{QMEGS}, gives
\begin{equation}
\begin{aligned}
    |\E^{\mathrm{bias}}_{l,r}(\theta)|
    &=\left|\int_{|t|>\sigma T}a_T(t)\bigl(h_{l,r}(0,\theta)-h_{l,r}(t,\theta)\bigr)\,dt\right|\\
    &\le2\int_{|t|>\sigma T}a_T(t)\,dt
    =\frac{4}{\sqrt{\pi}}\int_{\sigma/2}^{\infty}e^{-s^2}\,ds
    \le\frac{4}{\sigma\sqrt{\pi}}e^{-\sigma^2/4},\\
    \left|\frac{d}{d\theta}\E^{\mathrm{bias}}_{l,r}(\theta)\right|
    &=\left|\int_{|t|>\sigma T}a_T(t)\bigl(ith_{l,r}(t,\theta)\bigr)\,dt\right|
    \le\int_{|t|>\sigma T}|t|a_T(t)\,dt
    =\frac{2T}{\sqrt{\pi}}e^{-\sigma^2/4}.
\end{aligned}
\end{equation}
Summing the squared entrywise bounds proves the first two inequalities.

For the observable signal in Eq.~\eqref{eq:defn_F_O}, write
\begin{equation}
    \F^O_{l,r}(\theta)
    =
    \langle\phi_l|K(\theta)OK(\theta)|\psi_r\rangle,
    \qquad
    K(\theta)=e^{-(\theta-H)^2T^2/2}.
\end{equation}
The corresponding Gaussian representation and its truncation are
\begin{equation}\label{eq:defn_K}
    K=\int_{\mathbb{R}}a_{T/\sqrt{2}}(t)e^{i(\theta-H)t}\,dt,
    \qquad
    K_{\mathrm{trunc}}
    =\int_{\mathbb{R}}a_{T/\sqrt{2}}^{\mathrm{trunc}}(t)
    e^{i(\theta-H)t}\,dt.
\end{equation}
Here the scale $T/\sqrt{2}$ accounts for the factor $1/2$ in the exponent of $K$. By independence of the two sampled times and symmetry of $a_{T/\sqrt{2}}^{\mathrm{trunc}}$, the observable estimator satisfies
\begin{equation}
    \F^{O,\mathrm{trunc}}_{l,r}(\theta)
    =\langle\phi_l|
        K_{\mathrm{trunc}} O K_{\mathrm{trunc}}
      |\psi_r\rangle.
\end{equation}

Since truncation places the omitted mass at $t=0$, the difference $\Delta K=K_{\mathrm{trunc}}-K$ satisfies
\begin{equation}
\begin{aligned}
    \Delta K&=\int_{|t|>\sigma T/\sqrt{2}}a_{T/\sqrt{2}}(t)
    \bigl(I-e^{i(\theta-H)t}\bigr)\,dt,\\
    \Delta K'&=-i\int_{|t|>\sigma T/\sqrt{2}}
    t\,a_{T/\sqrt{2}}(t)e^{i(\theta-H)t}\,dt,
\end{aligned}
\end{equation}
where primes denote differentiation with respect to $\theta$. Unitarity and the preceding Gaussian tail estimates imply
\begin{equation}
\begin{aligned}
    \|\Delta K\|_2
    &\le 2\int_{|t|>\sigma T/\sqrt{2}}a_{T/\sqrt{2}}(t)\,dt
    =\mathcal{O}(e^{-\sigma^2/4}),\\
    \|\Delta K'\|_2
    &\le \int_{|t|>\sigma T/\sqrt{2}}
    |t|a_{T/\sqrt{2}}(t)\,dt
    =\sqrt{\frac{2}{\pi}}T e^{-\sigma^2/4}.
\end{aligned}
\end{equation}
Both $K$ and $K_{\mathrm{trunc}}$ are averages of unitary operators with probability weights. Thus $\|K\|_2,\|K_{\mathrm{trunc}}\|_2\le1$. Their derivatives satisfy
\begin{equation}
    \max\{\|K'\|_2,\|K_{\mathrm{trunc}}'\|_2\}
    \le\int_{\mathbb{R}}|t|a_{T/\sqrt{2}}(t)\,dt
    =\sqrt{\frac{2}{\pi}}T.
\end{equation}
Now set $\Delta K_O :=K_{\mathrm{trunc}}OK_{\mathrm{trunc}}-KOK =\Delta K\,O K_{\mathrm{trunc}}+KO\,\Delta K$. By the product rule and submultiplicativity,
\begin{equation}
\begin{aligned}
    \|\Delta K_O\|_2
    &\le 2\|O\|_2\|\Delta K\|_2
    =\mathcal{O}(\|O\|_2e^{-\sigma^2/4}),\\
    \|\Delta K_O'\|_2
    &\le \|O\|_2\left[
    2\|\Delta K'\|_2+
    \|\Delta K\|_2\bigl(\|K_{\mathrm{trunc}}'\|_2+\|K'\|_2\bigr)
    \right]\\
    &=\mathcal{O}(\|O\|_2T e^{-\sigma^2/4}).
\end{aligned}
\end{equation}
Since the initial states are normalized,
\begin{equation}
\begin{aligned}
    \|\E^{O,\mathrm{bias}}(\theta)\|_F
    &=\left(\sum_{l,r}|\langle\phi_l|\Delta K_O|\psi_r\rangle|^2\right)^{1/2}\\
    &\le\sqrt{LR}\|\Delta K_O\|_2
    \le\mathcal{O}\left(\|O\|_2\sqrt{LR}e^{-\sigma^2/4}\right),\\
    \left\|\frac{d}{d\theta}\E^{O,\mathrm{bias}}(\theta)\right\|_F
    &=\left(\sum_{l,r}|\langle\phi_l|\Delta K_O'|\psi_r\rangle|^2\right)^{1/2}\\
    &\le\sqrt{LR}\|\Delta K_O'\|_2
    \le\mathcal{O}\left(\|O\|_2T\sqrt{LR}e^{-\sigma^2/4}\right).
\end{aligned}
\end{equation}
This proves the remaining two inequalities.
\end{proof}

\begin{lemma}[Uniform Bounds on the Total Error Matrix over $\Theta$]
\label{lemma:error_term}
Let $\Theta$ be the finite evaluation grid. Suppose that $\sigma\ge\max\{2\sqrt{\log(LRN)},1\}$. For any $\tilde{\eta}\in(0,1)$, each of the following inequalities holds individually with probability at least $1-\tilde{\eta}$ under either sampling architecture:

\noindent\textbf{1. Entrywise:} Simultaneously for all $(l,r)\in[L]\times[R]$,
\begin{equation}
\label{eq:lemma_elem_error_bound}
    \sup_{\theta\in\Theta}
    |\E_{l,r}(\theta)|
    \le
    \widetilde{\mathcal{O}}\left(\frac{1}{\sqrt{N}}\right),
\end{equation}
and
\begin{equation}
\label{eq:lemma_elem_error_bound_diff}
    \sup_{\theta,\theta'\in\Theta,\ \theta\ne\theta'}
    \frac{|\E_{l,r}(\theta)-\E_{l,r}(\theta')|}
         {|\theta-\theta'|}
    \le
    \widetilde{\mathcal{O}}\left(\frac{\sigma T}{\sqrt{N}}\right).
\end{equation}
\noindent\textbf{2. Frobenius norm:}
\begin{equation}
\label{eq:lemma_frob_error_bound}
    \sup_{\theta\in\Theta}\|\E(\theta)\|_F
    \le
    \widetilde{\mathcal{O}}\left(\sqrt{\frac{LR}{N}}\right),
\end{equation}
and
\begin{equation}
\label{eq:lemma_frob_error_bound_diff}
    \sup_{\theta,\theta'\in\Theta,\,\theta\ne\theta'}
    \frac{\|\E(\theta)-\E(\theta')\|_F}{|\theta-\theta'|}
    \le
    \widetilde{\mathcal{O}}\left(
        \sigma T\sqrt{\frac{LR}{N}}
    \right).
\end{equation}
Each of the following additional inequalities also holds individually with probability at least $1-\tilde{\eta}$ under the respective sampling architectures.

\noindent\textbf{Independent Time Sampling:}
\begin{equation}
\label{eq:lemma_error_ind_diff_abs}
    \sup_{\theta,\theta'\in\Theta}
    \left|
        \|\E(\theta)\|_F^2-\|\E(\theta')\|_F^2+\frac{1}{N}\left(\|\F(\theta)\|_F^2-\|\F(\theta')\|_F^2\right)
    \right|
    \le
    \widetilde{\mathcal{O}}\left(\frac{\sqrt{LR}}{N}\right),
\end{equation}
\begin{equation}
\label{eq:lemma_error_ind_diff_lip}
    \sup_{\theta,\theta'\in\Theta,\,\theta\ne\theta'}
    \frac{
        \left|\|\E(\theta)\|_F^2-\|\E(\theta')\|_F^2
        +\frac{1}{N}\left(\|\F(\theta)\|_F^2-\|\F(\theta')\|_F^2\right)\right|
    }{|\theta-\theta'|}
    \le
    \widetilde{\mathcal{O}}\left(
        \frac{\sigma T\sqrt{LR}}{N}
    \right).
\end{equation}

\noindent\textbf{Shared Time Sampling:}
\begin{equation}
\label{eq:lemma_error_shared_diff_abs}
    \sup_{\theta,\theta'\in\Theta}|\|\E(\theta)\|_F^2-\|\E(\theta')\|_F^2|
    \le
    \widetilde{\mathcal{O}}\left(\frac{LR}{N}\right),
\end{equation}
\begin{equation}
\label{eq:lemma_error_shared_diff_lip}
    \sup_{\theta,\theta'\in\Theta,\,\theta\ne\theta'}
    \frac{|\|\E(\theta)\|_F^2-\|\E(\theta')\|_F^2|}{|\theta-\theta'|}
    \le
    \widetilde{\mathcal{O}}\left(\frac{\sigma TLR}{N}\right).
\end{equation}
Here $\widetilde{\mathcal{O}}$ hides logarithmic factors in the grid size, the matrix dimensions, and $\tilde{\eta}^{-1}$.
\end{lemma}

\begin{proof}
Recall from Eq.~\eqref{eq:defn_E_stat_bias} that $\E^{\mathrm{stat}}=G-\F^{\mathrm{trunc}}$ and $\E^{\mathrm{bias}}=\F^{\mathrm{trunc}}-\F$. Write $b_\sigma:=e^{-\sigma^2/4}$ for the Gaussian tail factor. We first establish concentration bounds for $\E^{\mathrm{stat}}$, and then transfer them to $\E=\E^{\mathrm{stat}}+\E^{\mathrm{bias}}$. Let each inequality hold with probability at least $1-\tilde{\eta}$.

\noindent\textbf{Part 1. Entrywise and Frobenius bounds.} For fixed $(l,r)$ and $\theta$,
\begin{equation}
    \E^{\mathrm{stat}}_{l,r}(\theta)
    =\frac1N\sum_{n=1}^N
    \left(
        Z_{l,r,n}e^{i\theta t_{l,r,n}}
        -\F^{\mathrm{trunc}}_{l,r}(\theta)
    \right).
\end{equation}
The summands are independent across $n$, centered, and bounded in magnitude by $\sqrt{2}+1$. Under the \textit{Shared Time Sampling} scheme, $t_{l,r,n}=t_n$, but independence across $n$ remains valid. Hoeffding's inequality, applied to the real and imaginary parts, gives, for $u>0$ and some constant $c>0$,
\begin{equation}\label{eq:S_bound}
    \Pr\left(|\E^{\mathrm{stat}}_{l,r}(\theta)|\ge u\right)\le \Pr\left(|\RE \E^{\mathrm{stat}}_{l,r}(\theta)|\ge \frac{u}{\sqrt{2}}\right) + \Pr\left(|\IM \E^{\mathrm{stat}}_{l,r}(\theta)|\ge \frac{u}{\sqrt{2}}\right) \le4e^{-cNu^2}.
\end{equation}

For distinct $\theta,\theta'$, the difference $\E^{\mathrm{stat}}_{l,r}(\theta)-\E^{\mathrm{stat}}_{l,r}(\theta')$ is the average of the centered variables associated with
\begin{equation}
    Z_{l,r,n}
    \left(e^{i\theta t_{l,r,n}}-e^{i\theta't_{l,r,n}}\right).
\end{equation}
Due to $|t_{l,r,n}|\le\sigma T$ and the Lipschitz continuity,
\begin{equation}
    \left|
        Z_{l,r,n}
        \left(e^{i\theta t_{l,r,n}}
             -e^{i\theta't_{l,r,n}}\right)
    \right|
    \le \sqrt{2}\sigma T|\theta-\theta'|.
\end{equation}
Centering increases this bound by at most a factor of two. Applying the same inequality to the centered differences gives, for $u>0$,
\begin{equation}\label{eq:S_diff_bound}
    \Pr\left(|\E^{\mathrm{stat}}_{l,r}(\theta)-\E^{\mathrm{stat}}_{l,r}(\theta')|
        \ge u\sigma T|\theta-\theta'|\right)\le4e^{-cNu^2}.
\end{equation}
Taking union bounds over all entries, grid points, and grid pairs, with failure probability at most $\tilde{\eta}/4$ for each of the following two bounds, gives
\begin{equation}\label{eq:S_bound_final}
    \sup_{\theta,l,r}|\E^{\mathrm{stat}}_{l,r}(\theta)|
    \le\widetilde{\mathcal{O}}(N^{-1/2}),
    \quad
    \sup_{\theta\ne\theta',l,r}
    \frac{|\E^{\mathrm{stat}}_{l,r}(\theta)-\E^{\mathrm{stat}}_{l,r}(\theta')|}
         {|\theta-\theta'|}
    \le\widetilde{\mathcal{O}}(\sigma TN^{-1/2}).
\end{equation}
Summing squared entrywise bounds yields the corresponding Frobenius bounds for $\E^{\mathrm{stat}}$.

The entrywise estimates in the proof of Lemma~\ref{lemma:truncation_bias} give, uniformly in $\theta$,
\begin{equation}
    |\E^{\mathrm{bias}}_{l,r}(\theta)|\le \mathcal{O}\left(b_\sigma\right),
    \quad
    |\partial_\theta \E^{\mathrm{bias}}_{l,r}(\theta)|\le \mathcal{O}\left(Tb_\sigma\right),
\end{equation}
Since $b_\sigma\le(LRN)^{-1}\le N^{-1/2}$, the triangle inequality and the mean value bound absorb these contributions into the preceding statistical bounds. Summing the squared entrywise bounds over the $LR$ entries and taking square roots gives the two global Frobenius bounds, with the additional factor $\sqrt{LR}$. This proves
Eqs.~\eqref{eq:lemma_elem_error_bound}--\eqref{eq:lemma_frob_error_bound_diff}.

\noindent\textbf{Part 2. Squared-norm differences under Independent Time Sampling.} Define
\begin{equation}\label{eq:defn_Y_S}
    \Delta Q^{\mathrm{stat}}(\theta,\theta')
    :=\|\E^{\mathrm{stat}}(\theta)\|_F^2-\|\E^{\mathrm{stat}}(\theta')\|_F^2.
\end{equation}
For fixed $(l,r)$ and $\theta$, the random variables $Z_{l,r,n}e^{i\theta t_{l,r,n}}-\F^{\mathrm{trunc}}_{l,r}(\theta)$, for all $n \in [N]$, are independent and centered. Hence the mixed terms vanish when expanding the squared magnitude of their average. Since $|Z_{l,r,n}|^2=2$, we obtain
\begin{equation}\label{eq:Ex_E_2}
\begin{aligned}
    \Ex[|\E^{\mathrm{stat}}_{l,r}(\theta)|^2]
    &=\frac{1}{N^2}\sum_{n=1}^N
        \Ex\left[
            \left|
                Z_{l,r,n}e^{i\theta t_{l,r,n}}
                -\F^{\mathrm{trunc}}_{l,r}(\theta)
            \right|^2
        \right]\\
    &=\frac{1}{N^2}\sum_{n=1}^N
        \left(
            \Ex[|Z_{l,r,n}|^2]
            -|\F^{\mathrm{trunc}}_{l,r}(\theta)|^2
        \right)\\
    &=\frac{2-|\F^{\mathrm{trunc}}_{l,r}(\theta)|^2}{N}.
\end{aligned}
\end{equation}
and therefore
\begin{equation}
    \Ex[\Delta Q^{\mathrm{stat}}(\theta,\theta')]
    =\frac{
        \|\F^{\mathrm{trunc}}(\theta')\|_F^2
        -\|\F^{\mathrm{trunc}}(\theta)\|_F^2
    }{N}.
\end{equation}

To control the squared fluctuations, let $\|\cdot\|_{\psi_2}$ and $\|\cdot\|_{\psi_1}$ denote the sub-Gaussian and sub-exponential norms in~\cite[Definitions 2.6.4 and 2.8.4]{vershynin2026high}. Explicitly, for $\alpha\in\{1,2\}$,
\begin{equation}
    \|X\|_{\psi_\alpha}:=\inf\left\{t>0:\Ex\left[\exp\left(\frac{|X|^\alpha}{t^\alpha}\right)\right]\le2\right\}.
\end{equation}
For complex random variables, we use the same definition through their magnitudes. Eq.~\eqref{eq:S_bound} and Eq.~\eqref{eq:S_diff_bound} in Part 1 give sub-Gaussian averages, with
\begin{equation}
\label{eq:total_lemma_stat_subgaussian}
\begin{aligned}
    \|\E^{\mathrm{stat}}_{l,r}(\theta)\|_{\psi_2}&\le \mathcal{O} \left(\frac{1}{\sqrt N}\right),\\
    \|\E^{\mathrm{stat}}_{l,r}(\theta)-\E^{\mathrm{stat}}_{l,r}(\theta')\|_{\psi_2}
    &\le \mathcal{O}\left(\frac{\sigma T|\theta-\theta'|}{\sqrt N}\right),
\end{aligned}
\end{equation}
By~\cite[Lemma 2.8.5]{vershynin2026high}, squaring a sub-Gaussian magnitude gives a sub-exponential variable, with $\||X|^2\|_{\psi_1}=\|X\|_{\psi_2}^2$.

For fixed $\theta,\theta'$, define the entrywise squared-error difference
\begin{equation}
    \Delta Q^{\mathrm{stat}}_{l,r}
    :=|\E^{\mathrm{stat}}_{l,r}(\theta)|^2-|\E^{\mathrm{stat}}_{l,r}(\theta')|^2.
\end{equation}
The variables $\Delta Q^{\mathrm{stat}}_{l,r}$ are independent across $(l,r)$. The triangle inequality and Eq.~\eqref{eq:total_lemma_stat_subgaussian} therefore give
\begin{equation}
    \|\Delta Q^{\mathrm{stat}}_{l,r}\|_{\psi_1}\le\|\E^{\mathrm{stat}}_{l,r}(\theta)\|_{\psi_2}^2+\|\E^{\mathrm{stat}}_{l,r}(\theta')\|_{\psi_2}^2\le \mathcal{O}\left(\frac{1}{N}\right).
\end{equation}
Moreover, using
\begin{equation}
    \big||z|^2-|w|^2\big| \le |z-w|(|z|+|w|)
\end{equation}
and the product bound in~\cite[Lemma 2.8.6]{vershynin2026high}, we have
\begin{equation}
\begin{aligned}
    \|\Delta Q^{\mathrm{stat}}_{l,r}\|_{\psi_1}
    &\le \|\E^{\mathrm{stat}}_{l,r}(\theta)-\E^{\mathrm{stat}}_{l,r}(\theta')\|_{\psi_2}
    \left(\|\E^{\mathrm{stat}}_{l,r}(\theta)\|_{\psi_2}+\|\E^{\mathrm{stat}}_{l,r}(\theta')\|_{\psi_2}\right)\\
    &\le \mathcal{O}\left(\frac{\sigma T|\theta-\theta'|}{N}\right).
\end{aligned}
\end{equation}
Define the centered entrywise difference
\[
\widehat{\Delta Q}^{\mathrm{stat}}_{l,r}
:=\Delta Q^{\mathrm{stat}}_{l,r}-\Ex[\Delta Q^{\mathrm{stat}}_{l,r}].
\]
The centering inequality~\cite[Eq.~(2.26)]{vershynin2026high} gives $\|\widehat{\Delta Q}^{\mathrm{stat}}_{l,r}\|_{\psi_1}\le C\|\Delta Q^{\mathrm{stat}}_{l,r}\|_{\psi_1}$ for some constant $C$. In particular,
\begin{equation}
    \max_{l,r}\|\widehat{\Delta Q}^{\mathrm{stat}}_{l,r}\|_{\psi_1}\le \mathcal{O}\left(1/N\right), \quad \sum_{l,r}\|\widehat{\Delta Q}^{\mathrm{stat}}_{l,r}\|_{\psi_1}^2\le \mathcal{O}\left(LR/N^2\right).
\end{equation}
Recall $\Delta Q^{\mathrm{stat}}$ defined in Eq.~\eqref{eq:defn_Y_S}. We apply Bernstein's inequality~\cite[Theorem 2.9.1]{vershynin2026high} to the independent centered variables $\widehat{\Delta Q}^{\mathrm{stat}}_{l,r}$ and use $\Delta Q^{\mathrm{stat}}-\Ex[\Delta Q^{\mathrm{stat}}]=\sum_{l,r}\widehat{\Delta Q}^{\mathrm{stat}}_{l,r}$, which implies
\begin{equation}
    \Pr\left(
        |\Delta Q^{\mathrm{stat}}-\Ex[\Delta Q^{\mathrm{stat}}]|\ge u
    \right)
    \le
    2\exp\left[
        -c\min\left\{
            \frac{N^2u^2}{LR},Nu
        \right\}
    \right],
\end{equation}
where $c>0$ is an absolute constant. For $\theta\ne\theta'$, it also implies
\begin{equation}
    \Pr\left(
        \frac{|\Delta Q^{\mathrm{stat}}-\Ex[\Delta Q^{\mathrm{stat}}]|}{|\theta-\theta'|}
        \ge u
    \right)
    \le
    2\exp\left[
        -c'\min\left\{
            \frac{N^2u^2}{LR(\sigma T)^2},
            \frac{Nu}{\sigma T}
        \right\}
    \right],
\end{equation}
with another absolute constant $c'>0$. Union bounds over the finite set of grid pairs yield each of the following two bounds individually with failure probability at most $\tilde{\eta}/2$:
\begin{equation}\label{eq:Y_S_diff}
    \sup_{\theta,\theta'}
    |\Delta Q^{\mathrm{stat}}-\Ex[\Delta Q^{\mathrm{stat}}]|
    \le\widetilde{\mathcal{O}}\left(\frac{\sqrt{LR}}N\right), \quad \sup_{\theta\ne\theta'}
    \frac{|\Delta Q^{\mathrm{stat}}-\Ex[\Delta Q^{\mathrm{stat}}]|}{|\theta-\theta'|}
    \le\widetilde{\mathcal{O}}\left(
        \frac{\sigma T\sqrt{LR}}N
    \right).
\end{equation}

We now replace $\E^{\mathrm{stat}}$ by $\E$ and $\F^{\mathrm{trunc}}$ by $\F$. Define the signal squared-norm bias and the resulting bias remainder by
\begin{equation}\label{eq:defn_squared_bias_remainder}
\begin{aligned}
    \Delta\F_2(\theta)
    &:=\|\F^{\mathrm{trunc}}(\theta)\|_F^2
       -\|\F(\theta)\|_F^2,\\
    \mathcal{R}_{\mathrm{bias}}(\theta)
    &:=2\RE\langle \E^{\mathrm{stat}}(\theta),\E^{\mathrm{bias}}(\theta)\rangle
       +\|\E^{\mathrm{bias}}(\theta)\|_F^2-\frac{\Delta\F_2(\theta)}N.
\end{aligned}
\end{equation}
The following identity is exact:
\begin{equation}
    \|\E(\theta)\|_F^2+\frac{\|\F(\theta)\|_F^2}{N}
    =
    \|\E^{\mathrm{stat}}(\theta)\|_F^2
    +\frac{\|\F^{\mathrm{trunc}}(\theta)\|_F^2}{N}
    +\mathcal{R}_{\mathrm{bias}}(\theta).
\end{equation}
Using the Gaussian integral representation of $\F_{l,r}$ in Eq.~\eqref{eq:F_Gaussian}, we have
\begin{equation}\label{eq:F_entry_bound}
    |\F_{l,r}(\theta)| \le \int_{\mathbb{R}}a_T(t)\,dt=1, \quad
    |\partial_\theta\F_{l,r}(\theta)|
    \le \int_{\mathbb{R}}|t|a_T(t)\,dt
    =\frac{2T}{\sqrt{\pi}}.
\end{equation}
Summing the squared entrywise bounds gives
\begin{equation}\label{eq:F_bound}
    \sup_\theta\|\F(\theta)\|_F\le\sqrt{LR},
    \quad
    \sup_\theta\|\partial_\theta\F(\theta)\|_F
    \le\frac{2T}{\sqrt{\pi}}\sqrt{LR}.
\end{equation}
Since $\F^{\mathrm{trunc}}=\F+\E^{\mathrm{bias}}$, we have
\begin{equation}
    \Delta\F_2(\theta)
    =2\RE\langle\F(\theta),\E^{\mathrm{bias}}(\theta)\rangle
    +\|\E^{\mathrm{bias}}(\theta)\|_F^2.
\end{equation}
By the Cauchy--Schwarz inequality and Lemma~\ref{lemma:truncation_bias}, we establish
\begin{equation}
\begin{aligned}
    |\Delta\F_2(\theta)|
    &\le 2\|\F(\theta)\|_F\|\E^{\mathrm{bias}}(\theta)\|_F
       +\|\E^{\mathrm{bias}}(\theta)\|_F^2\\
    &\le \mathcal{O}\left(LR(b_\sigma+b_\sigma^2)\right)
    \le \mathcal{O}\left(N^{-1}\right),
\end{aligned}
\end{equation}
where we used $b_\sigma=e^{-\sigma^2/4}\le (LRN)^{-1}$. Differentiating the same identity and applying the Cauchy--Schwarz inequality again gives
\begin{equation}
\begin{aligned}
    |\Delta\F_2'(\theta)|
    &\le 2\|\partial_\theta\F(\theta)\|_F
             \|\E^{\mathrm{bias}}(\theta)\|_F+2\|\F(\theta)\|_F
             \|\partial_\theta\E^{\mathrm{bias}}(\theta)\|_F\\
    &\quad+2\|\E^{\mathrm{bias}}(\theta)\|_F
             \|\partial_\theta\E^{\mathrm{bias}}(\theta)\|_F\le \mathcal{O}\left(TN^{-1}\right).
\end{aligned}
\end{equation}
Consequently, the mean value theorem yields
\begin{equation}
    \sup_{\theta\ne\theta'}
    \frac{|\Delta\F_2(\theta)-\Delta\F_2(\theta')|}{|\theta-\theta'|}
    \le \sup_\theta|\Delta\F_2'(\theta)|
    \le \mathcal{O}\left(TN^{-1}\right).
\end{equation}
Recall the bias remainder $\mathcal{R}_{\mathrm{bias}}$ in Eq.~\eqref{eq:defn_squared_bias_remainder} and $b_\sigma=e^{-\sigma^2/4}$. According to Eq.~\eqref{eq:S_bound_final} and the Cauchy--Schwarz inequality, we obtain
\begin{equation}\label{eq:R_b_bound}
    \sup_{\theta\in\Theta}|\mathcal{R}_{\mathrm{bias}}(\theta)|
    \le
    \widetilde{\mathcal{O}}\left(
        LR\left(
            \frac{b_\sigma}{\sqrt N}
            +b_\sigma^2+\frac{b_\sigma}{N}
        \right)
    \right) \le \widetilde{\mathcal{O}}\left(N^{-3/2}\right).
\end{equation}
For the difference bound, expand
\begin{equation}
\begin{aligned}
    \langle \E^{\mathrm{stat}}(\theta),\E^{\mathrm{bias}}(\theta)\rangle
    -\langle \E^{\mathrm{stat}}(\theta'),\E^{\mathrm{bias}}(\theta')\rangle
    &=
    \langle \E^{\mathrm{stat}}(\theta)-\E^{\mathrm{stat}}(\theta'),\E^{\mathrm{bias}}(\theta)\rangle\\
    &\quad+
    \langle \E^{\mathrm{stat}}(\theta'),\E^{\mathrm{bias}}(\theta)-\E^{\mathrm{bias}}(\theta')\rangle.
\end{aligned}
\end{equation}
The same estimates give
\begin{equation}\label{eq:R_b_diff_bound}
    \sup_{\theta\ne\theta'}
    \frac{|\mathcal{R}_{\mathrm{bias}}(\theta)-\mathcal{R}_{\mathrm{bias}}(\theta')|}{|\theta-\theta'|}
    \le
    \widetilde{\mathcal{O}}\left(
        \sigma T N^{-3/2}
    \right).
\end{equation}
Since $L,R,N\ge1$, we have $N^{-3/2}\le\sqrt{LR}/N$. Thus both bias-remainder bounds are absorbed into the corresponding statistical bounds. For each of the two conclusions, combine the corresponding bound in Eq.~\eqref{eq:Y_S_diff} with the joint event in Eq.~\eqref{eq:S_bound_final}, on which Eqs.~\eqref{eq:R_b_bound} and~\eqref{eq:R_b_diff_bound} hold. Each event has failure probability at most $\tilde{\eta}/2$. A union bound therefore proves Eqs.~\eqref{eq:lemma_error_ind_diff_abs} and~\eqref{eq:lemma_error_ind_diff_lip}, each individually with probability at least $1-\tilde{\eta}$.

\noindent\textbf{Part 3. Squared-norm differences under Shared Time Sampling.} On the joint event in Eq.~\eqref{eq:S_bound_final}, both total-error Frobenius bounds from Part 1 hold. This event has probability at least $1-\tilde{\eta}/2$. The total-error Frobenius bounds from Part 1 (Eq.~\eqref{eq:lemma_frob_error_bound}) directly give
\begin{equation}
    |\|\E(\theta)\|_F^2-\|\E(\theta')\|_F^2|
    \le \|\E(\theta)\|_F^2+\|\E(\theta')\|_F^2
    \le\widetilde{\mathcal{O}}\left(\frac{LR}{N}\right),
\end{equation}
which proves Eq.~\eqref{eq:lemma_error_shared_diff_abs}. Also,
\begin{equation}
    |\|\E(\theta)\|_F^2-\|\E(\theta')\|_F^2|
    \le
    \left(\|\E(\theta)\|_F+\|\E(\theta')\|_F\right)
    \|\E(\theta)-\E(\theta')\|_F.
\end{equation}
Dividing both sides by $|\theta-\theta'|$ and using Eq.~\eqref{eq:lemma_frob_error_bound} and Eq.~\eqref{eq:lemma_frob_error_bound_diff}, we establish Eq.~\eqref{eq:lemma_error_shared_diff_lip}. This completes the proof.
\end{proof}

\begin{lemma}[Cross Term Bounds for the Total Error]
\label{lemma:cross_terms}
Let $\E(\theta)=G(\theta)-\F(\theta)$, and suppose that $\sigma\ge\max\{2\sqrt{\log(LRN)},1\}$. For any $\tilde{\eta}\in(0,1)$, each of the following inequalities holds individually with probability at least $1-\tilde{\eta}$ under the respective sampling architectures.

\noindent\textbf{Independent Time Sampling:}
\begin{equation}
\label{eq:lemma_cross_indep}
    \sup_{\theta\in\Theta,\,\F(\theta)\ne0}
    \frac{|2\RE\langle\F(\theta),\E(\theta)\rangle|}
         {\|\F(\theta)\|_F}
    \le
    \widetilde{\mathcal{O}}\left(\frac1{\sqrt N}\right),
\end{equation}
\begin{equation}
\label{eq:lemma_cross_indep_diff}
\begin{aligned}
    \sup_{\theta,\theta'\in\Theta,\,\F(\theta)\ne\F(\theta')}
    \frac{
        |2\RE\langle\F(\theta)-\F(\theta'),\E(\theta)\rangle|
    }{\|\F(\theta)-\F(\theta')\|_F}
    &\le
    \widetilde{\mathcal{O}}\left(\frac1{\sqrt N}\right),\\
    \sup_{\theta,\theta'\in\Theta,\ \theta\ne\theta',\,\F(\theta')\ne0}
    \frac{
        |2\RE\langle\F(\theta'),
            \E(\theta)-\E(\theta')\rangle|
    }{|\theta-\theta'|\|\F(\theta')\|_F}
    &\le
    \widetilde{\mathcal{O}}\left(\frac{\sigma T}{\sqrt N}\right).
\end{aligned}
\end{equation}

\noindent\textbf{Shared Time Sampling:}
\begin{equation}
\label{eq:lemma_cross_shared}
    \sup_{\theta\in\Theta,\,\F(\theta)\ne0}
    \frac{|2\RE\langle\F(\theta),\E(\theta)\rangle|}
         {\|\F(\theta)\|_F}
    \le
    \widetilde{\mathcal{O}}\left(\sqrt{\frac{LR}{N}}\right),
\end{equation}
\begin{equation}
\label{eq:lemma_cross_shared_diff}
\begin{aligned}
    \sup_{\theta,\theta'\in\Theta,\,\F(\theta)\ne\F(\theta')}
    \frac{
        |2\RE\langle\F(\theta)-\F(\theta'),\E(\theta)\rangle|
    }{\|\F(\theta)-\F(\theta')\|_F}
    &\le
    \widetilde{\mathcal{O}}\left(\sqrt{\frac{LR}{N}}\right),\\
    \sup_{\theta,\theta'\in\Theta,\ \theta\ne\theta',\,\F(\theta')\ne0}
    \frac{
        |2\RE\langle\F(\theta'),
            \E(\theta)-\E(\theta')\rangle|
    }{|\theta-\theta'|\|\F(\theta')\|_F}
    &\le
    \widetilde{\mathcal{O}}\left(
        \sigma T\sqrt{\frac{LR}{N}}
    \right).
\end{aligned}
\end{equation}
Here $\widetilde{\mathcal{O}}$ hides logarithmic factors in the grid size, the matrix dimensions, and $\tilde{\eta}^{-1}$.
\end{lemma}

\begin{proof}
Recall $\E=\E^{\mathrm{stat}}+\E^{\mathrm{bias}}$, where $\E^{\mathrm{stat}}=G-\F^{\mathrm{trunc}}$ and $\E^{\mathrm{bias}}=\F^{\mathrm{trunc}}-\F$; see Eq.~\eqref{eq:defn_E_stat_bias}. For a fixed deterministic matrix $A\in\mathbb{C}^{L\times R}$, define
\begin{equation}
    \mathcal{C}^{\mathrm{stat}}_A(\theta):=2\RE\langle A,\E^{\mathrm{stat}}(\theta)\rangle.
\end{equation}

\noindent\textbf{Part 1. }\textbf{Independent Time Sampling.} We have
\begin{equation}
    \mathcal{C}^{\mathrm{stat}}_A(\theta)
    =\RE\sum_{l,r,n}\frac{2}{N}\overline{A_{l,r}}
    \left(
        Z_{l,r,n}e^{i\theta t_{l,r,n}}
        -\F^{\mathrm{trunc}}_{l,r}(\theta)
    \right).
\end{equation}
The real summands are independent and centered. For a bounded real random variable $X$, its range length is $\sup(X)-\inf(X)$ (i.e., the maximum possible value minus the minimum possible value when both are attained). The range length of the $(l,r,n)$ summand is at most $\mathcal{O}(|A_{l,r}|/N)$. Thus the sum of squared range lengths is at most
\begin{equation}
    \mathcal{O}\left(\frac1N\sum_{l,r}|A_{l,r}|^2\right)
    =\mathcal{O}\left(\frac{\|A\|_F^2}{N}\right).
\end{equation}
Hoeffding's inequality gives, for $A\ne0$ and an absolute constant $c>0$,
\begin{equation}\label{eq:A_S}
    \Pr\left(
        |\mathcal{C}^{\mathrm{stat}}_A(\theta)|\ge u\|A\|_F
    \right)
    \le 2e^{-cNu^2}.
\end{equation}

Likewise, $2\RE\langle A,\E^{\mathrm{stat}}(\theta)-\E^{\mathrm{stat}}(\theta')\rangle$ is a sum of independent centered real variables. Using $|e^{i\theta t}-e^{i\theta't}|\le\sigma T|\theta-\theta'|$, their squared range lengths sum to at most
\begin{equation}
    \mathcal{O}\left(\frac{\|A\|_F^2}{N}(\sigma T|\theta-\theta'|)^2\right).
\end{equation}
Consequently,
\begin{equation}\label{eq:A_S_diff}
    \Pr\left(
        |2\RE\langle A,\E^{\mathrm{stat}}(\theta)-\E^{\mathrm{stat}}(\theta')\rangle|
        \ge u\sigma T|\theta-\theta'|\|A\|_F
    \right)
    \le2e^{-cNu^2}.
\end{equation}
For a given $\tilde{\eta}$, setting $u=\sqrt{\frac{\log(6|\Theta|^2/\tilde{\eta})}{cN}} =\widetilde{\mathcal O}(N^{-1/2})$ ensures that each resulting uniform inequality holds individually with probability at least $1-\tilde{\eta}$ after a union bound over its grid points. Apply the first estimate Eq.~\eqref{eq:A_S} with $A=\F(\theta)$ and $A=\F(\theta)-\F(\theta')$, and the second estimate Eq.~\eqref{eq:A_S_diff} with $A=\F(\theta')$. All these matrices are deterministic. Union bounds over the corresponding grid points and grid pairs establish the independent-sampling conclusions (Eq.~\eqref{eq:lemma_cross_indep} and Eq.~\eqref{eq:lemma_cross_indep_diff}) with $\E^{\mathrm{stat}}$ in place of $\E$.

\noindent\textbf{Part 2. }\textbf{Shared Time Sampling.} Recall $\mathcal{C}^{\mathrm{stat}}_A(\theta)=2\RE\langle A,\E^{\mathrm{stat}}(\theta)\rangle$. Group the summands by their common sample index:
\begin{equation}
    \mathcal{C}^{\mathrm{stat}}_A(\theta)=\frac1N\sum_{n=1}^N \mathcal{C}^{\mathrm{stat}}_{A,n},
\end{equation}
where
\begin{equation}
    \mathcal{C}^{\mathrm{stat}}_{A,n}
    :=2\RE\sum_{l,r}\overline{A_{l,r}}
    \left(
        Z_{l,r,n}e^{i\theta t_n}
        -\F^{\mathrm{trunc}}_{l,r}(\theta)
    \right).
\end{equation}
The variables $\mathcal{C}^{\mathrm{stat}}_{A,n}$ are independent across $n$ and centered. By the Cauchy--Schwarz inequality, their range lengths are at most
\begin{equation}
    \mathcal{O}\left(\sum_{l,r}|A_{l,r}|\right)
    = \mathcal{O}\left(\sqrt{LR}\|A\|_F\right).
\end{equation}
Hence the squared range lengths of $\mathcal{C}^{\mathrm{stat}}_{A,n}/N$ sum to at most $\mathcal{O}(LR\|A\|_F^2/N)$. Hoeffding's inequality gives, for some absolute constant $c>0$,
\begin{equation}\label{eq:A_S_shared}
    \Pr\left(
        |\mathcal{C}^{\mathrm{stat}}_A(\theta)|\ge u\|A\|_F
    \right)
    \le2\exp\left(-\frac{cNu^2}{LR}\right).
\end{equation}
The same argument for frequency differences adds a factor $\sigma T|\theta-\theta'|$ to each range length, giving
\begin{equation}\label{eq:A_S_shared_diff}
    \Pr\left(
        |2\RE\langle A,\E^{\mathrm{stat}}(\theta)-\E^{\mathrm{stat}}(\theta')\rangle|
        \ge u\sigma T|\theta-\theta'|\|A\|_F
    \right)
    \le2\exp\left(-\frac{cNu^2}{LR}\right).
\end{equation}
For a given $\tilde{\eta}$, setting $u=\sqrt{\frac{LR\log(6|\Theta|^2/\tilde{\eta})}{cN}} =\widetilde{\mathcal O}\left(\sqrt{LR/N}\right)$ ensures that each resulting uniform inequality holds individually with probability at least $1-\tilde{\eta}$ after a union bound over its grid points. Using Eq.~\eqref{eq:A_S_shared} with $A=\F(\theta)$ and $A=\F(\theta)-\F(\theta')$, and Eq.~\eqref{eq:A_S_shared_diff} with $A=\F(\theta')$, and taking union bounds establishes the shared-sampling conclusions for $\E^{\mathrm{stat}}$ in Eqs.~\eqref{eq:lemma_cross_shared} and~\eqref{eq:lemma_cross_shared_diff}.

\noindent\textbf{Part 3. Absorbing the truncation bias.} By Lemma~\ref{lemma:truncation_bias}, with $b_\sigma=e^{-\sigma^2/4}$,
\begin{equation}
    \sup_\theta\|\E^{\mathrm{bias}}(\theta)\|_F
    \le \mathcal{O}\left(\sqrt{LR}\,b_\sigma\right),
    \quad
    \sup_\theta\|\partial_\theta \E^{\mathrm{bias}}(\theta)\|_F
    \le \mathcal{O}\left(T\sqrt{LR}\,b_\sigma\right).
\end{equation}
Thus, for every deterministic $A$,
\begin{equation}
    |2\RE\langle A,\E^{\mathrm{bias}}(\theta)\rangle|
    \le \mathcal{O}\left(\|A\|_F\sqrt{LR}\,b_\sigma\right),
\end{equation}
and
\begin{equation}
    |2\RE\langle A,\E^{\mathrm{bias}}(\theta)-\E^{\mathrm{bias}}(\theta')\rangle|
    \le \mathcal{O}\left(T|\theta-\theta'|\|A\|_F\sqrt{LR}\,b_\sigma\right).
\end{equation}
The truncation condition implies
\begin{equation}
    \sqrt{LR}\,b_\sigma
    \le\frac1{\sqrt{LR}N}
    \le\frac1{\sqrt N}.
\end{equation}
Under the stated lower bound on $\sigma$, these deterministic contributions are absorbed into the bounds for $\E^{\mathrm{stat}}$ in Eqs.~\eqref{eq:lemma_cross_indep}--\eqref{eq:lemma_cross_shared_diff}, including those for $\E^{\mathrm{stat}}(\theta)-\E^{\mathrm{stat}}(\theta')$. Since the bias bounds are deterministic, adding $\E^{\mathrm{bias}}$ requires no additional failure probability. Thus, under each sampling architecture, each corresponding inequality for $\E=\E^{\mathrm{stat}}+\E^{\mathrm{bias}}$ holds individually with probability at least $1-\tilde{\eta}$.
\end{proof}

\begin{lemma}[Generalized Projection Bounds for the Total Error]
\label{lemma:spectral_norm_bound}
Let $P\in\mathbb{C}^{d_1\times L}$ and $Q\in\mathbb{C}^{R\times d_2}$ be fixed deterministic matrices. Suppose that $\sigma\ge\max\{2\sqrt{\log(LRN)},1\}$. Under the \textit{Independent Time Sampling} scheme, for any $\tilde{\eta}\in(0,1)$, the total error $\E(\theta)=G(\theta)-\F(\theta)$ satisfies
\begin{equation}
\label{eq:lemma_error_spectral}
    \sup_{\theta\in\Theta}\|P\E(\theta)Q\|_2
    \le
    \widetilde{\mathcal{O}}\left(
        \frac{
            \|P\|_F\|Q\|_2+\|P\|_2\|Q\|_F
        }{\sqrt N}
    \right)
\end{equation}
with probability at least $1-\tilde{\eta}$.

For the observable estimator, assume that the sample blocks $(t_{l,r,n},t'_{l,r,n},Z^O_{l,r,n})$ are independent across $(l,r,n)$, and the measurement outcome satisfies $|Z^O_{l,r,n}|\le C_O\|O\|_2$ for an absolute constant $C_O$. Then for any $\tilde{\eta}\in(0,1)$, the total error $\E^O(\theta)=G^O(\theta)-\F^O(\theta)$ satisfies
\begin{equation}
\label{eq:lemma_error_spectral_observable}
    \sup_{\theta\in\Theta}\|P\E^O(\theta)Q\|_2
    \le
    \widetilde{\mathcal{O}}\left(
        \|O\|_2
        \frac{
            \|P\|_F\|Q\|_2+\|P\|_2\|Q\|_F
        }{\sqrt N}
    \right)
\end{equation}
with probability at least $1-\tilde{\eta}$.

Here $\widetilde{\mathcal{O}}$ hides logarithmic factors in the grid size, the matrix dimensions, and $\tilde{\eta}^{-1}$.
\end{lemma}

\begin{proof}
We first bound the centered error $\E^{\mathrm{stat}}(\theta)=G(\theta)-\F^{\mathrm{trunc}}(\theta)$ from Eq.~\eqref{eq:defn_E_stat_bias}. For fixed $\theta\in\Theta$,
\begin{equation}
    P\E^{\mathrm{stat}}(\theta)Q=\sum_{l,r,n}\Xi_{l,r,n}(\theta),
\end{equation}
where
\begin{equation}
    \Xi_{l,r,n}(\theta)
    :=
    \frac{
        Z_{l,r,n}e^{i\theta t_{l,r,n}}
        -\F^{\mathrm{trunc}}_{l,r}(\theta)
    }{N}
    P_{:,l}Q_{r,:}.
\end{equation}
Under independent sampling, the matrices $\Xi_{l,r,n}(\theta)$ are independent across $(l,r,n)$ and satisfy $\Ex[\Xi_{l,r,n}(\theta)]=0$. Since the centered scalar residual $\xi_{l,r,n} :=Z_{l,r,n}e^{i\theta t_{l,r,n}} -\F^{\mathrm{trunc}}_{l,r}(\theta)$ satisfies $|\xi_{l,r,n}|\le \sqrt{2}+1$, we define the uniform summand bound
\begin{equation}\label{eq:defn_b_sum}
    b_{\mathrm{sum}}
    :=\frac{\sqrt{2}+1}{N}\|P\|_2\|Q\|_2,
\end{equation}
so that $\|\Xi_{l,r,n}(\theta)\|_2\le b_{\mathrm{sum}}$ almost surely for all $l,r,n,\theta$. Moreover, according to Eq.~\eqref{eq:Ex_E_2},
\begin{equation}
    \Ex[|\xi_{l,r,n}|^2]
    =2-|\F^{\mathrm{trunc}}_{l,r}(\theta)|^2
    \le2.
\end{equation}
Since $P_{:,l}P_{:,l}^{\dagger}$ is positive semidefinite and $\Xi_{l,r,n}=N^{-1}\xi_{l,r,n}P_{:,l}Q_{r,:}$, we obtain
\begin{equation}\label{eq:Ex_Xi_1}
\begin{aligned}
    \sum_{l,r,n}\Ex[
        \Xi_{l,r,n}\Xi_{l,r,n}^{\dagger}]
    &=\frac1{N^2}\sum_{l,r,n}
        \Ex[|\xi_{l,r,n}|^2]
        \|Q_{r,:}\|_2^2P_{:,l}P_{:,l}^{\dagger}\\
    &\preceq\frac2N\sum_{l,r}
        \|Q_{r,:}\|_2^2P_{:,l}P_{:,l}^{\dagger}=\frac2N\|Q\|_F^2PP^{\dagger}.
\end{aligned}
\end{equation}
Similarly,
\begin{equation}\label{eq:Ex_Xi_2}
    \sum_{l,r,n}\Ex
    [\Xi_{l,r,n}^{\dagger}\Xi_{l,r,n}]
    \preceq
    \frac2N\|P\|_F^2Q^{\dagger}Q.
\end{equation}
The matrix variance parameter is therefore bounded by
\begin{equation}\label{eq:defn_var_Bern}
    \sigma_{\mathrm{Bern}}^2:=
    \frac2N\max\left\{
        \|P\|_2^2\|Q\|_F^2,
        \|P\|_F^2\|Q\|_2^2
    \right\}.
\end{equation}
The rectangular matrix Bernstein inequality~\cite[Theorem 6.1.1]{tropp2015matrix}, applied with the summand bound $b_{\mathrm{sum}}$ and variance bound $\sigma_{\mathrm{Bern}}^2$ above, gives
\begin{equation}\label{eq:PEQ_Hoeff}
    \Pr\left(\|P\E^{\mathrm{stat}}(\theta)Q\|_2\ge t\right)
    \le
    (d_1+d_2)
    \exp\left(
        -\frac{t^2/2}{\sigma_{\mathrm{Bern}}^2+b_{\mathrm{sum}}t/3}
    \right).
\end{equation}
Let
\begin{equation}
    \ell:=
    \log\left(
        \frac{4(d_1+d_2)\max\{1,|\Theta|\}}{\tilde{\eta}}
    \right).
\end{equation}
A union bound over $\Theta$ yields
\begin{equation}
    \sup_{\theta\in\Theta}\|P\E^{\mathrm{stat}}(\theta)Q\|_2
    \le \mathcal{O}\left(\sqrt{\sigma_{\mathrm{Bern}}^2\ell}+b_{\mathrm{sum}}\ell\right)
\end{equation}
with failure probability at most $\tilde{\eta}$. Since $N\ge1$ and the Frobenius norm dominates the spectral norm, both terms are bounded by
\begin{equation}\label{eq:PEQ_final}
    \sup_{\theta\in\Theta}\|P\E^{\mathrm{stat}}(\theta)Q\|_2
    \le
    \widetilde{\mathcal{O}}\left(
        \frac{
            \|P\|_F\|Q\|_2+\|P\|_2\|Q\|_F
        }{\sqrt N}
    \right).
\end{equation}

We next bound the truncation bias $\E^{\mathrm{bias}}(\theta)$. Lemma~\ref{lemma:truncation_bias} implies
\begin{equation}\label{eq:PEQ_bias}
\begin{aligned}
    \sup_\theta\|P\E^{\mathrm{bias}}(\theta)Q\|_2
    &\le
    \|P\|_2\|Q\|_2
    \sup_\theta\|\E^{\mathrm{bias}}(\theta)\|_F\\
    &\le
    \mathcal{O}\left(\|P\|_2\|Q\|_2\sqrt{LR}\,e^{-\sigma^2/4}\right)\\
    &\le
    \mathcal{O}\left(\frac{\|P\|_2\|Q\|_2}{\sqrt N}\right),
\end{aligned}
\end{equation}
given $\sigma\ge\max\{2\sqrt{\log(LRN)},1\}$. This is absorbed into the statistical bound above. The triangle inequality proves Eq.~\eqref{eq:lemma_error_spectral}.

For the observable estimator, write $\E^O=\E^{O,\mathrm{stat}}+\E^{O,\mathrm{bias}}$ and use the centered scalar residual
\begin{equation}
    \xi^O_{l,r,n}
    :=Z^O_{l,r,n}e^{i\theta t_{l,r,n}}
      e^{-i\theta t'_{l,r,n}}
      -\F^{O,\mathrm{trunc}}_{l,r}(\theta).
\end{equation}
We assumed that $|Z^O_{l,r,n}|=\mathcal{O}(\|O\|_2)$. Moreover, with $K$ and $K_{\mathrm{trunc}}$ defined in Eq.~\eqref{eq:defn_K}, $\F^{O,\mathrm{trunc}}_{l,r}(\theta)$ is a matrix element of $K_{\mathrm{trunc}}OK_{\mathrm{trunc}}$ between normalized initial states. Since $\|K_{\mathrm{trunc}}\|_2\le1$, we have $|\F^{O,\mathrm{trunc}}_{l,r}(\theta)|\le\|O\|_2$. Therefore,
\begin{equation}
\begin{aligned}
    |\xi^O_{l,r,n}|
    &\le |Z^O_{l,r,n}|
       +|\F^{O,\mathrm{trunc}}_{l,r}(\theta)|
    =\mathcal{O}(\|O\|_2),\\
    \Ex[|\xi^O_{l,r,n}|^2]
    &=\Ex[|Z^O_{l,r,n}|^2]
      -|\F^{O,\mathrm{trunc}}_{l,r}(\theta)|^2
    \le\mathcal{O}(\|O\|_2^2).
\end{aligned}
\end{equation}
Replacing $\xi_{l,r,n}$ by $\xi^O_{l,r,n}$ in the matrix summands therefore changes the uniform summand bound defined in Eq.~\eqref{eq:defn_b_sum} by a factor of $\mathcal{O}(\|O\|_2)$ and the variance bound defined in Eq.~\eqref{eq:defn_var_Bern} by a factor of $\mathcal{O}(\|O\|_2^2)$. Applying the same matrix Bernstein argument and taking a union bound over $\Theta$ yields
\begin{equation}
    \sup_{\theta\in\Theta}
    \|P\E^{O,\mathrm{stat}}(\theta)Q\|_2
    \le
    \widetilde{\mathcal{O}}\left(
        \|O\|_2
        \frac{
            \|P\|_F\|Q\|_2+\|P\|_2\|Q\|_F
        }{\sqrt N}
    \right)
\end{equation}
with failure probability at most $\tilde{\eta}$. Similarly, Eq.~\eqref{eq:PEQ_bias} becomes
\begin{equation}
\begin{aligned}
    \sup_\theta\|P\E^{O,\mathrm{bias}}(\theta)Q\|_2
    &\le \|P\|_2\|Q\|_2
         \sup_\theta\|\E^{O,\mathrm{bias}}(\theta)\|_F\\
    &\le\mathcal{O}\left(
        \|O\|_2\|P\|_2\|Q\|_2\sqrt{LR}\,e^{-\sigma^2/4}
    \right)\\
    &\le\mathcal{O}\left(
        \frac{\|O\|_2\|P\|_2\|Q\|_2}{\sqrt N}
    \right).
\end{aligned}
\end{equation}
Again, this term is absorbed into the statistical bound above. The triangle inequality proves Eq.~\eqref{eq:lemma_error_spectral_observable}.
\end{proof}

\begin{lemma}[Perturbation Bound for Generalized Eigenvalue Problems]\label{lemma:gen_eig_perturbation}
Let $A,B\in\mathbb C^{m\times m}$ be exact square matrices where $B$ is nonsingular. Define $M=B^{-1}A$, and suppose $M$ is diagonalizable such that $M=V\Lambda V^{-1}$, where $\Lambda=\diag(\lambda_1,\dots,\lambda_m)$ contains the exact generalized eigenvalues. Consider the perturbed matrices $\tilde A=A+\Delta A$ and $\tilde B=B+\Delta B$. If $\|\Delta B\|_2<s_{\min}(B)$, then, counting both spectra with algebraic multiplicity, there exists a permutation $\rho\in S_m$ such that the perturbed generalized eigenvalues $\tilde\lambda_1,\dots,\tilde\lambda_m$ satisfy
\begin{equation}\label{eq:lemma_gen_eig_matching}
    \max_{k\in[m]}|\tilde\lambda_k-\lambda_{\rho(k)}|
    \le
    \frac{(2m-1)\kappa_2(V)}
    {s_{\min}(B)-\|\Delta B\|_2}
    \left(\|\Delta A\|_2+\|\Delta B\|_2\|M\|_2\right),
\end{equation}
where $\kappa_2(V)=\|V\|_2\|V^{-1}\|_2$ is the spectral condition number of the eigenbasis, and $S_m$ denotes the permutations of $[m]$.
\end{lemma}

\begin{proof}
Let $\tilde\lambda_k$ be an arbitrary perturbed generalized eigenvalue with its corresponding eigenvector $\tilde w\ne0$, such that
\begin{equation}
    (A+\Delta A)\tilde w
    =\tilde\lambda_k(B+\Delta B)\tilde w.
\end{equation}
Since $\|\Delta B\|_2<s_{\min}(B)$, $\tilde B=B+\Delta B$ is invertible. Left-multiplying by $\tilde B^{-1}$, we obtain the equivalent standard eigenvalue problem
\begin{equation}
    \tilde M\tilde w=\tilde\lambda_k\tilde w,
    \qquad
    \tilde M=(B+\Delta B)^{-1}(A+\Delta A).
\end{equation}

If $\tilde\lambda_k$ is exactly equal to some eigenvalue $\lambda_j$ of $M$, its distance to the exact spectrum is zero. Otherwise,
\begin{equation}
    (\tilde\lambda_k I-M)\tilde w
    =(\tilde M-M)\tilde w.
\end{equation}
Since $\tilde\lambda_k$ is not an eigenvalue of $M$, the matrix $\tilde\lambda_k I-M$ is invertible. Left-multiplying by its inverse yields
\begin{equation}
    \tilde w
    =(\tilde\lambda_k I-M)^{-1}(\tilde M-M)\tilde w.
\end{equation}
Taking the spectral norm on both sides and dividing by $\|\tilde w\|_2>0$, we have
\begin{equation}\label{eq:lemma_resolvent_ineq}
    1\le
    \|(\tilde\lambda_k I-M)^{-1}\|_2
    \|\tilde M-M\|_2.
\end{equation}
Substituting the exact eigendecomposition $M=V\Lambda V^{-1}$, the norm of this inverse matrix is bounded by
\begin{equation}\label{eq:lemma_resolvent_norm}
\begin{aligned}
    \|(\tilde\lambda_k I-M)^{-1}\|_2
    &=\|V(\tilde\lambda_k I-\Lambda)^{-1}V^{-1}\|_2\\
    &\le
    \|V\|_2\|V^{-1}\|_2
    \|(\tilde\lambda_k I-\Lambda)^{-1}\|_2\\
    &=\kappa_2(V)\max_j
    \frac{1}{|\tilde\lambda_k-\lambda_j|}.
\end{aligned}
\end{equation}
Combining Eqs.~\eqref{eq:lemma_resolvent_ineq} and~\eqref{eq:lemma_resolvent_norm}, and including the case where $\tilde\lambda_k$ is an exact eigenvalue, gives
\begin{equation}\label{eq:lemma_spectral_radius}
    \min_j|\tilde\lambda_k-\lambda_j|
    \le\kappa_2(V)\|\tilde M-M\|_2.
\end{equation}

Next, we bound the matrix deviation $\|\tilde M-M\|_2$. Recall that $M=B^{-1}A$ and $\tilde M=(B+\Delta B)^{-1}(A+\Delta A)$. We expand the exact difference as follows:
\begin{equation}
\begin{aligned}
    \tilde M-M
    &=(B+\Delta B)^{-1}(A+\Delta A)-B^{-1}A\\
    &=(B+\Delta B)^{-1}
    \left[A+\Delta A-(B+\Delta B)B^{-1}A\right]\\
    &=(B+\Delta B)^{-1}
    \left(\Delta A-\Delta B\cdot B^{-1}A\right)\\
    &=(B+\Delta B)^{-1}(\Delta A-\Delta B\cdot M).
\end{aligned}
\end{equation}
Applying the triangle inequality and submultiplicativity of the spectral norm, we obtain
\begin{equation}\label{eq:lemma_M_diff_bound}
    \|\tilde M-M\|_2
    \le
    \|(B+\Delta B)^{-1}\|_2
    \left(\|\Delta A\|_2+\|\Delta B\|_2\|M\|_2\right).
\end{equation}
By Weyl's inequality for singular values \cite[Theorem~3.3.16]{horn1991topics},
\begin{equation}
    s_{\min}(B+\Delta B)
    \ge s_{\min}(B)-\|\Delta B\|_2>0.
\end{equation}
Thus,
\begin{equation}\label{eq:lemma_B_inv_bound}
    \|(B+\Delta B)^{-1}\|_2
    =\frac{1}{s_{\min}(B+\Delta B)}
    \le\frac{1}{s_{\min}(B)-\|\Delta B\|_2}.
\end{equation}
Define
\begin{equation}\label{eq:lemma_gen_eig_bound}
    r:=
    \frac{\kappa_2(V)}
    {s_{\min}(B)-\|\Delta B\|_2}
    \left(\|\Delta A\|_2+\|\Delta B\|_2\|M\|_2\right).
\end{equation}
Substituting Eq.~\eqref{eq:lemma_M_diff_bound} and Eq.~\eqref{eq:lemma_B_inv_bound} into Eq.~\eqref{eq:lemma_spectral_radius} gives
\begin{equation}\label{eq:lemma_lambda_dis_r}
    \min_j|\tilde\lambda_k-\lambda_j|
    \le\kappa_2(V)\|\tilde M-M\|_2
    \le r.
\end{equation}

To establish a one-to-one matching, set $E=\tilde M-M$ and consider $M(s)=M+sE$ for $s\in[0,1]$. If $r=0$, then $E=0$ and the conclusion is immediate. Otherwise, applying the same argument as Eq.~\eqref{eq:lemma_spectral_radius} with perturbation $sE$ shows that every eigenvalue $\mu$ of $M(s)$ satisfies
\begin{equation}
    \min_j|\mu-\lambda_j|
    \le s\kappa_2(V)\|E\|_2
    \le r.
\end{equation}
Thus every eigenvalue of $M(s)$ remains in the fixed union of closed disks
\begin{equation}
    \mathcal U
    :=\bigcup_{j=1}^m
    \{z\in\mathbb C:|z-\lambda_j|\le r\}.
\end{equation}

$\mathcal U$ can be partitioned into connected components. Counting algebraic multiplicity, choose continuous eigenvalue paths $\mu_j(s)$ of $M(s)$ such that $\mu_j(0)=\lambda_j$. Their endpoints $\mu_j(1)$ give all perturbed eigenvalues, counting algebraic multiplicity, and hence define a permutation $\rho\in S_m$ pairing each $\tilde\lambda_k$ with $\lambda_{\rho(k)}$. Since each path remains in $\mathcal U$, its two endpoints belong to the same connected component.

Consider a component containing $q$ exact eigenvalues, counting algebraic multiplicity. According to Eq.~\eqref{eq:lemma_lambda_dis_r}, each perturbed eigenvalue $\tilde\lambda_k$ in this component is within distance $r$ of some disk center $\lambda_a$. The centers $\lambda_a$ and $\lambda_{\rho(k)}$ can be joined by a chain of at most $q$ disks in this component, with consecutive disks intersecting. Since the centers of two intersecting disks are at distance at most $2r$, we obtain
\begin{equation}
    |\tilde\lambda_k-\lambda_{\rho(k)}|
    \le r+2(q-1)r
    \le(2m-1)r.
\end{equation}
Substituting the definition of $r$ in Eq.~\eqref{eq:lemma_gen_eig_bound} proves Eq.~\eqref{eq:lemma_gen_eig_matching}.
\end{proof}

\begin{remark}
Lemma~\ref{lemma:gen_eig_perturbation} is a Bauer--Fike-type perturbation bound; see, for example, \cite[Theorem~2.20]{DemmelDumitriuSchneider2026}. We give a self-contained proof that keeps the dependence on $\|\Delta A\|_2$ and $\|\Delta B\|_2$ separate and strengthens spectral inclusion to a one-to-one matching, at the cost of the factor $2m-1$, as needed in our analysis.
\end{remark}

\bibliographystyle{quantum}
\bibliography{refs}

\end{document}